\documentclass[%
 aip,
 sd,%
 amsmath,amssymb,
preprint,%
]{revtex4-1}

\usepackage{graphicx}
\usepackage{dcolumn}
\usepackage{bm}

\usepackage[utf8]{inputenc}
\usepackage[T1]{fontenc}
\usepackage{mathptmx}
\usepackage{ulem}
\usepackage[dvipsnames]{xcolor}
\usepackage{float}

\usepackage{CJKutf8}

\newcommand\Rey{\mbox{\textit{Re}}}  

\newcommand\eg{e.g., }

\newcommand\B{{\textsf B}}
\newcommand\E{{\textsf E}}
\newcommand\A{{\textsf A}}

\begin{document}
\begin{CJK*}{UTF8}{gbsn}
\preprint{AIP/123-QED}

\title{Numerical simulation of rolling pad instability in cuboid liquid metal batteries}
\author{Linyan Xiang（相林言） and Oleg Zikanov}
 \email{zikanov@umich.edu}
\affiliation{University of Michigan - Dearborn, 4901 Evergreen Rd., Dearborn, MI, 48128, USA}%
\date{\today}

\begin{abstract}
The rolling pad instability is caused by electromagnetic interactions in systems of horizontal layers with strongly different electric conductivities. We analyze the instability for a simplified model of a liquid metal battery (LMB), a promising device for large-scale stationary energy storage. Numerical simulations of the flow and the dynamics of electromagnetically coupled interfacial waves are performed using OpenFOAM. The work confirms the earlier conclusions that the instability is a significant factor affecting battery's operation. The critical role played by the ratio between the density differences across the two interfaces is elucidated. It is found that the ratio determines the stability characteristics and the type (symmetrically or antisymmetrically coupled) of dominant interfacial waves.
\end{abstract}

\pacs{47.35.Bb, 47.65.-d}
\keywords{ Liquid metal battery, Interfacial waves, Instability }
\maketitle

\end{CJK*}
\section{Introduction}
\label{sec:intro}

\begin{figure}
\begin{center}
\includegraphics[width=0.49\textwidth]{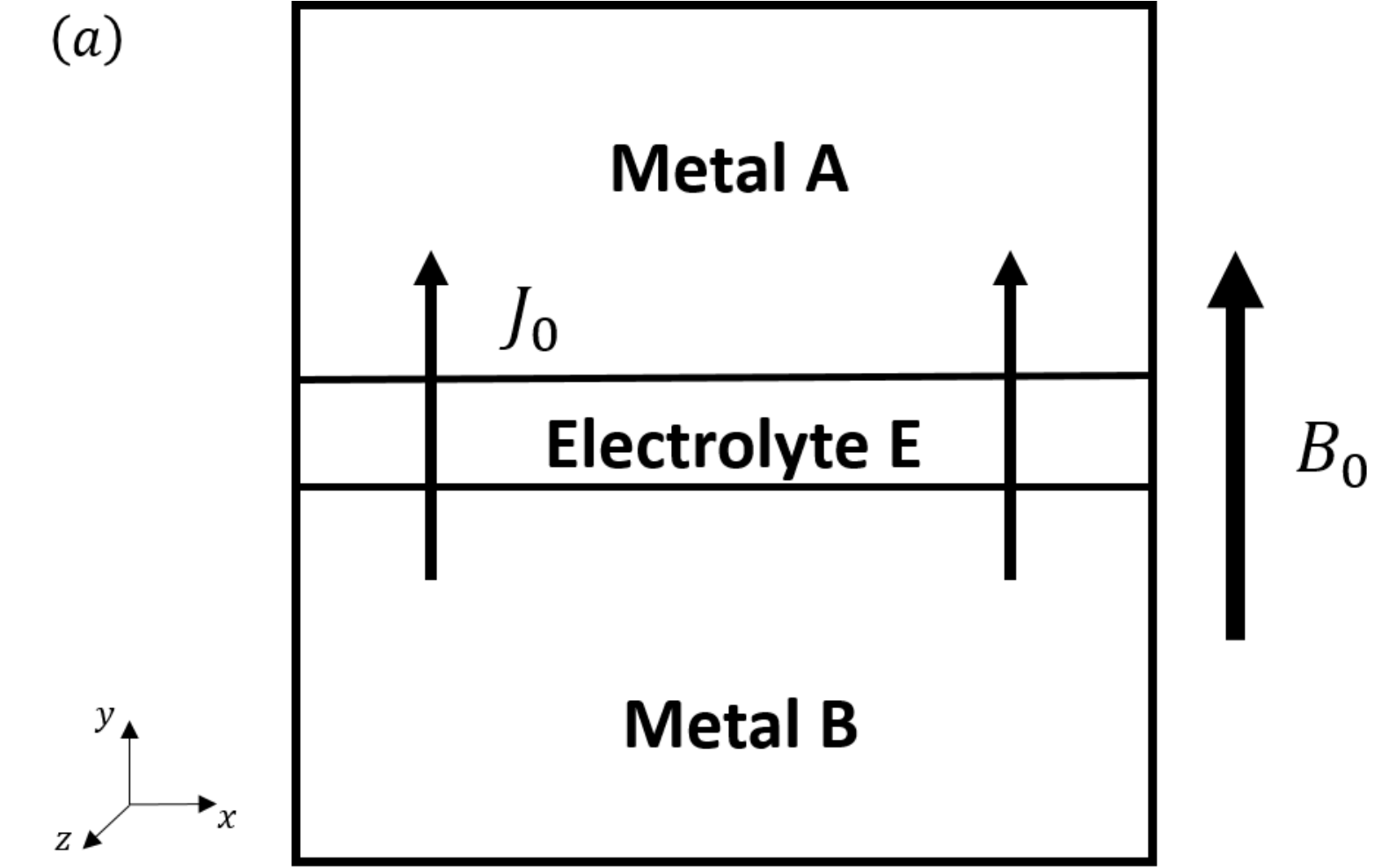}\includegraphics[width=0.49\textwidth]{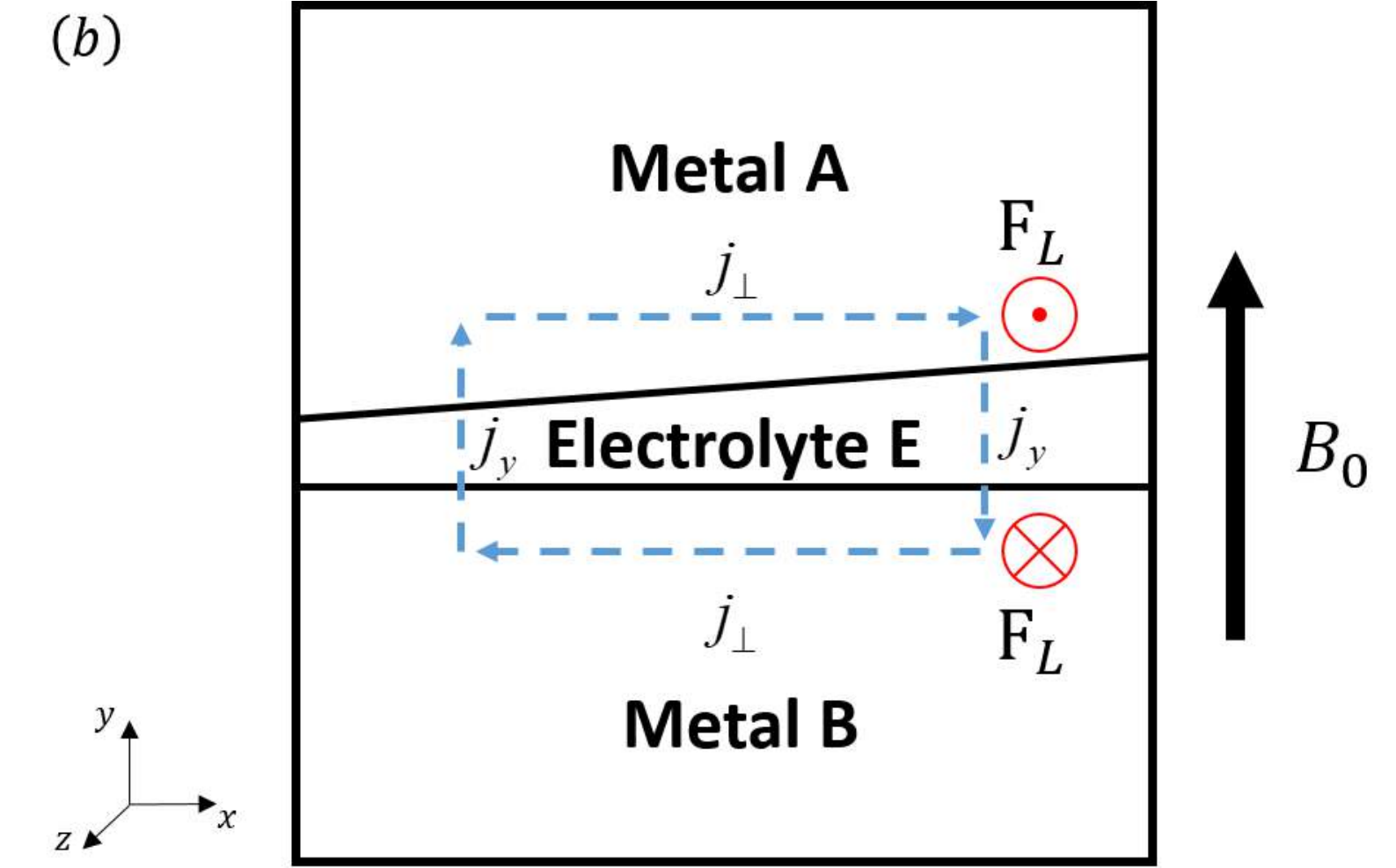}\\
\caption{\emph{(a)} Model of liquid metal battery considered in the study; \emph{(b)} Schematic representation of the physical mechanism of the rolling pad instability discussed in the text.}
\label{fig:geom}
\end{center}
\end{figure}

This paper presents the results of numerical simulations of electromagnetically modified interfacial waves in a liquid metal battery (LMB) -- a promising new device for large-scale stationary energy storage. The principle of the battery's operation and the complex electrochemical, materials, technological, and economical aspects are discussed in  specialized literature, e.g., in Ref.~\onlinecite{Kim:2013}. The following brief description and the illustration in Fig.~\ref{fig:geom} are sufficient for our analysis. The version of the battery considered in the paper can be, in a simplified way, viewed as a cuboid vessel filled with three liquid layers: heavy metalloid {\B }  (\eg Bi, Sb, Zn, or PbSb) at the bottom, molten salt electrolyte {\E }  in the middle, and light metal {\A } (\eg Na, Li, Ca, or Mg) at the top. The system is stably stratified by density, with $\rho_{A}<\rho_{E}<\rho_{B}$. The electrolyte is immiscible with the metals but conductive to positive ions of the light metal. The energy stored in the battery is the difference between the Gibbs free energies of metal {\A } in its pure state and in the state of alloy with the metal \B. 

During discharging, the ions of {\A }  pass through the electrolyte and form the alloy within the bottom layer. During charging, the ions are electrochemically reduced from the alloy and pass into the top layer. The sidewalls of the vessel are electrically insulated. The top and bottom walls contain current collectors for the electrons released or consumed during the charging/discharging reactions. The resulting vertical currents flowing through the battery are quite strong, with the density $J\sim 10^4$ A/m$^2$. The system is maintained at a temperature above the melting point of the three materials, which is facilitated by the strong Joule heating of the electrically poorly conducting electrolyte.

The system presents significant challenges related to hydrodynamic effects, especially when transition from small-scale laboratory prototypes to economically preferable large-scale ($\sim 1$ m$^3$) devices is considered (see Ref.~\onlinecite{Kelley:2018} for a  review). One such challenge is the need for mixing of reactants in the bottom layer. Others are identified as hydrodynamic instabilities expected in large batteries. When considering the instabilities, it is important to realize that if the deformation of the electrolyte-metal interfaces is strong enough to bring the two metals into contact with each other, it creates short circuit between the metal layers and, thus, requires immediate stop of the battery's operation. Any instability leading to such a deformation must be avoided. 

Several instability mechanisms have been identified in recent studies: the Tayler instability,\cite{Seilmayer:2012,Weber:2014,Herreman:2015} thermal convection caused by the Joule heating of the electrolyte\cite{Shen:2016} or by bottom heating,\cite{Kelley:2014}  the Marangoni instability at the electrolyte-metal interfaces,\cite{Koellner:2017} and the rolling pad (also called metal pad) instability.\cite{Zikanov:2015,Weber:2016,Horstmann:2018,Zikanov:2018shallow,Tucs:2018,Molokov:2018,Herreman:2019} The first three instabilities, while interesting and definitively affecting the battery's dynamics, are not expected to result in the just described disruption of operation. Estimates\cite{Herreman:2015,Shen:2016} show that, at realistic battery parameters, the forces leading to the instability are balanced by the gravity force due to the density stratification between the layers so that the growing perturbations saturate at a low amplitude of interface deformation. The situation is less clear in the case of the rolling pad instability considered in this paper.

The mechanism of the rolling pad instability is known from the operation of the Hall-H\'{e}roult aluminum reduction cells, each of which can, in a schematic and drastically simplified way, be presented as the cell in Fig.~\ref{fig:geom} minus the top metal layer (see, \eg Ref.~\onlinecite{Davidson:2016}). Many substantial features of the Hall-H\'{e}roult cell differentiate it from the liquid metal battery: two layers instead of three, a structured top current collector, the effects of gas bubbles and turbulent flow of the melts, low aspect ratio at very large (several m) horizontal dimensions, and so on. Both the systems, however, have the same key feature necessary for the rolling pad instability, namely strong electric currents passing through layers of vastly different electric conductivities. The conductivity of molten salt electrolytes is about four orders of magnitude lower than the conductivity of liquid metals. This means that even a small-amplitude deformation of the metal-electrolyte interface causing a small-amplitude variation of the local electrolyte thickness along the current path leads to significant perturbations of currents and, thus, to significant additional Lorentz forces acting on the melts. 

For the Hall-H\'{e}roult cells, it has been long understood that the Lorentz forces may create melt flows enhancing the interface deformation and leading to an instability in the form of a growing interfacial wave rotating around the cell (the ``rolling pad''). The key mechanism has been identified as the interaction between the  perturbation currents primarily closing within the metal due to its high electric conductivity and the vertical component of the magnetic field always present in the cell.\cite{Sele:1977,Urata:1985,Bojarevics:1994,Sneyd:1994,Davidson:1998,Zikanov:2000,Sun:2004,Sun:2005}

The possibility of a similar scenario in the case of liquid metal battery has been a subject of active research recently. The first approach\cite{Zikanov:2015} was based on the mechanical analogy, in which sloshing of two metal layers was modeled by oscillations of two rectangular solid pendula suspended in an electrolyte. The main result was a demonstration, albeit in the framework of a strongly simplified model, that the rolling pad instability mechanism is realizable in a three-layer system. This conclusion was confirmed using a much more realistic model in the three-dimensional simulations of a cylindrical battery cell.\cite{Weber:2016} The instability was developing at the upper interface, where the density jump was much smaller than at the lower one. It was concluded that the non-dimensional instability parameter derived for the Hall-H\'{e}roult cells \cite{Sele:1977} was also a good indicator of the instability in the battery system. Further analysis was performed  using two-dimensional shallow water models.\cite{Zikanov:2018shallow,Tucs:2018} While limited in their applicability to systems with low aspect ratio and low-to-moderate amplitudes of interface deformation, such models provided the advantage of easier numerical solution and stability analysis than for the three-dimensional model and allowed the authors to study various aspects of the instability. The two other recent studies remain to be mentioned: the linear stability analysis,\cite{Molokov:2018} where the critical role of the non-conducting sidewalls was elucidated, the theoretical and numerical analysis,\cite{Horstmann:2018} where the hydrodynamic coupling between the waves at the two interfaces was analyzed, and the perturbation theory analysis of waves in a cylindrical cell with viscous, Joule dissipation, and capillary effects.\cite{Herreman:2019}

This paper presents the results of the three-dimensional simulations of the rolling pad instability in a liquid metal battery of cubic shape. As in the other studies of the instability,\cite{Zikanov:2015,Weber:2016,Horstmann:2018,Zikanov:2018shallow,Tucs:2018,Molokov:2018,Herreman:2019} a simplified physical model is used, in which the instability effect is separated from the effects of thermal convection, Marangoni, and Tayler instabilities. The model is presented in detail in section \ref{sec:model}. The numerical method described in section \ref{sec:method} is based on the OpenFoam\cite{OpenFOAM} and, in its many features, similar to the method applied in Ref.~\onlinecite{Weber:2016}. The simulation procedure is described in section IV. The results presented in section \ref{sec:results} are quite different from those in Ref.~\onlinecite{Weber:2016} in that they pertain to a cuboid rather than a cylindrical cell (a substantial difference for sloshing flows), extend the analysis to systems of various ratios between the density differences across the interfaces, and apply a broader array of diagnostics tools thus putting the study into the context established by the recent works.\cite{Horstmann:2018,Zikanov:2018shallow,Tucs:2018,Molokov:2018}

\section{Physical model}
\label{sec:model}
The liquid metal battery is a complex system. Its operation involves  various physical mechanisms, some of which are unknown. Therefore, a reasonable first step toward understanding the physics of the battery is to investigate each mechanism individually, isolated from the rest, and in the framework of a simplified model. This approach has been applied to the Tayler instability,\cite{Seilmayer:2012,Weber:2014,Herreman:2015} thermal convection,\cite{Shen:2016,Kelley:2014}   Marangoni instability,\cite{Koellner:2017} and the rolling pad instability.\cite{Zikanov:2015,Weber:2016,Horstmann:2018,Zikanov:2018shallow,Tucs:2018,Molokov:2018,Herreman:2019}

We apply the approach to the rolling pad instability. The simplified physical model is, in its many aspects, similar to that used in Ref.~\onlinecite{Weber:2016} and based on the following main assumptions.
\begin{description}
\item[Simplified geometry] The geometry of the battery is simplified in our case to a cuboid with electrically perfectly insulating sidewalls and the top and bottom walls serving as equipotential current collectors (see Fig.~\ref{fig:geom}a). 
\item[Simplified model of magnetic field] The typical magnetic field within an operating battery is three-dimensional and possibly unsteady. A part of the field, which is primarily horizontal, is induced by the currents flowing through the battery. The other part is truly three-dimensional and is generated by the currents flowing in the supply lines and neighboring battery cells. By the analogy with aluminum reduction cells\cite{Sele:1977,Urata:1985,Bojarevics:1994,Sneyd:1994,Davidson:1998,Zikanov:2000,Sun:2004} and based on the recent results concerning the liquid metal batteries,\cite{Zikanov:2015,Weber:2016,Horstmann:2018,Zikanov:2018shallow,Tucs:2018,Molokov:2018,Herreman:2019} we assume  that the rolling pad instability is caused by the vertical component of the magnetic field. For this reason, we neglect all components but the vertical one, which is approximated by a constant. In addition to removing the effects of the Tayler instability and the electro-vortex flow, this assumption also means that we ignore  the background flow of the melts that would be induced by the interaction between the variable component of the electric current and the three-dimensional magnetic field even when the  interfaces are flat. Only the electromagnetic interactions directly participating in the rolling pad instability are retained in the model.
\item[Absence of solid phase] During the discharge stage, poor mixing at insufficiently high temperature can cause  formation of intermetallic solid in the bottom layer just below the interface.\cite{Kelley:2018} This effect is ignored in our study under the assumption that proper care is taken to avoid this phenomenon  generally detrimental for battery's operation.
\item[Immiscibility] The liquid metals forming the top and bottom layers of the liquid metal battery are assumed to be immiscible with the electrolyte forming the middle layer.
\item[Constant temperature] The liquid metals and the electrolyte are assumed to be maintained at the same constant temperature. Accordingly, the effects of thermal and Marangoni convection are ignored. The approximation is justified by the recent results\cite{Shen:2016,Koellner:2017} showing that under typical operational conditions  the thermal and Marangoni convection are unlikely to perturb the interfaces strongly and to have any significant effect beyond additional mixing within each layer.
\item[Constant fluid properties] Each fluid is assumed to be incompressible, Newtonian and having a constant electrical conductivity. The properties are, however, different among the three fluids. The coefficients of surface tension at the two interfaces are assumed to be constant.
\item[Ignoring mass transport through interfaces] The typical time scale of the processes analyzed in this paper is tens to hundreds of seconds, which is much smaller than the typical time of charging or discharging the battery. This allows us to ignore the change of mass and physical properties of the metal layers due to the possible cross-interface diffusion and  the transport of ions through the electrolyte. The only manifestation of the ion motion is the vertical electric current.
\item[Quasi-static approximation of electromagnetic fields] Using the expected typical velocity $U$ below $0.1$ m/s, the typical size $L=0.1$ m, and the typical maximum electric conductivity $\sigma=4\times 10^6$ S/m, we estimate the magnetic Reynolds number as $\Rey_m\equiv UL\sigma\mu_0=4.5\times 10^{-2}$ (here $\mu_0$ is the magnetic permeability of vacuum). Even at  larger $UL$, the number remains substantially smaller than one. We can apply the quasi-static approximation, according to which the  magnetic field induced by fluid motion is neglected in comparison with the imposed field in the expressions for the Ohm's law and Lorentz force (see, \eg Ref.~\onlinecite{Davidson:2016}).
\end{description}

Further simplifications, such as the shallowness of the system\cite{Zikanov:2018shallow,Tucs:2018} or small amplitude of perturbations\cite{Molokov:2018,Herreman:2019} are not made in our model. Similarly to Ref.~\onlinecite{Weber:2016}, the problem is solved as three-dimensional, non-linear, and time-dependent. 

The assumptions made above imply that in the unperturbed state, in which the interfaces between the layers are perfectly horizontal, the electric current is purely vertical and has constant density:
\begin{equation}
\label{base_current}
\bm{J}_0=J_0\bm{e}_y.
\end{equation}
As will be evident in the following discussion of the governing equations, the body forces and, thus, the melt velocity are identically zero in this case. A deformation of the interfaces causes a variation of the local thickness of the electrolyte, which leads to current perturbations $\bm{j}$, Lorentz forces, and melt flows (see Fig.~\ref{fig:geom}b).

The governing equations are based on the one-fluid approach in its volume-of-fluid version (see, e.g., Ref.~\onlinecite{Prosperetti:2009}). This is an Eulerian method, in which the kinematic and dynamic conditions at the interface are implemented on a fixed grid via direct balances of  fluxes of mass, momentum, and, in the case of our system, electric charge.  The presence of each of the three liquids at a given location is determined by its phase fraction $0\le \alpha_i(\bm{x},t)\le 1$, which is defined as the volume fraction of the liquid $i$ in a computational cell. Only two phase fractions are independent, since at every location we have 
\begin{equation}
\label{phase_sum}
\sum_{i=1}^3 \alpha_i = 1,
\end{equation}
or, in terms of our notation,
\[
\alpha_A+\alpha_E+\alpha_B=1.
\]
The physical properties of the mixture are defined as weighted averages. For the density, dynamic viscosity and electrical conductivity, we have, respectively,
\begin{equation}
\label{properties}
\rho=\sum_{i=1}^3 \alpha_i\rho_i, \: \mu=\sum_{i=1}^3 \alpha_i\mu_i, \: \sigma=\sum_{i=1}^3 \alpha_i\sigma_i. 
\end{equation}

The mass conservation in the absence of diffusion and mass transfer between the phases requires that each phase fraction satisfies
\begin{equation}
\label{phase_eq}
 \frac{{\partial \alpha _i}}{{\partial t}} + \nabla  \cdot (\alpha _i {\bm{U}}) = 0,
\end{equation}
where $\bm{U}$ is the velocity of the mixture assumed to be shared by all fluids. 

The momentum equations is
\begin{equation}
\label{momentum}
\frac{{\partial \left(\rho {\bm{U}}\right)}}{{\partial t}} + \nabla  \cdot (\rho {\bm{UU}}) =  - \nabla p_d + \nabla  \cdot \left[ {\mu (\nabla {\bm{U}} + \nabla {{\bm{U}}^T})} \right] + \bm{f}_L + \bm{f}_{st},
\end{equation}       
where 
\begin{equation}
\label{Lorentz}
    \bm{f}_L = \bm{J} \times \bm{B}
\end{equation}
is the Lorentz force discussed in detail below, and  
\begin{equation}
\label{tension}
\bm{f}_{st}=-\sum_j \sum_{i\ne j} \gamma_{ij}\nabla\cdot\left(\frac{\alpha_j\nabla\alpha_i-\alpha_i\nabla\alpha_j}{\left| \alpha_j\nabla\alpha_i-\alpha_i\nabla\alpha_j \right|} \right)\left( \alpha_j\nabla\alpha_i-\alpha_i\nabla\alpha_j \right)
\end{equation}
is the interface tension force approximated as a volumetric force active in the interface area,\cite{Brackbill:1992} with $\gamma_{ij}$ being the constant coefficient of interfacial tension.

The velocity is subject to the incompressibility constraint:
\begin{equation}
\label{incompr}
\nabla \cdot \bm{U} =0.
\end{equation}
The constraint is satisfied in the standard manner via solution of the Poisson equation for the modified pressure $p_d=p+\rho gy$. 

As we have already mentioned, only the vertical component of the magnetic field is included:
\begin{equation}
\label{magnot}
\bm{B}=B_0\bm{e}_y, \: B_0=const.
\end{equation}
The electric current density is found using the Ohm's law
\begin{equation}
\label{Omh}
    {\bm{J}} = \sigma ( - \nabla \phi  + {\bm{U}} \times {\bm{B}})
\end{equation}
and the charge conservation condition
\begin{equation}
\label{chargecons}
\nabla\cdot \bm{J} = 0,
\end{equation}
which results in the elliptic equation for the electric potential $\phi$:
\begin{equation}
\label{potential}
    \nabla  \cdot (\sigma \nabla \phi ) = \nabla  \cdot (\sigma ({\bm{U}} \times {\bm{B}})).
\end{equation}
The electric current can be viewed as the sum of the base current (\ref{base_current}) and the current perturbations caused by the deformation of the interfaces and the flow of conducting fluids in a constant magnetic field.

The no-slip boundary conditions are imposed at the walls:
\begin{equation}
\label{noslip}
\bm{U}=0 \textrm{ at } x=0,L_x, \:\: y=0,L_y, \:\: z=0,L_z.
\end{equation}
The conditions on the electric potential include those of perfect insulation of the sidewalls:
\begin{equation}
\label{potside}
\frac{\partial \phi}{\partial n}=0 \textrm{ at } x=0,L_x, \:\: z=0,L_z
\end{equation}
and those of equipotential top and bottom surfaces:
\begin{equation}
\label{pothor}
\phi=0 \textrm{ at } y=0, \:\: \phi=\phi_0=const \textrm{ at } y=L_y.
\end{equation}
The potential drop is calculated for the unperturbed state of the battery with constant thicknesses $H_A^0$, $H_E^0$, $H_B^0$ of the layers as 
\begin{equation}
\label{potzero}
\phi_0=-J_0\left(\sigma_A H_A^0+\sigma_E H_E^0+\sigma_B H_B^0 \right).
\end{equation}

The condition (\ref{pothor}) is an approximation based on the assumption that the perturbations of the electric current form loops closing entirely within the battery cell. The same condition was used in Ref.~\onlinecite{Weber:2016}. It is slightly different from the condition applied, \eg in Ref.~\onlinecite{Zikanov:2018shallow}, where the potential drop was adjusted continuously to account for the change of the total electric resistance of the cell. The difference only appears at strong deformations of the interfaces and has no critical impact on the instability even in that case.

\section{Numerical method}
\label{sec:method}
Time-dependent three-dimensional solutions of the problem are computed using the finite-volume CFD tool OpenFOAM.\cite{OpenFOAM} The method is similar to that applied to the rolling pad instability in a cylindrical battery cell in Ref.~\onlinecite{Weber:2016}. It is based on the solver for simulations of incompressible multiphase flows \textit{multiphaseInterFoam}\cite{OpenFOAM,Ubbink:1997,Deshpande:2012} modified to include  electromagnetic fields and account for  sharp variations of the electrical conductivity.

The finite-volume technique is described in detail in, \eg Ref.~\onlinecite{Deshpande:2012}, so only a brief description is given here. The spatial discretization is of the second order. The first-order simple implicit scheme is applied for time discretization. The PISO algorithm\cite{Issa:1986} is applied to iteratively solve the momentum and pressure equations. The advection equation for the phase fraction (\ref{phase_eq}) is solved using the implicit Euler time discretization and the spatial discretization by the MULES scheme\cite{marquez:2013} based on  the upwind interpolation outside the interface areas and a delimiter-based combination of the linear and upwind interpolations inside it. The use of the compressive fluxes allows the solver to restrict the numerically diffused interface thickness to a few (typically two or three in our simulations) cell widths.

The discretization of the electromagnetic equations is based on the conservative scheme.\cite{Ni1:2007,Ni2:2007} The potential equation (\ref{potential}) integrated over the finite volume cell $\Omega$ with faces $\bm{A}_f$, $f=1,\ldots,n$ is written as
 \begin{equation}
 \label{pot_fv}
      \sum\limits_{f = 1}^n {[{\sigma _f}{{\bm{A}}_f} \cdot {{(\nabla \phi )}_f}} ] = \sum\limits_{f = 1}^n {[{\sigma _f}{{\bm{A}}_f} \cdot {{({\bm{U}} \times {\bm{B}})}_f}]}, 
 \end{equation}
 where the subscript $f$ indicates the values at the midpoint of the cell faces $f$, $(\nabla \phi )_f$ is the face-normal gradient evaluated by central differences, and ${({\bm{U}} \times {\bm{B}})}_f$ is computed using linear interpolation. The face-normal components of the electric current are calculated at the cell faces using the Ohm's law as
\begin{equation}
\label{Ohm_fv}
    \bm{A}_f \cdot \bm{J}_f = \sigma_f \left[ - \bm{A}_f \cdot (\nabla \phi )_f  + \bm{A}_f \cdot (\bm{U} \times \bm{B})_f\right]
\end{equation}
and then interpolated linearly to the cell's centers to be used in the evaluation of the Lorentz force (\ref{Lorentz}). 

The extreme (four orders of magnitude) difference between the electrical conductivities of the electrolyte and the metals presents a challenge to accurate calculation of the electric potential and currents, in particular to satisfying the charge conservation condition (\ref{chargecons}) in the interface area. We have determined that the difficulties are resolved by two adjustments. One is the use of harmonic interpolation for the evaluation of the electric conductivity of the multiphase mixture $\sigma$ (see (\ref{properties})) at the cell faces in (\ref{pot_fv}) and (\ref{Ohm_fv}):
\begin{equation}
\label{harmonic}
{\sigma _f} = {\left( {\frac{{{\raise0.7ex\hbox{${{d_N}}$} \!\mathord{\left/
 {\vphantom {{{d_N}} d}}\right.\kern-\nulldelimiterspace}
\!\lower0.7ex\hbox{$d$}}}}{{{\sigma _N}}} + \frac{{{\raise0.7ex\hbox{${{d_P}}$} \!\mathord{\left/
 {\vphantom {{{d_P}} d}}\right.\kern-\nulldelimiterspace}
\!\lower0.7ex\hbox{$d$}}}}{{{\sigma _P}}}} \right)^{ - 1}},
\end{equation}
where $P$ and $N$ are the two neighboring cell centers, $d=|PN|$ is the distance between them, and $d_N$ and $d_P$ are the distances from cell centers to the midpoint $f$ of the cell face separating them from each other ($d_N=|fN|$, $d_P=|fP|$). As discussed in the context of heat conduction in Ref.~\onlinecite{Patankar:1980}, the harmonic interpolation avoids the unacceptably large error in the evaluation of fluxes at cell faces in the situation of strong variation of the transport coefficient within the cell. 

We have also found that satisfying the charge conservation and avoiding unacceptably large spurious electric currents near the interfaces requires convergence of the iterative solution of (\ref{pot_fv}) to very low tolerance, at the level of $10^{-20}$. In the simulations presented in section \ref{sec:results} of this paper, this  typically required about 500 iterations. 

The numerical model was verified by repeating several results presented in the earlier three-dimensional simulations of the rolling pad instability in a cylindrical battery.\cite{Weber:2016} Consistently good agreement was found. In particular, we were able to accurately reproduce the period of interfacial waves in an unstable system and the value of the threshold magnetic field above which the system experiences short circuit (see Fig. 10b of Ref.~\onlinecite{Weber:2016}).

\section{Simulation procedure}
\label{sec:procedure}

\subsection{Problem setting and diagnostics}
\label{sec:setting}
The interior of the battery is modeled as a cube of side length $L_x=L_y=L_z=0.1$ m (see Fig.~\ref{fig:geom}). This is approximately the size of the currently tested commercial LMB prototypes, but substantially smaller than the size ($\sim 1$ m) of the envisioned large-scale batteries.  Various values of the unperturbed thickness of the electrolyte layer $H_E^0$ are used in the simulations. The layer is always located in the middle of the cell, so the unperturbed thicknesses of the metal layers are  $H_A^0=H_B^0=\left(L_y-H_E^0\right)/2$.

The square horizontal cross-section of the battery cell differentiates our work from the recent studies, where cells of circular\cite{Weber:2016} or rectangular\cite{Zikanov:2018shallow,Tucs:2018} cross-sections were considered. The  analogy with the Hall-H\'{e}roult reduction cells leads us to expect the instability characterized by \emph{(i)} the threshold much lower than for rectangular cells, and \emph{(ii)} possible spatial complexity and multitude of instability modes.\cite{Davidson:1998} We will return to these expectations in the following analysis.

The physical properties of the materials corresponding to the Mg-Sb battery with NaCl-KCl-MgCl$_2$ electrolyte\cite{Kim:2013} (see table \ref{table1}) are used in some of the simulations. This relatively well studied\cite{Zikanov:2015,Weber:2016,Horstmann:2018,Zikanov:2018shallow,Tucs:2018,Molokov:2018} combination of material properties is an example of the system with small density difference across the upper interface $\Delta \rho_A\equiv\rho_E-\rho_A\ll \Delta \rho_B\equiv\rho_B-\rho_E$. In order to see how the instability changes in batteries with other ratios between $\Delta \rho_A$ and $\Delta \rho_B$, the electrolyte density $\rho_E$ is varied in the other simulations, while the rest of the properties are kept the same.

The imposed vertical current $J_0=7850$ A/m$^2$ is used in all the simulations. The value $\phi_0$ of the potential at the top wall is computed according to (\ref{potzero}). The positive value of $J_0$ implies that the process of battery charging is simulated. Based on the physical mechanism established in earlier studies\cite{Zikanov:2015,Weber:2016,Horstmann:2018,Zikanov:2018shallow,Tucs:2018,Molokov:2018} and confirmed in our work, the instability considered in the framework of our simplified model is expected to have the same behavior in the course of charging, except for the opposite direction of the precession of interfacial waves.\cite{Zikanov:2018shallow}

\begin{table}
\centering
\begin{tabular}{c|c|c|c}
                           &   $\rho$ & $\nu$ & $\sigma$ \\ \hline
unit                       &  kg/m$^{3}$ & m$^{2}$/s     & S/m      \\ \hline
Metal A    &  1577  & $6.70 \times {10^{ - 7}}$ & $3.62 \times {10^{  6}}$ \\ \hline
Electrolyte E        &  1715  & $6.80\times {10^{ - 7}}$ & 80       \\ \hline
Metal B &  6270  & $1.96\times {10^{ - 7}}$ & $8.66\times {10^{  5}}$
\end{tabular}
\caption{Physical properties of a typical Mg-Sb LMB\cite{Kim:2013} used in the simulations. Density $\rho$, kinematic viscosity $\nu=\mu/\rho$ and electric conductivity $\sigma$ are shown for each layer. As discussed in the text, the electrolyte density $\rho_E$ is different from the table value in some simulations. The other parameters used in the simulations include the dimensions of the cell $L_x=L_y=L_z=0.1$ m, the unperturbed current density $J_0$=7850 A/m$^2$ and the interfacial tension coefficients $\gamma_{A-E}=0.19$ N/m, $\gamma_{B-E}=0.095$ N/m.}
\label{table1}
\end{table}

Each simulation starts with the initial state, in which $\bm{U}=0$, $\bm{J}=J_0\bm{e}_y$, and $\phi$ is given by the piecewise-linear function of $y$ derived as a solution of (\ref{potential}) with the boundary conditions (\ref{potside})--(\ref{potzero}). A weak perturbation of the electrolyte thickness is added to initiate the flow. It has the shape
\begin{equation}\label{deta_1}
    \Delta \eta^A=0.02 z,
\end{equation}
which implies the maximum amplitude of 2 mm. Each simulation is continued forward in time until the behavior of the flow becomes clear.

To investigate the effect of the initial conditions, we have repeated several simulations of unstable systems with different shapes of the initial perturbations as well as with the initial perturbations applied to the bottom rather than top interface. As discussed and illustrated in section \ref{sec:results_param}, these changes affect the first stage of the flow evolution, but not the qualitative and quantitative characteristics of the fully developed waves. 

The solutions are visualized using the snapshots of the interfaces, distributions of velocity and Lorentz force fields. The profiles of the upper and lower interfaces $y=\eta^A(x,z,t)$ and $y=\eta^B(x,z,t)$ are determined during the simulations as the locations where, respectively,  $\alpha_B=0.5$ and $\alpha_A=0.5$. We also use the distributions of the vertical electric current density $J_y(x,z,t)$ in the mid-plane of the electrolyte layer $y=0.05$ m, which is nearly reciprocal to the local thickness of the electrolyte.  

The quantitative characteristics of the interfacial waves (period, growth or decay of amplitude) are determined from the time signals of $\eta^A$, $\eta^B$ and of  the velocity and  electric current components at several points in the metal and electrolyte layers. The points are chosen along the vertical line at $x$=0.05 m, $z=0.016$ m, although any other line would give similar characteristics. 

As in the previous simulations of the instability,\cite{Weber:2016,Zikanov:2018shallow,Tucs:2018} the problem is solved in dimensional units. This is justified by the multitude of relevant physical properties and the general desire to keep the analysis focused on the realistic values of these properties. At the same time, it is useful and interesting to consider the possible forms of the non-dimensional parameters controlling the instability. From the three-dimensional analysis of the cylindrical cells\cite{Weber:2016} and the shallow water analysis of rectangular cells\cite{Zikanov:2018shallow} we conclude that in the case of small density difference across the upper interface the key parameters are likely to be the shape of the horizontal cross-section of the cell and the parameter\cite{Sele:1977}
\begin{equation}
\label{beta}
\beta=\frac{IB_0}{g\Delta \rho_A H_E^0 H_A^0},
\end{equation}
where $I$ is the total current through the battery ($=J_0L_xL_z$ in our system). This non-dimensional parameter was originally introduced by Sele\cite{Sele:1977} to characterize the metal pad rolling in aluminum reduction cells. It estimates the typical ratio between the destabilizing Lorentz force and the stabilizing effect of density stratification. A similar parameter with $\Delta \rho_B$ in place of $\Delta \rho_A$ can be formally introduced in the case of small density difference across the lower interface. There is no reason to expect the parameter to accurately describe the instability in the cases where the density differences across the two interfaces are comparable to each other.

The role of the other non-dimensional parameters, such as the Reynolds, Hartmann, and Weber numbers remains largely unknown. We will leave this question to future studies, but note that the expected large typical length scale of growing perturbations suggests that these parameters somewhat affect the growth rate and the shape of the unstable modes, but do not change them in a profound way.

The results reported in this paper can be recalculated into non-dimensional form using, for example, the velocity of a large-scale gravitational wave on the upper interface
\begin{equation}
\label{velocity}
U_0=\left[ g\left(\rho_E-\rho_A\right)\left(\rho_A\left(H_A^0\right)^{-1}+\rho_E\left(H_E^0\right)^{-1}\right)^{-1}  \right]^{1/2}
\end{equation}
as the typical velocity scale, and $L_x$, $L_x/U_0$, $\rho_EU_0^2$, $J_0$, and $J_0L_x\sigma_A^{-1}$ as the typical scales of length, time, pressure, electric current density, and electric potential, respectively.\cite{Zikanov:2018shallow}

\subsection{Computational grid and grid sensitivity tests}
\label{sec:grid}
Constant time step and a structured Cartesian grid are used in all the simulations. The design of the grid is based on the following considerations. Firstly, we expect particularly strong vertical gradients of velocity, electric potential, and other variables in the areas around the interfaces, especially when the interfaces are significantly deformed. Secondly and more importantly, accuracy of determining the position of the interfaces is critical for the overall accuracy of the model. An error in the evaluation of the local relative change of $H_E$ leads to an approximately $\sigma_A/\sigma_E\sim \sigma_B/\sigma_E \gg 1$ times bigger error in the evaluation of the  perturbation currents and, thus, of the Lorentz forces that drive the instability. 
An accurate volume-of-fluid simulation of the interface movement requires fine resolution in the direction perpendicular to the interface. Since long waves are expected to dominate in the flows created by the rolling pad instability, this means a requirement of fine resolution in the vertical direction. 

Constant grid steps $\Delta x=\Delta z$ are used in the horizontal plane. In the $y$-direction, a constant small grid step $\Delta y_{min}$ is used within the horizontal layer that 
extends on both sides sufficiently far beyond the area of expected locations of the two interfaces. The position of this layer is not necessarily symmetric with respect to the middle of the cell, since in some of the simulations, only one interface is expected to move significantly. In the metal domains below and above this layer, $\Delta y$ increases in  geometric progression with the expansion rate chosen so that it reaches $\Delta y_{max}=1$ mm at the bottom and top boundaries of the cell.

\begin{figure}
\begin{center}
\includegraphics[width=0.45\textwidth]{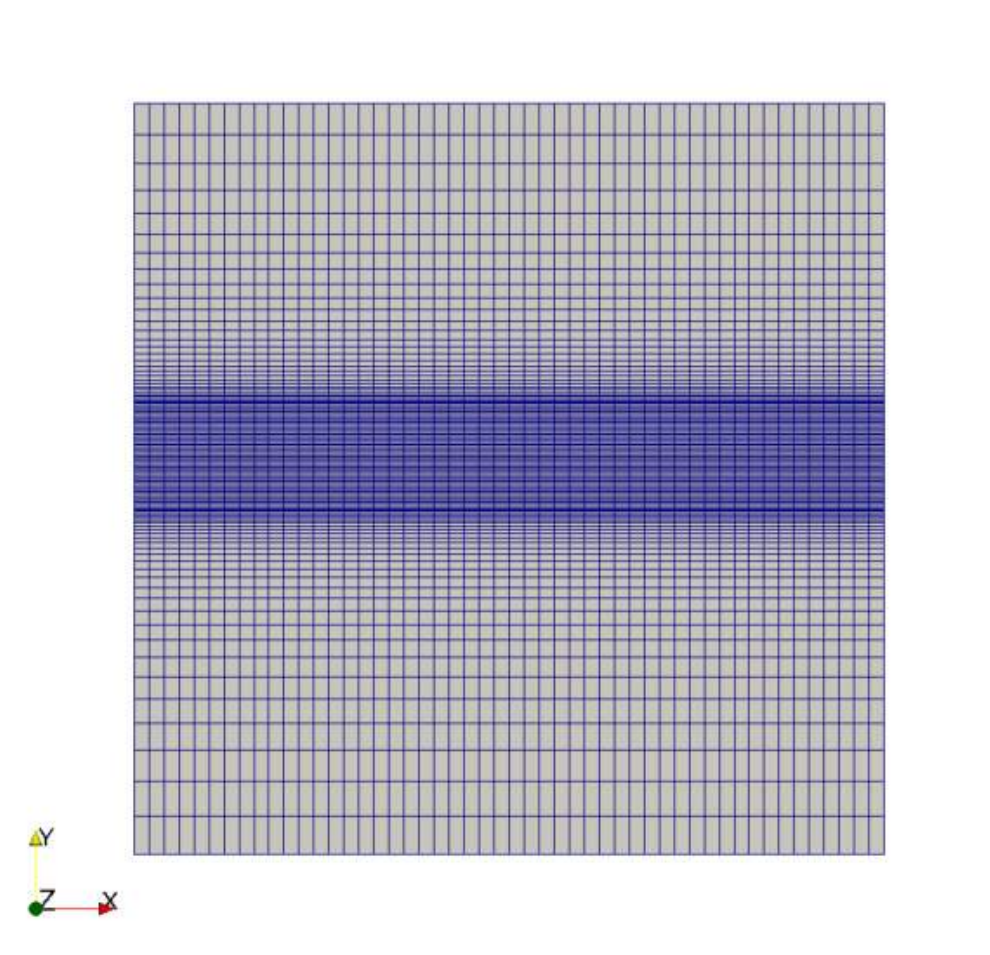}\\
\caption{An example of the vertical cross-section of the optimal grid determined in the course of the grid sensitivity study.\cite{Xiang:2018} The zone of refinement is positioned so that it fully covers the electrolyte layer and the areas of interface movement. The grid in the figure was used for  the simulations of the case 5 (see table II).}
\label{fig2}
\end{center}
\end{figure}

Grid sensitivity tests were performed to determine the optimal values of $\Delta x=\Delta z$, $\Delta y_{min}$, and the time step $\Delta t$.  The grid was deemed sufficient if further refinement did not result in significant changes of the time period and peak values of the signals of velocity and electric current density. A detailed description of the results can be found in Ref.~\onlinecite{Xiang:2018}. In summary, the spatial discretization with $\Delta y_{min}=0.2$ mm and $\Delta x=\Delta z=2$ mm is sufficient for accurate simulations (see an example in Fig.~\ref{fig2}). The situation with the effect of the time step $\Delta t$ is more complex. We have found that, at the spatial grid described above, $\Delta t=4\times 10^{-3}$ s is sufficient to avoid numerical instabilities and accurately predict the velocity field. At the same time, the signals of the vertical electric current $J_y$ inside and near the electrolyte, which is very sensitive to the accuracy of reproduction of interface movement, only become truly $\Delta t$-independent at $\Delta t$ below  $2\times 10^{-3}$ s. Furthermore, $\Delta t$ must be decreased in strongly unstable regimes, when the amplitudes of the interface deformation, electric current perturbations, velocities and velocity gradients become large, and the electrolyte layer is about to be ruptured. The strategy in the simulations was to use $\Delta t=4\times 10^{-3}$ s, while periodically conducting verification runs at $\Delta t=2\times 10^{-3}$ s and reducing $\Delta t$ in the cases of rapidly growing perturbations.

\section{Results}
\label{sec:results}
\subsection{Summary of completed simulations}
\label{sec:results_summary}
The completed simulations are summarized in Table II. Each case is identified by the values of the parameters varied in the study: the imposed vertical magnetic field $B_0$, the unperturbed thickness of electrolyte $H^0_E$, and the electrolyte density  $\rho_E$. We also show the values of the non-dimensional parameter $\beta$ (see (\ref{beta})) and the main period of the wave if it can be determined from the time signals of the flow variables. The last three columns show the characteristics of the flow regimes introduced later in the text: the stability identification, the type of wave coupling, and the typical ratio between the wave amplitudes at the upper and lower interfaces.

\begin{table}
\centering
\label{table2}
\begin{tabular}{c|c|c|c|c|c|c|c|c|c}
 Case & $B_0\ \left[ mT \right]$ &  $ H_E^0\ \left[ mm \right]$   &  $ \rho _E\ \left[ {kg}/{m^3} \right]$      &  $\beta$     &   $T\ \left[ s \right] $  & stability & coupling & ${\Delta \eta^A}/{\Delta \eta^B}$        \\ \hline
  1    & 1     & 5         & 1715        & 0.244 &      & stable & A &  40\\
  2    & 3     & 5         & 1715        & 0.732  &     &  stable & A &   40\\
  3    & 5     & 5         & 1715        & 1.22   & 3.3   & saturation & A & 40    \\
  4    & 7.5   & 5         & 1715        & 1.833  & 2.5    & saturation & A & 15  \\
  5    & 10    & 5         & 1715        & 2.444 & 3.3     & saturation & A & 32   \\
  6    & 15    & 5         & 1715       & 3.666   & 2.9    & saturation & A & 32    \\
  7    & 20    & 5         & 1715        & 4.888  & 3.3    & rupture   & A & 40   \\
  8    & 26    & 5         & 1715        & 6.354   & 3.4   & rupture   & A & 32    \\
  9    & 15    & 3         & 1715       & 5.984   & 4.4    & rupture & A & 24     \\ 
  10   & 15    & 7         & 1715        & 2.675   & 3.0  & rupture & A & 31    \\
  11   & 15    & 15        & 1715        & 1.496  & 2.3   & rupture & A & 38      \\
  12   & 15    & 20        & 1715        & 1.122  & 2.0   & rupture & A & 57       \\ 
  13   & 15    & 5         & 1646        & 7.332  & 4.0    & rupture   & A & 40       \\
  14   & 15    & 5         & 3452.2    & 0.27     & 0.5   & stable & S & 1.0         \\  
  15   & 15    & 3         & 3452.2    & 0.45     & 0.5   & stable & S & 1.0         \\
  16   & 50    & 3         & 3452.2     & 1.5    & 0.32   & stable & S & 1.0         \\
  17   & 200   & 3         & 3452.2      & 6    &   0.24   & stable & S & 1.0         \\
  18   & 7.5   & 5         & 5994        & 0.917  & 0.5/4.7    & saturation & S/A & 1.3         \\
  19   & 15    & 5         & 5994       & 1.833  & 0.5/4.6     & saturation & S/A & 1.2         \\
  20   & 30    & 5         & 5994       & 3.666    & 0.5/4.5   & rupture & S/A & 1.1/0.25         \\
  21   & 50    & 5         & 5994      & 6.11     & 0.5/4.4    & rupture   & S/A & 1.1/0.10       
\end{tabular}
\caption{Summary of completed simulations. Values of the imposed magnetic field $B_0$, unperturbed electrolyte thickness $H_E^0$, and electrolyte density $\rho_E$ are shown. The other physical properties remain the same in all simulations (see Table I). Also shown in the table are the value of the non-dimensional parameter (\ref{beta}) (computed with $\Delta \rho_A$ in runs 1-17, and $\Delta \rho_B$ in runs 18-21)  and the characteristics of the flow: time period of the unstable wave, stability identification, and the type of coupling (`A' denotes the antisymmetric coupling, while `S' denotes the symmetric, in-phase coupling, both being described in detail in the text) and the typical ratio of amplitudes of the waves at the upper and lower interfaces. Two digits are shown for the amplitude ratios, but in some cases, in particular in flows with rupture, only the first digit can be reliably determined.  }
\end{table}

\subsection{Examples of flow behavior}
\label{sec:results_examples}
In this section, we show three examples typifying the three principal scenarios possible in the system. The electrolyte density is $\rho_E=1715$ kg/m$^3$ (a Mg-Sb battery), which corresponds to $\Delta \rho_A\ll \Delta \rho_B$, and allows us to consider the solutions in the context of earlier studies.\cite{Weber:2016,Zikanov:2018shallow,Tucs:2018,Molokov:2018}

\begin{figure}
\begin{center}
\includegraphics[width=0.5\textwidth]{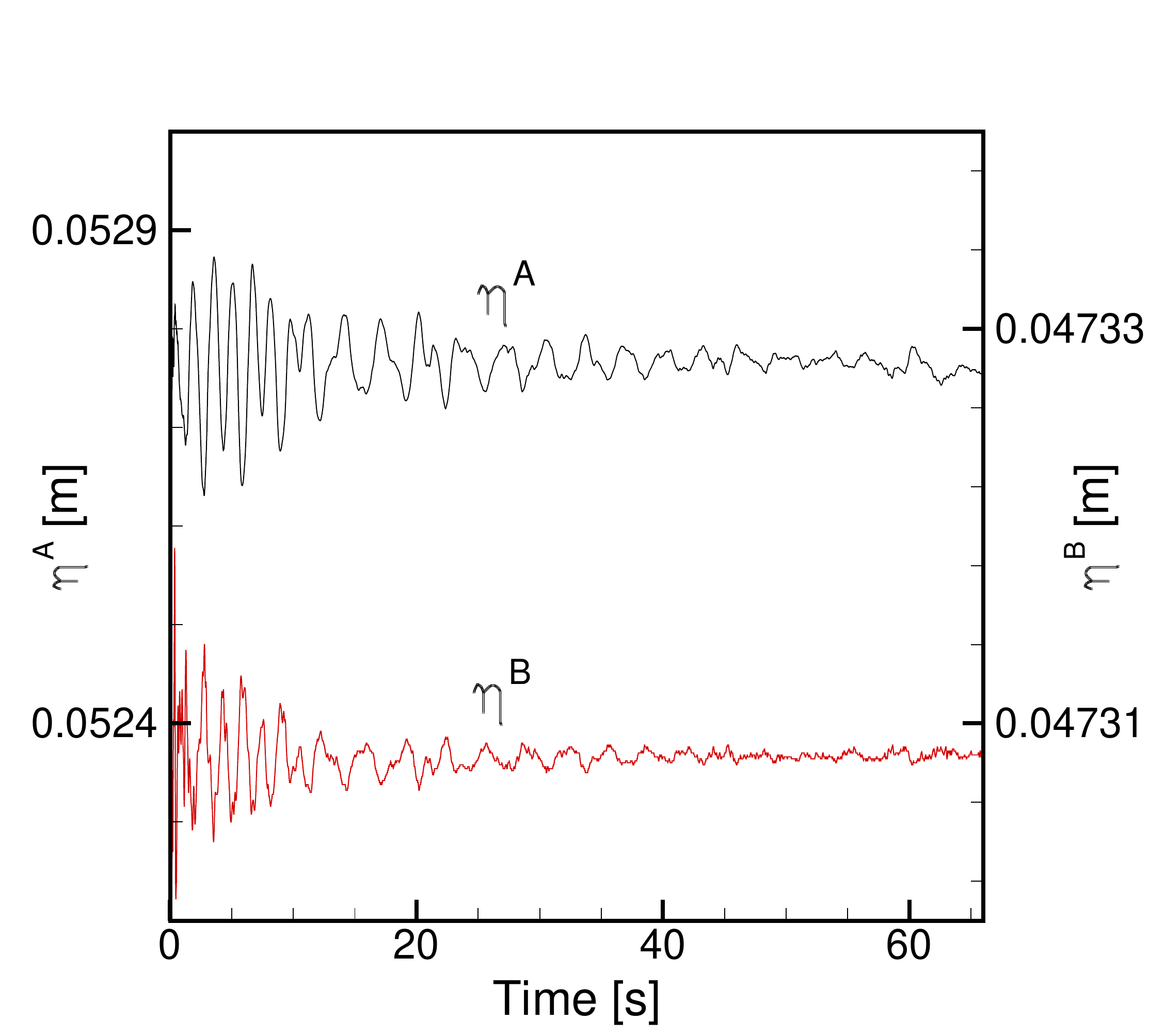}\\
\caption{Stable flow at $H_E^0=5$ mm, $B_0=1$ mT, $\rho_E=1715$ kg/m$^3$ (case 1 in table II). The locations of the upper ($\eta^A$) and lower ($\eta^B$) interfaces at $x=0.05$ m, $z=0.016$ are shown as functions of time. Note that the scales used for $\eta^A$ and $\eta^B$ are strongly different.}
\label{fig3}
\end{center}
\end{figure}
The first example is the  flow at $H_E^0=5$ mm, $B_0=1$ mT, and $\rho_E=1715$ kg/m$^3$ (case 1 in table II). We see in Fig.~\ref{fig3} that the amplitudes of the interfacial waves created by the initial perturbation decay with time. Similar decay is observed for all the other flow variables, such as the kinetic energy, perturbations of electric currents, etc. This allows us to identify the system as stable.

\begin{figure}
\begin{center}
\includegraphics[width=0.4\textwidth]{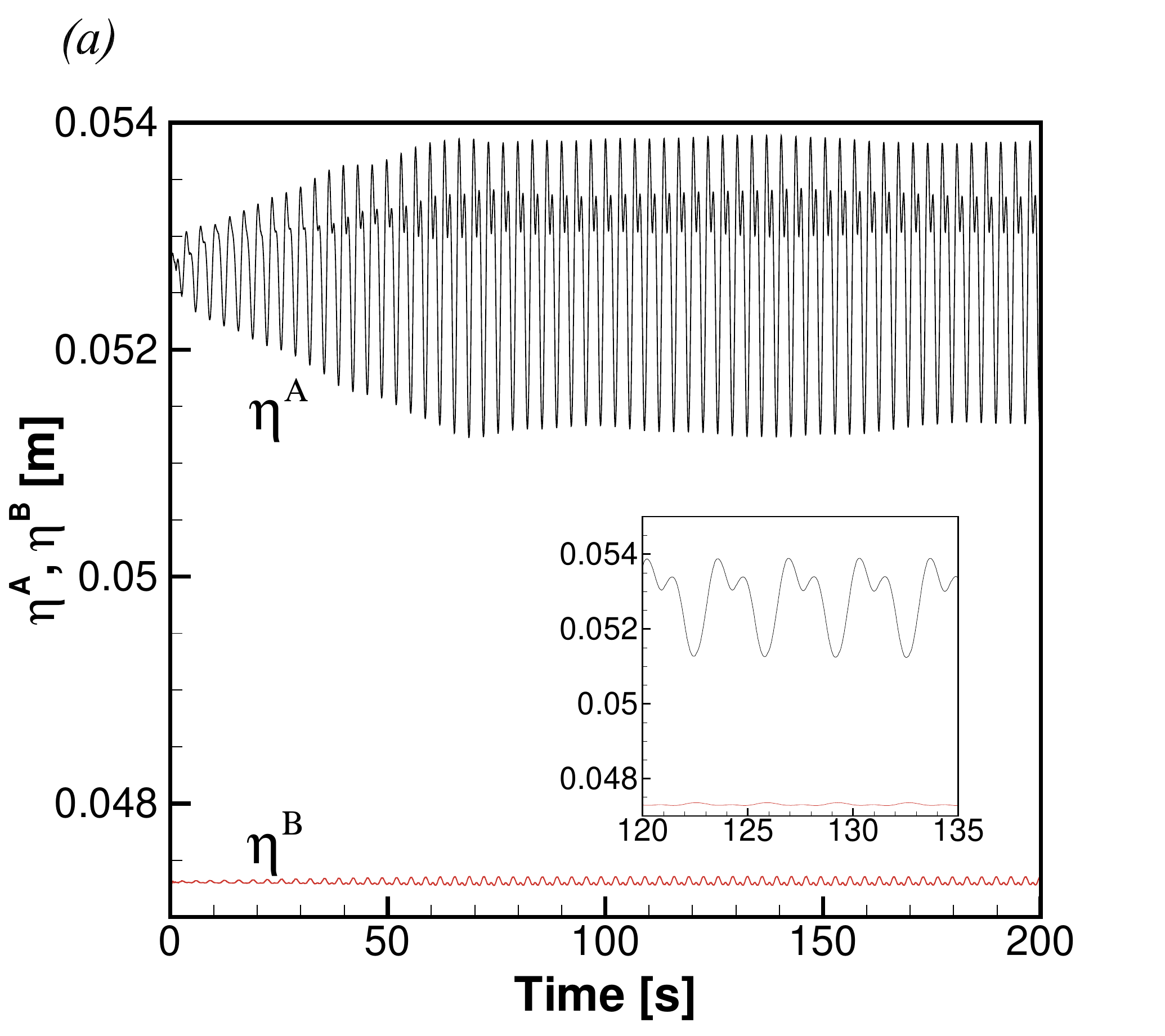}\includegraphics[width=0.4\textwidth]{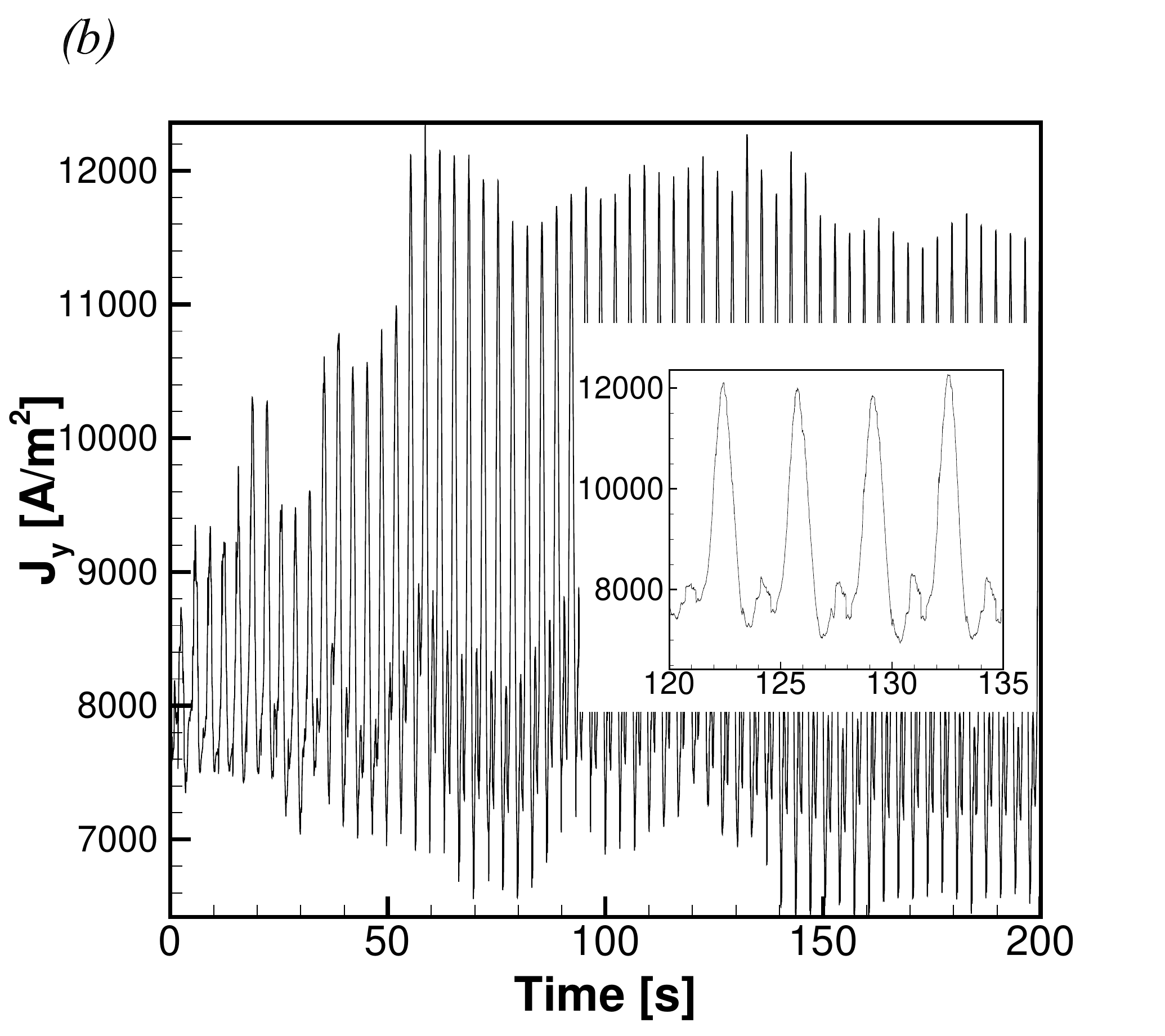}\\
\includegraphics[width=0.4\textwidth]{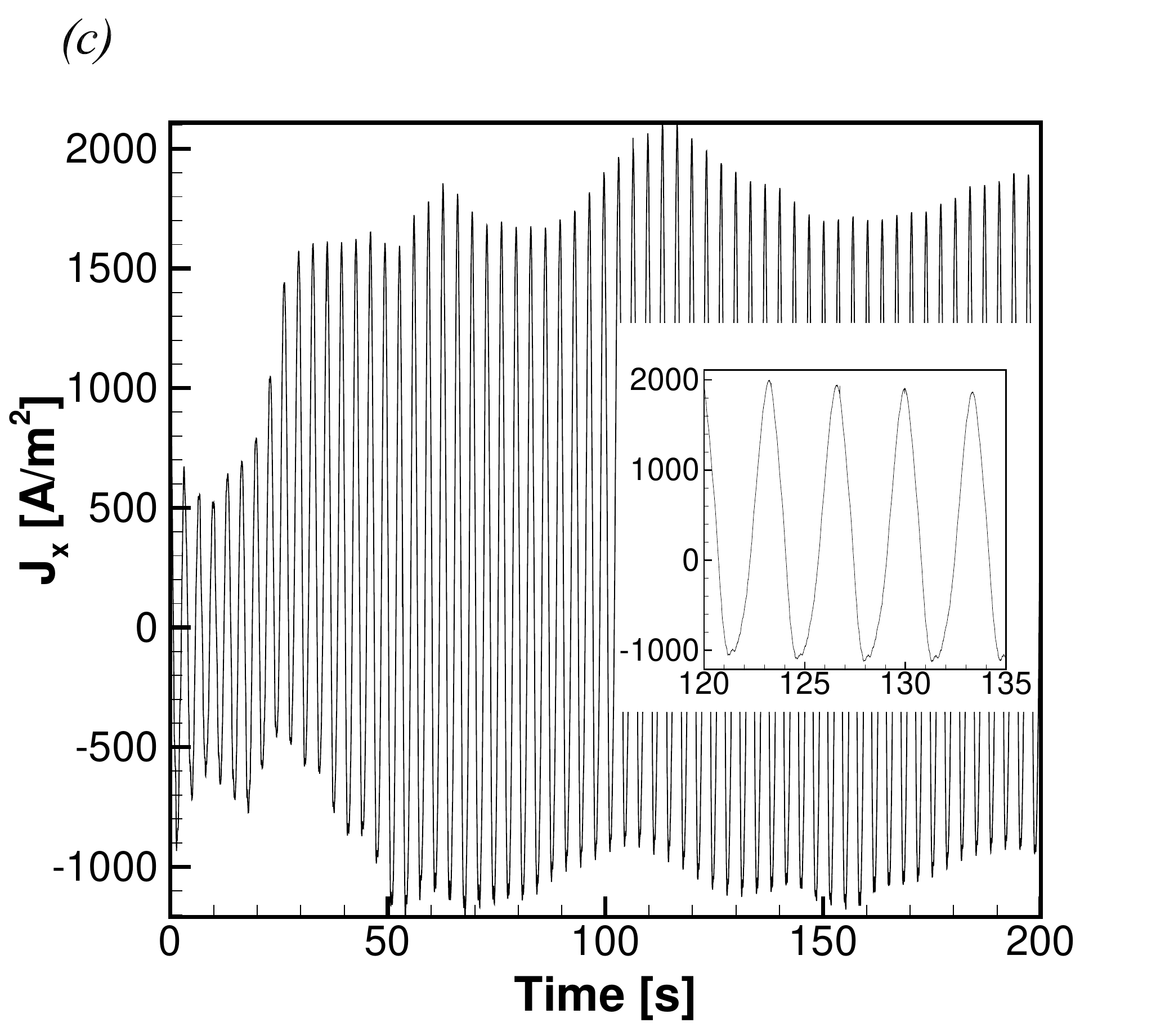}\includegraphics[width=0.4\textwidth]{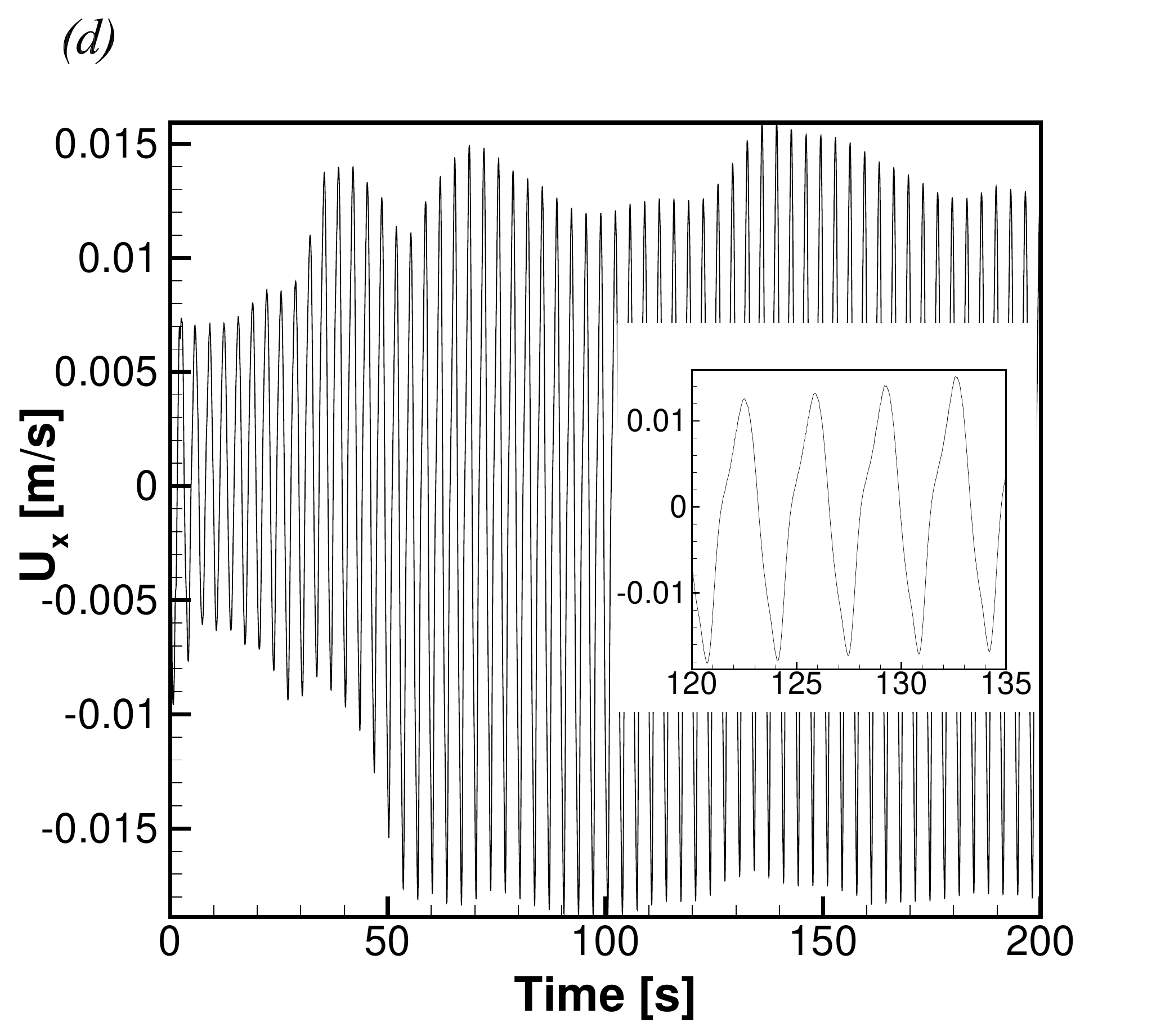}\\
\includegraphics[width=0.4\textwidth]{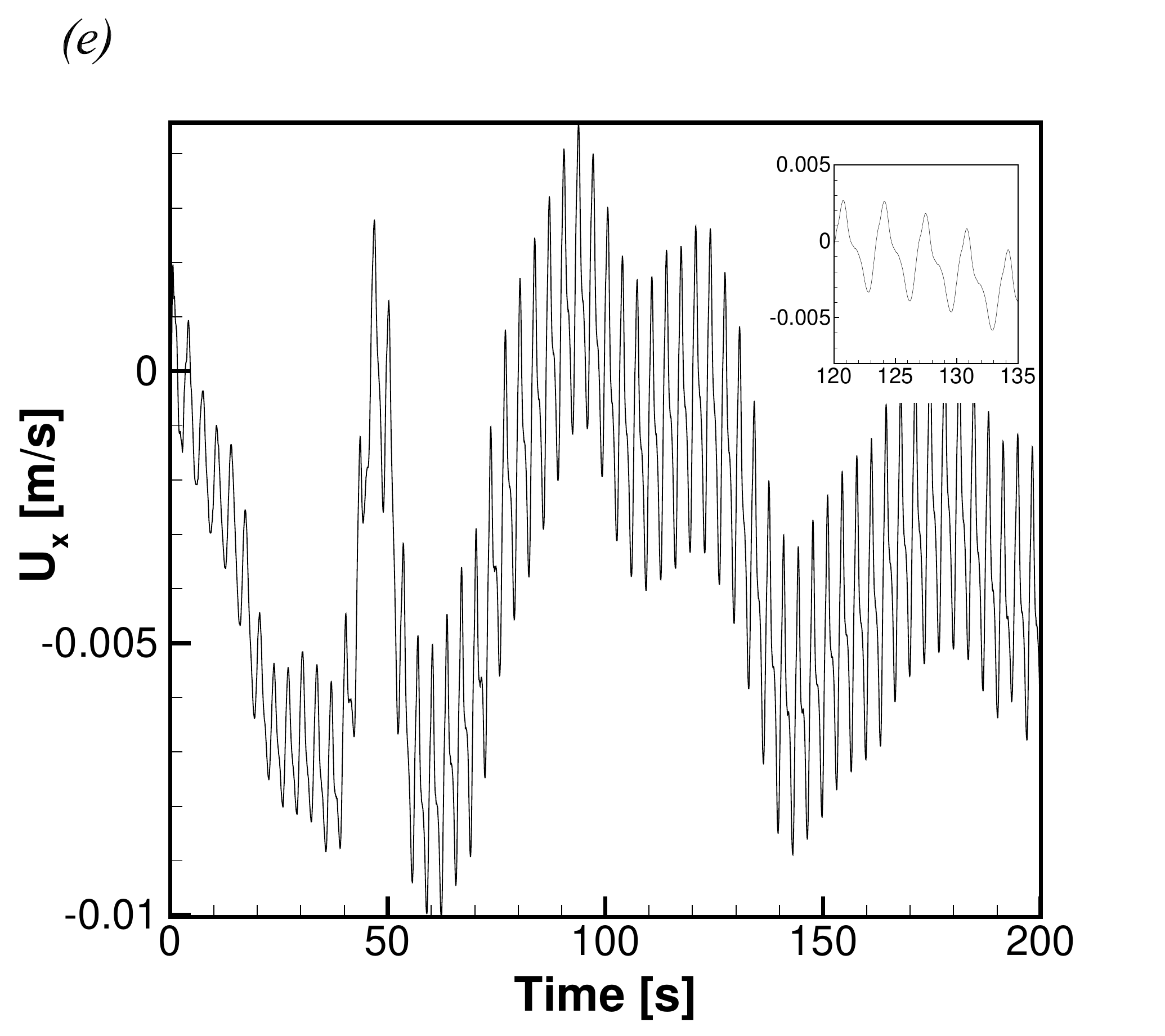}\includegraphics[width=0.4\textwidth]{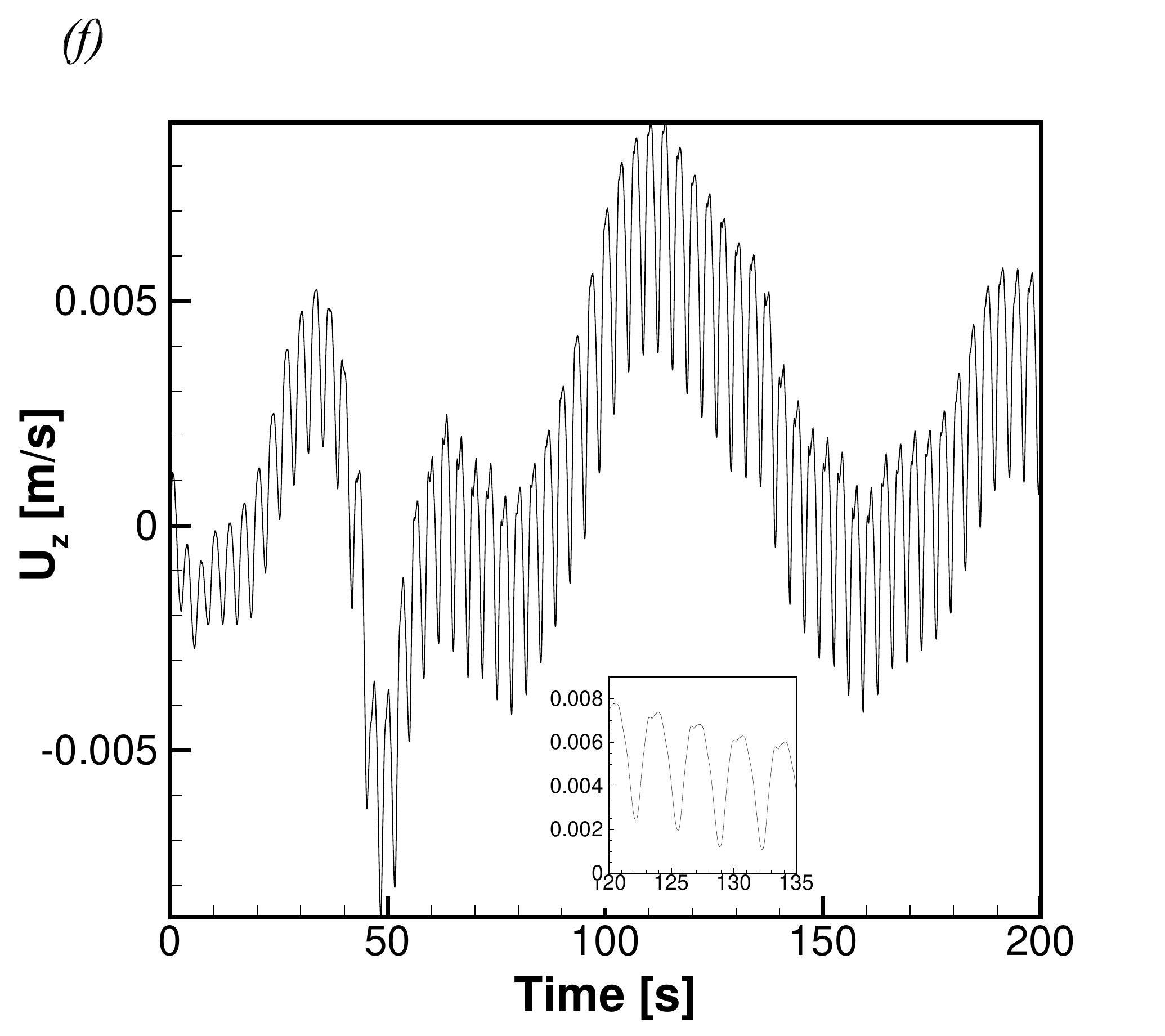}\\
\caption{Unstable flow with saturation at $H_E^0=5$ mm, $B_0=10$ mT, $\rho_E=1715$ kg/m$^3$ (case 5 in table II). Time signals of flow variables recorded at $x=0.05$ m, $z=0.016$ m during the entire simulation are shown. \emph{(a)}, locations of the upper ($\eta^A$) and lower ($\eta^B$) interfaces;  \emph{(b)}, vertical component $J_y$ of electric current  in the electrolyte at $y=0.05$ m;  \emph{(c)}, horizontal component $J_x$ of electric current  in the top metal layer at $y=0.06$ m;  \emph{(d)}, horizontal component $U_x$ of velocity in the electrolyte at $y=0.05$ m;  \emph{(e)} and \emph{(f)}, horizontal components $U_x$ and $U_z$ of velocity in the top metal layer at $y=0.06$ m.}
\label{fig4}
\end{center}
\end{figure}

\begin{figure}
\begin{center}
\includegraphics[width=0.45\textwidth]{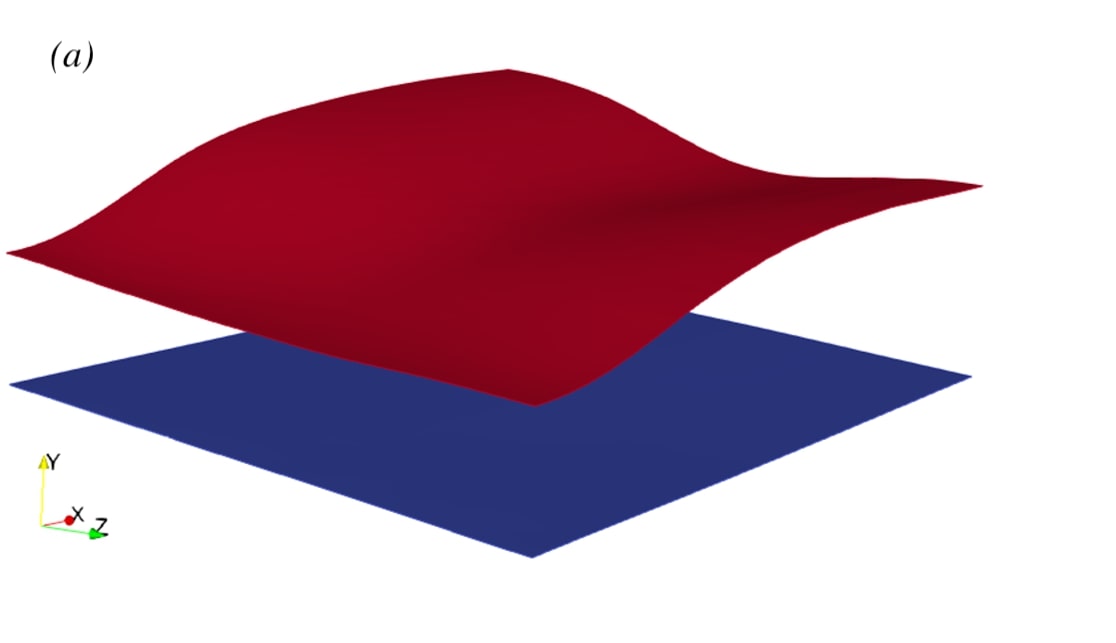}\includegraphics[width=0.45\textwidth]{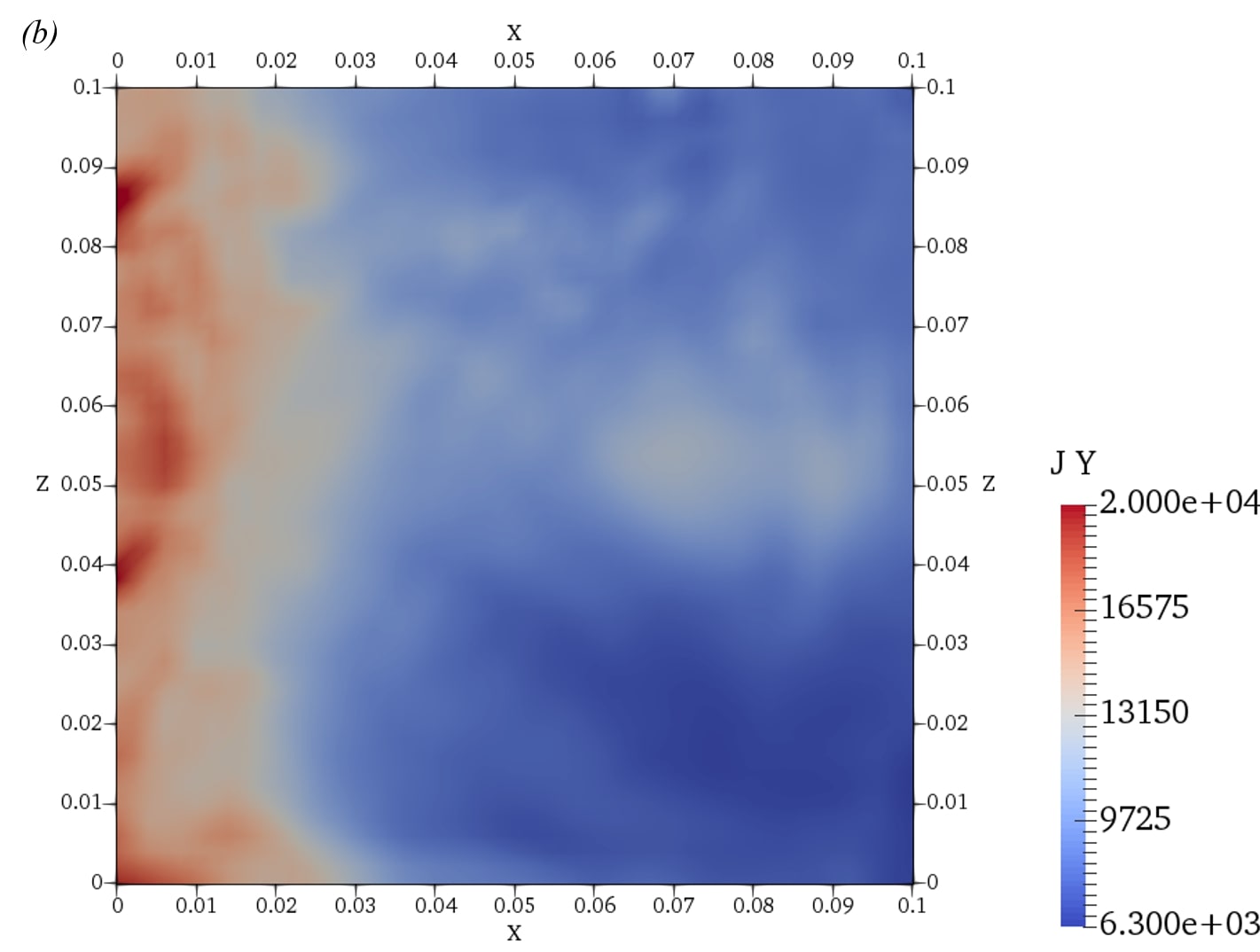}\\
\includegraphics[width=0.45\textwidth]{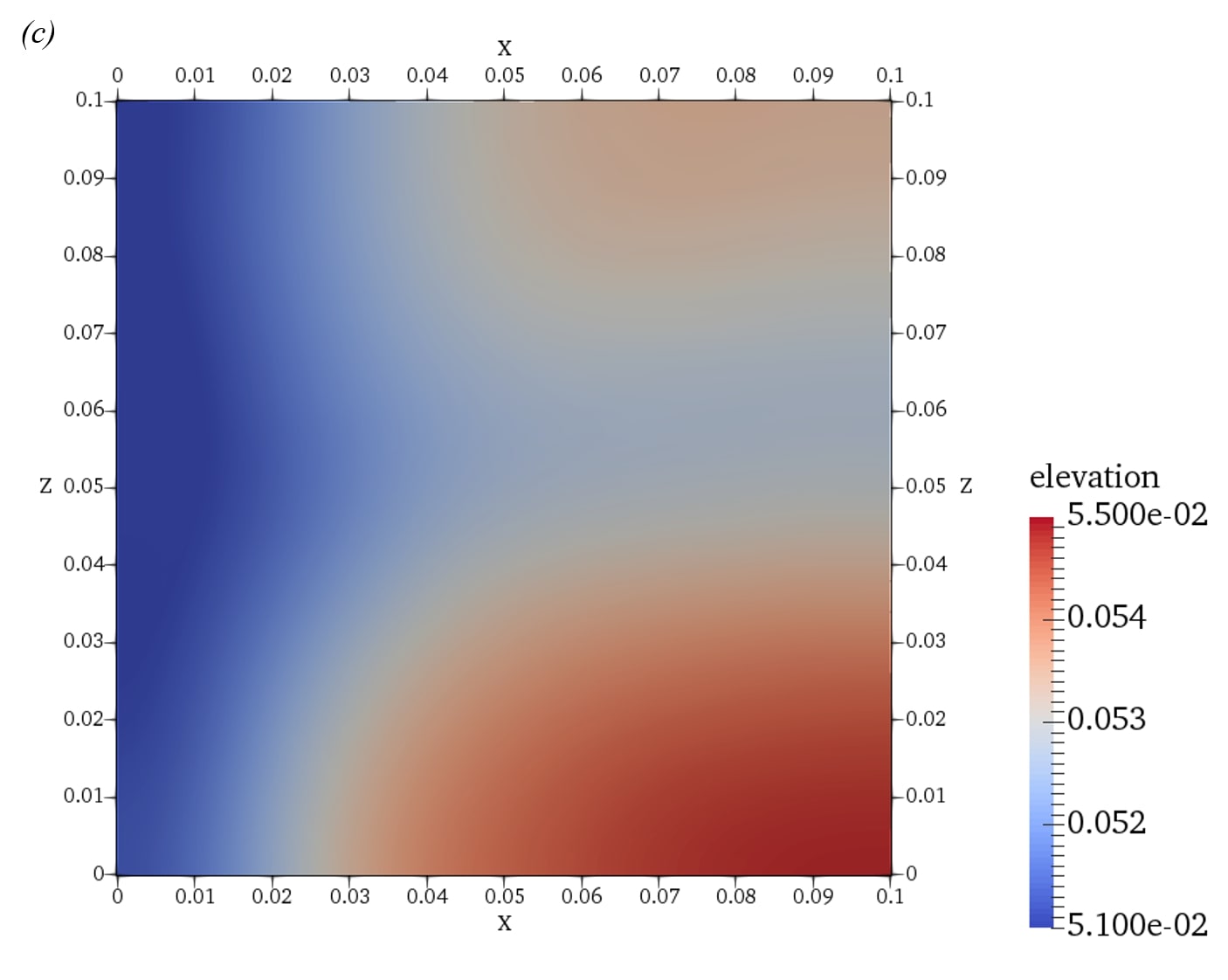}\includegraphics[width=0.45\textwidth]{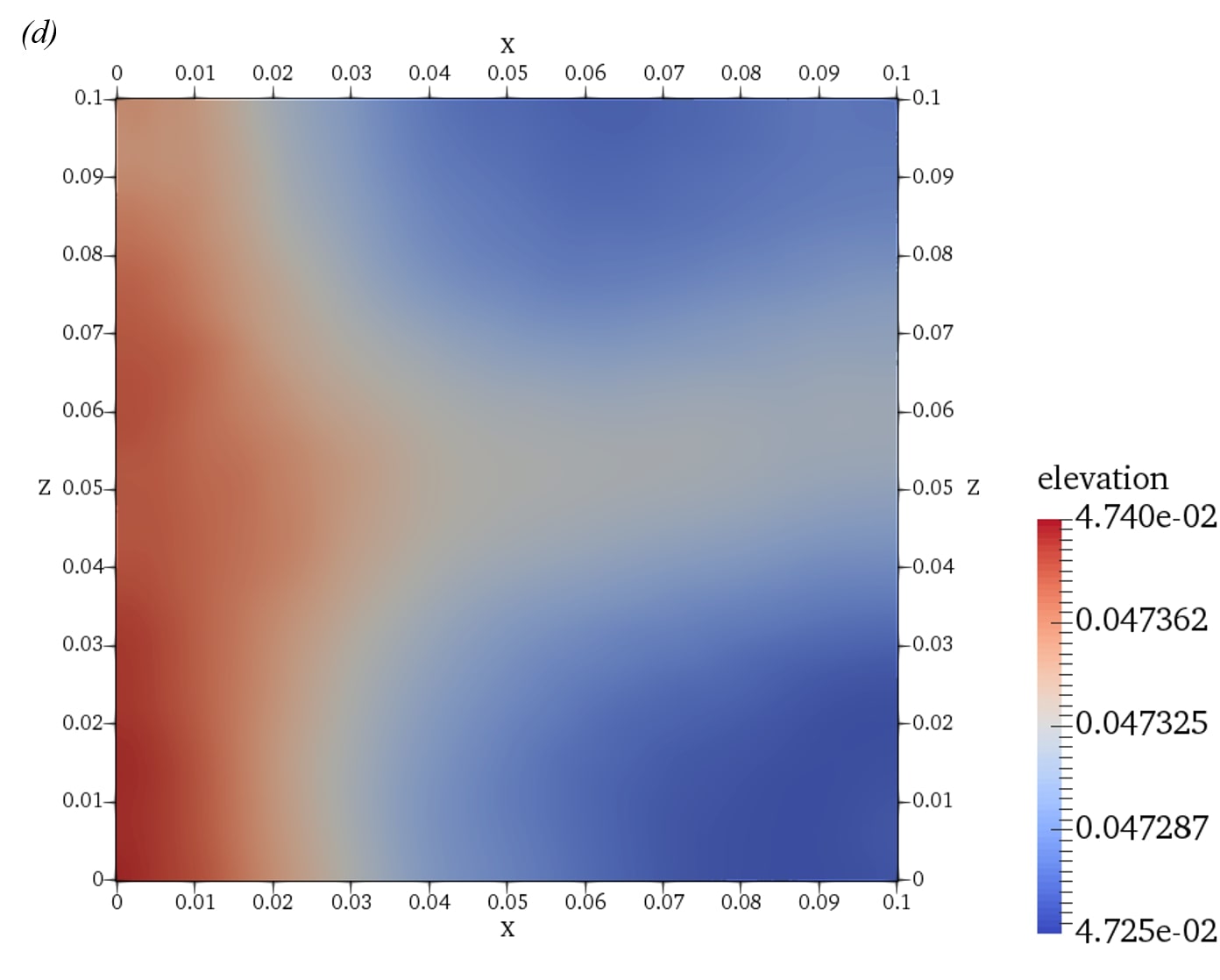}\\
\includegraphics[width=0.45\textwidth]{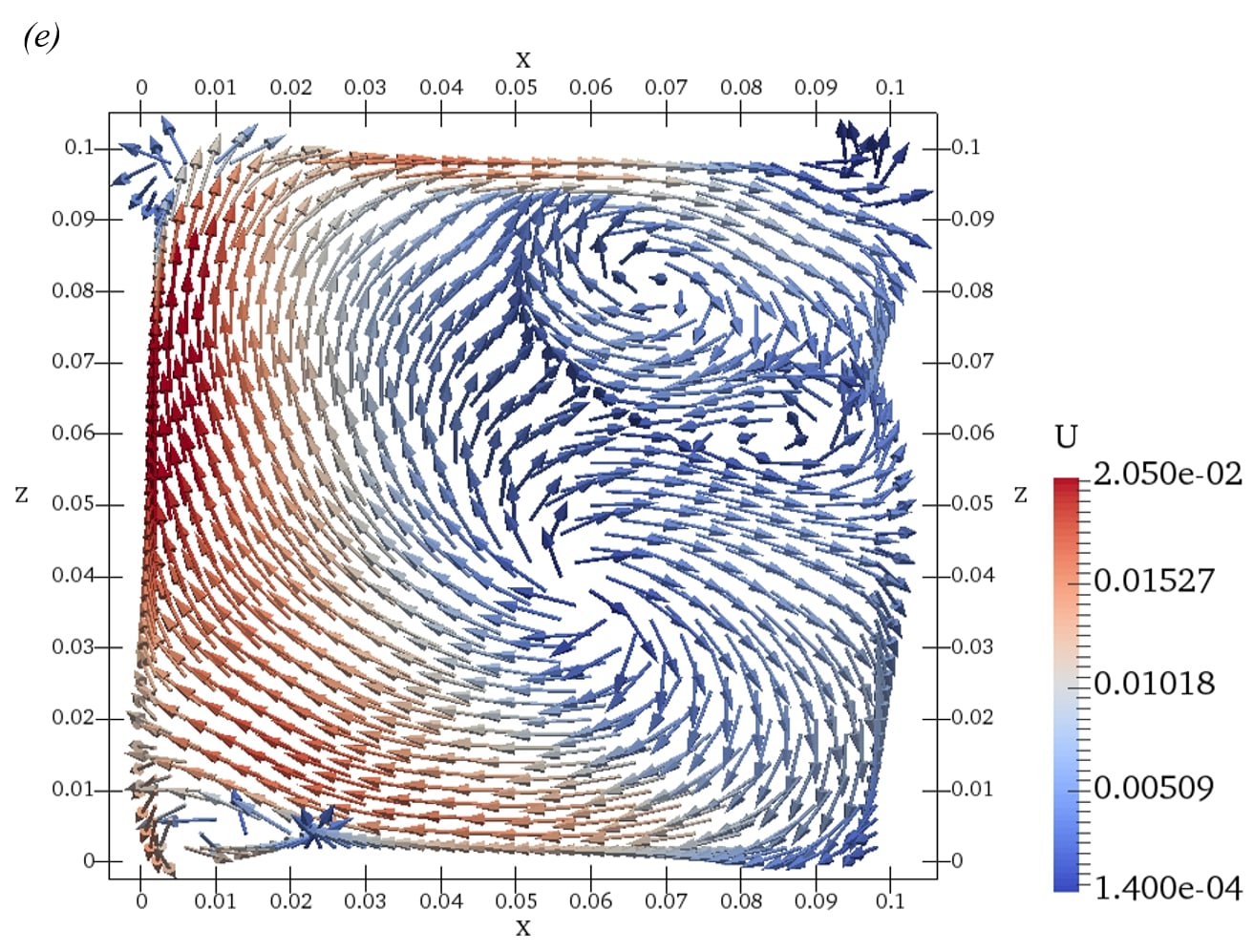}\includegraphics[width=0.45\textwidth]{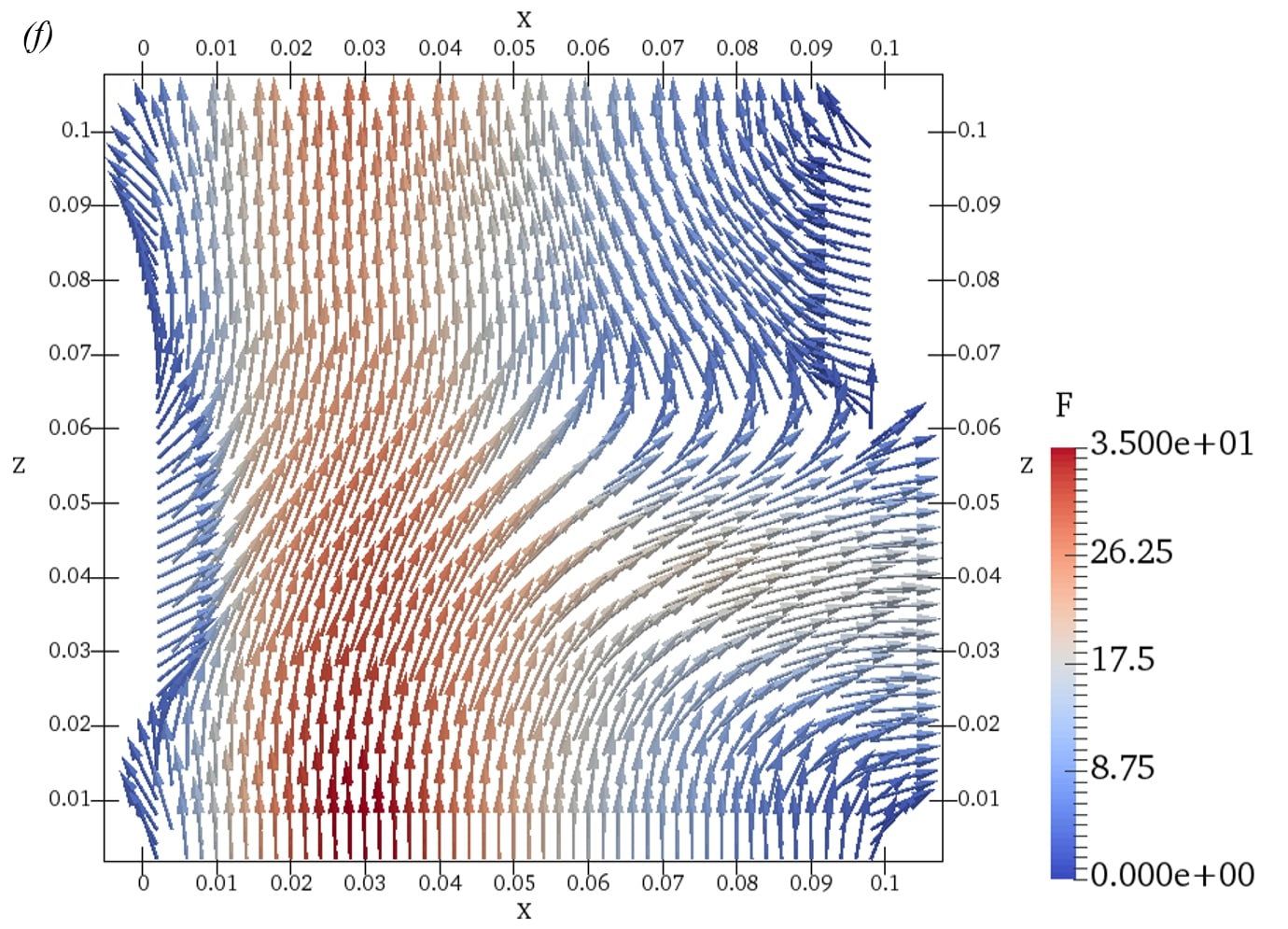}\\
\caption{Typical structure of the saturated unstable flow at $H_E^0=5$ mm, $B_0=10$ mT, $\rho_E=1715$ kg/m$^3$ (the case 5 in table II). Instantaneous distributions of variables at $t=137$ s are shown.  \emph{(a)}, profiles of upper ($\eta^A$) and lower ($\eta^B$) interfaces (scaled in the $y$-direction for better visibility);  \emph{(b)}, vertical component $J_y$ of electric current  in the electrolyte at $y=0.05$ m;  \emph{(c)}, profile of  the upper interface $\eta^A(x,z)$;  \emph{(d)}, profile of  the lower interface $\eta^B(x,z)$;  \emph{(e)}, velocity vector $(U_x,U_y,U_z)$, with color indicating the velocity magnitude, in the electrolyte at $y=0.05$ m;  \emph{(f)}, Lorentz force vector $(f_{L,x},f_{L,y},f_{L,z})$, with  color indicating the force magnitude, in the electrolyte at $y=0.05$ m.}
\label{fig5}
\end{center}
\end{figure}

The second example illustrated in Figs.~\ref{fig4} and \ref{fig5} is the flow of the case 5, in which the system parameters are the same as in the case 1, except that the strength of the vertical magnetic field $B_0$ is increased to 10 mT. Fig.~\ref{fig4} clearly shows that there are two stages of the flow evolution. The first stage, lasting for about 60 s, is characterized by growth of the fluctuation amplitudes of all variables. During the second stage, which continues  for the rest of the simulated flow evolution, the amplitudes remain approximately constant. The situation is identifiable as that of an instability leading to nonlinear saturation at some level of finite-amplitude perturbations. 

The quantitative characteristics of the interfacial wave in the saturated flow regime can be determined from the plots in Fig.~\ref{fig4} and other simulation data not shown here. The dominant period of the wave is about 3.3 s. The amplitude of the fluctuation  is slightly less than 2.5 mm for the upper interface, and much smaller, about 0.07 mm for the lower interface.  As we see in, respectively, Figs.~\ref{fig4}b and \ref{fig4}c, the resulting variations of the local thickness of electrolyte result in strong (up to 5000 A/m$^2$ in amplitude) fluctuations of the vertical current and strong (up to 2000 A/m$^2$)  horizontal currents in the  metal layer. The melt flows associated with the wave are not strong. The typical magnitude of the horizontal velocity components is $\sim 1$ cm/s in the electrolyte and smaller than that in the metal layers (see Figs.~\ref{fig4}d-f). It is interesting that the velocity in the electrolyte fluctuates around zero, while the signals of velocity in the metal layers show additional strong variations on the typical time scale of many tens of s. Our analysis of the evolution of the velocity structure indicates that this behavior is a reflection of a slow transformation of the rotating pattern of horizontal velocity associated with the wave.

We need to stress that the just mentioned fluctuation amplitudes refer to the signals recorded at the specific locations used in Fig.~\ref{fig4}. Signals recorded at other points would produce different values. The order of magnitude and the general characteristics of the wave, such as the lengths of growth and saturation stages, time period, and the qualitative shape of the signal curves would, however, remain the same.

The structure of the flow with a fully developed wave (the saturation stage) is illustrated in Fig.~\ref{fig5}. The interfacial waves presented as three-dimensional (in Fig.~\ref{fig5}a) and two-dimensional (in Figs.~\ref{fig5}c and d) plots of $\eta^A(x,z)$ and $\eta^B(x,z)$ are qualitatively similar to the typical waves observed in the two layer systems (aluminum reduction cells)\cite{Zikanov:2000,Sun:2004,Lukyanov:2001} and three-layer systems with $\Delta \rho^A\ll \Delta \rho^B $.\cite{Weber:2016,Horstmann:2018,Zikanov:2018shallow,Tucs:2018,Molokov:2018,Herreman:2019} The waves are characterized by the wavelength comparable with the horizontal size of the cell and the zone of large-amplitude interface deformation located near the sidewall. The pattern rotates clockwise if viewed from above (see animations\cite{anime}). The rotation is responsible for the nearly periodic oscillations of $\eta^A$, $\eta^B$ and other variables shown in Fig.~\ref{fig4}. Two interesting and useful for our further discussion observations can be made. One is that the wave at the upper interface has much larger amplitude than the lower interface wave. Another observation is that the two waves are antisymmetrically (with the phase shift of 180 degree) coupled. 

The distribution of the vertical electric current $J_y$ in the electrolyte (see Fig.~\ref{fig5}b) follows closely the variation of the local electrolyte thickness, thus the high-amplitude interfacial wave at the upper interface. The distributions of other variables, such as flow velocity (see Fig.~\ref{fig5}e) or Lorentz force (see Fig.~\ref{fig5}f) cannot be explained that simply but are evidently parts of the flow associated with the electromagnetically coupled interfacial waves.

\begin{figure}[h]
\begin{center}
\includegraphics[width=0.8\textwidth]{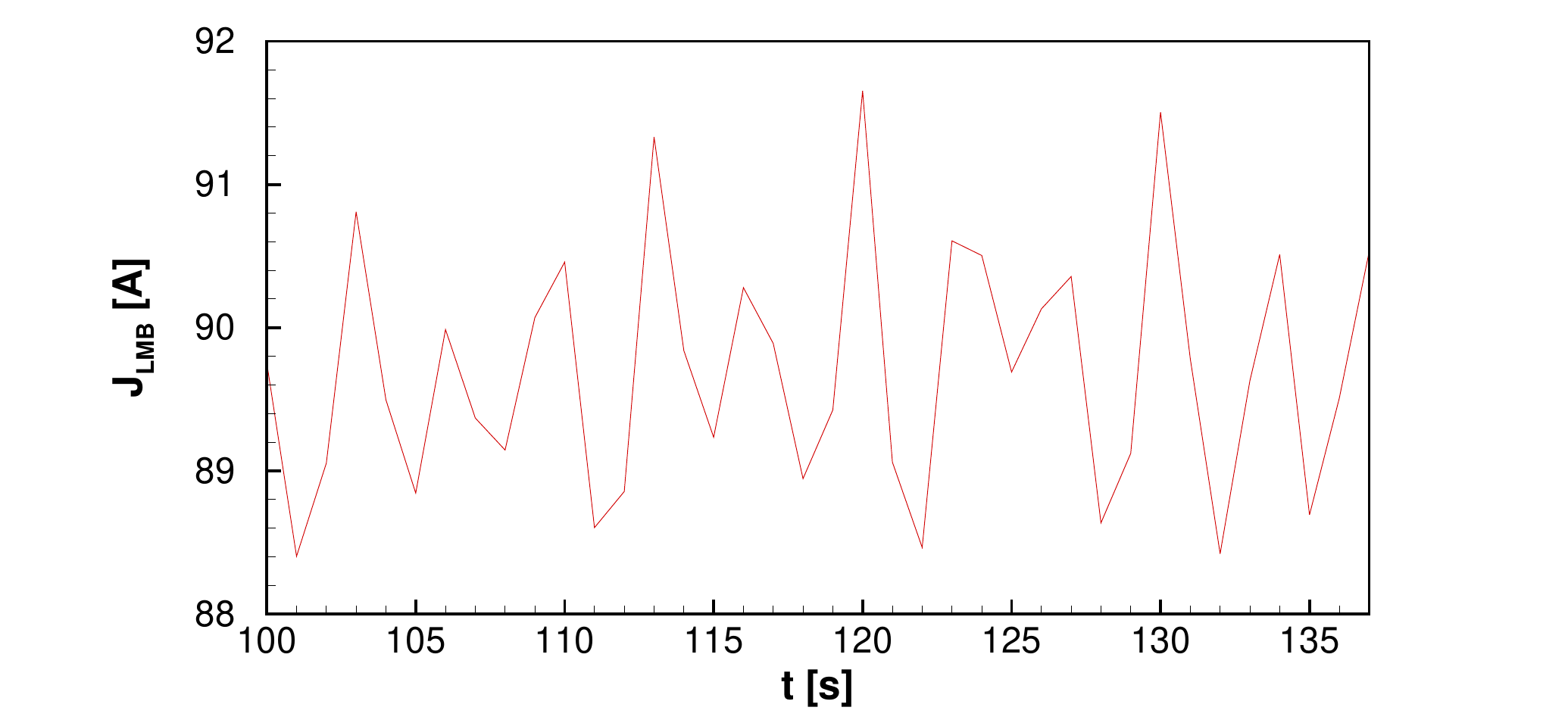}\\
\caption{Unstable flow with saturation at $H_E^0=5$ mm, $B_0=10$ mT, $\rho_E=1715$ kg/m$^3$ (case 5 in table II). Time signal of the integral of vertical electric current $J_{LMB}$ during the sloshing stage is shown. The sampling frequency is one second.}
\label{fig:impedance}
\end{center}
\end{figure}

In addition to causing local current perturbations, the variation of the electrolyte thickness caused by interfacial waves also changes the overall impedance $R_{LMB}$ of the battery. This affects the energy efficiency. In our model simulations, in which the voltage drop across the battery remains constant, the impedance changes are manifested as the reciprocal variations of the total (integrated over a horizontal cross-section) vertical current $J_{LMB}$. Furthermore, the mass conservation of liquid phases implies that the integral of the perturbation of  electrolyte thickness and, thus, of the perturbation of  local electric resistance are zero. One should expect decreased overall impedance and increased total current in this case.

An example shown in Fig.~\ref{fig:impedance} illustrates the typical behavior found in the simulated flows with instability. The time-averaged overall current is about 89.6 A, which is noticeably larger than 78.5 A for the unperturbed battery. The interfacial wave significantly reduces the overall impedance of the cell. We also see fluctuations correlating with the evolution of the wave. This is to be expected for a wave of finite amplitude, which changes its shape in the course of its progression around the battery. 

We note that in the case 5 as in all the other cases identified in table II as those with saturation, the interface deformation never grows as strong as to cause a rupture of the electrolyte layer and short circuit between the metal layer. The system is hydrodynamically unstable, but this does not result in operational failure of the battery.

\begin{figure}
\begin{center}
\includegraphics[width=0.5\textwidth]{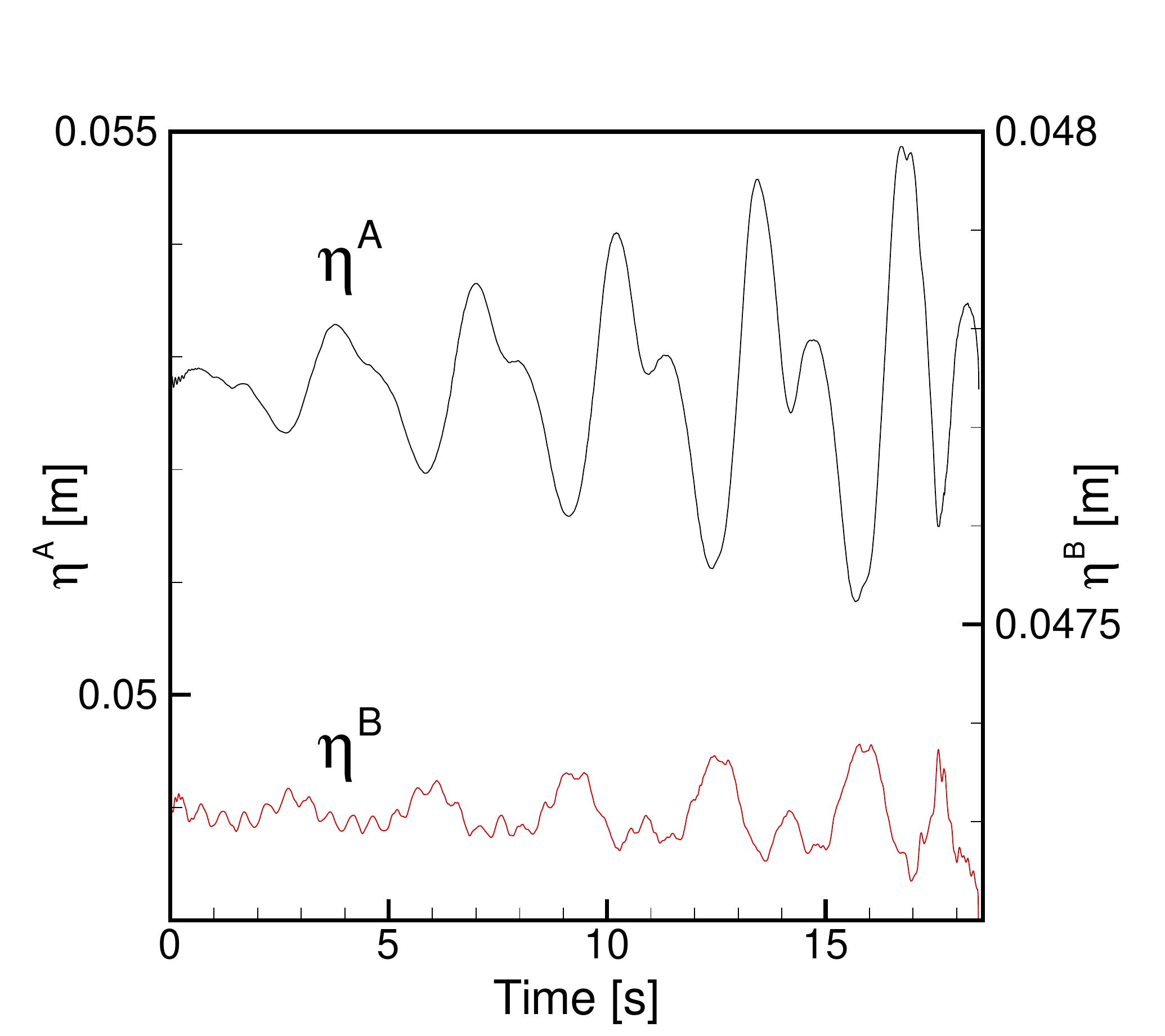}\\
\caption{Unstable flow with rupture of electrolyte layer at $H_E^0=5$ mm, $B_0=20$ mT, $\rho_E=1715$ kg/m$^3$ (case 7 in table II). Time signals of the locations of the upper ($\eta^A$) and lower ($\eta^B$) interfaces at $x=0.05$ m, $z=0.016$ m during the entire simulation are shown.}
\label{fig6}
\end{center}
\end{figure}
Our third example is the case 7 in table II. The system's parameters are the same as in the cases 1 and 5, but the vertical magnetic field $B_0$ is now increased to 20 mT. We see in Fig.~\ref{fig6} that the system is unstable and that the outcome of the instability is different from that observed in the previous example. The perturbations do not saturate but continue to grow until the electrolyte layer ruptures at some point. This implies zero overall impedance of the electrolyte layer and a short circuit between the metal layers, which makes further operation of the battery impossible. The situation can be identified as that of not just hydrodynamic, but also operational instability.

\begin{figure}
\begin{center}
\includegraphics[width=0.5\textwidth]{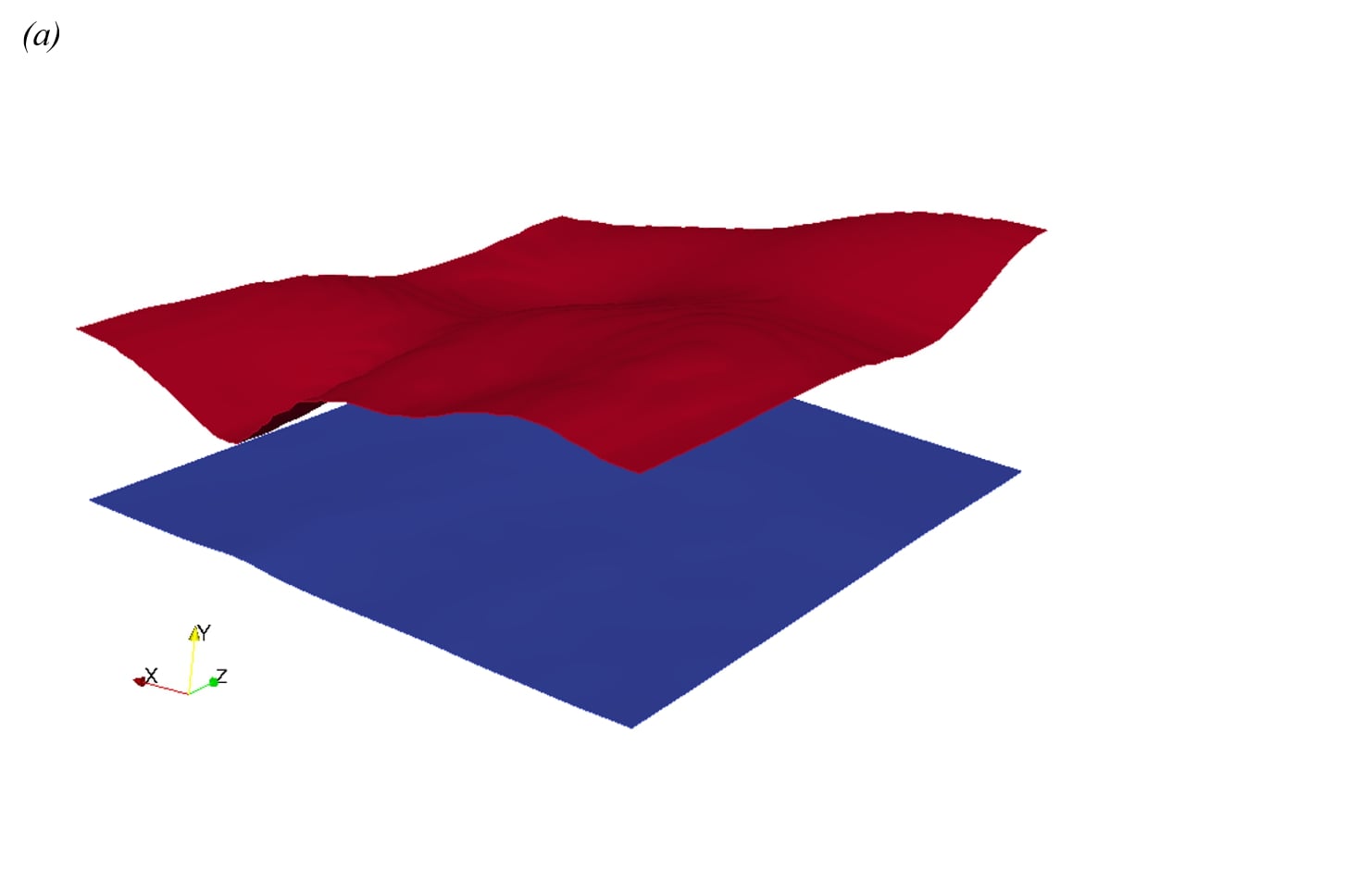}\includegraphics[width=0.5\textwidth]{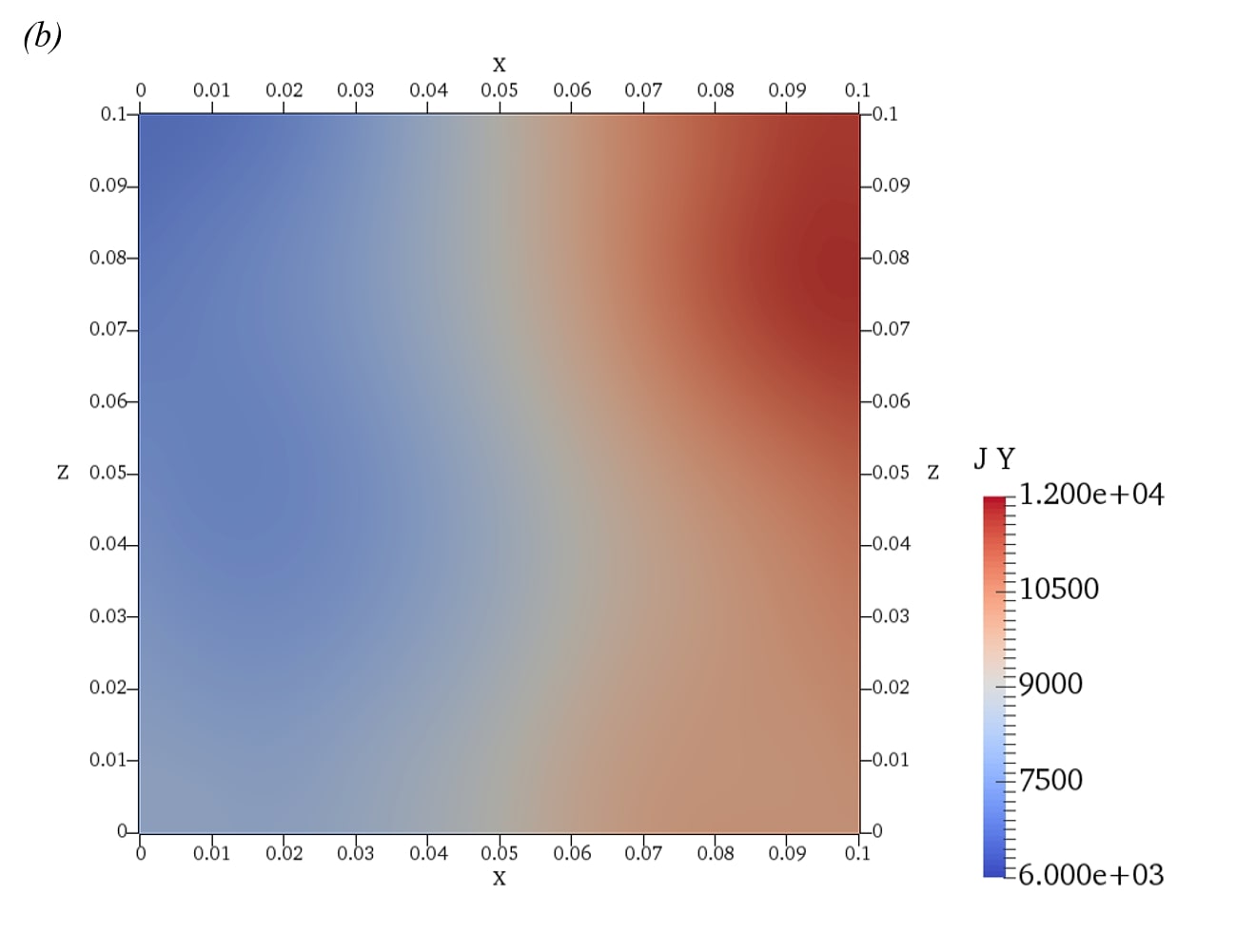}\\
\includegraphics[width=0.5\textwidth]{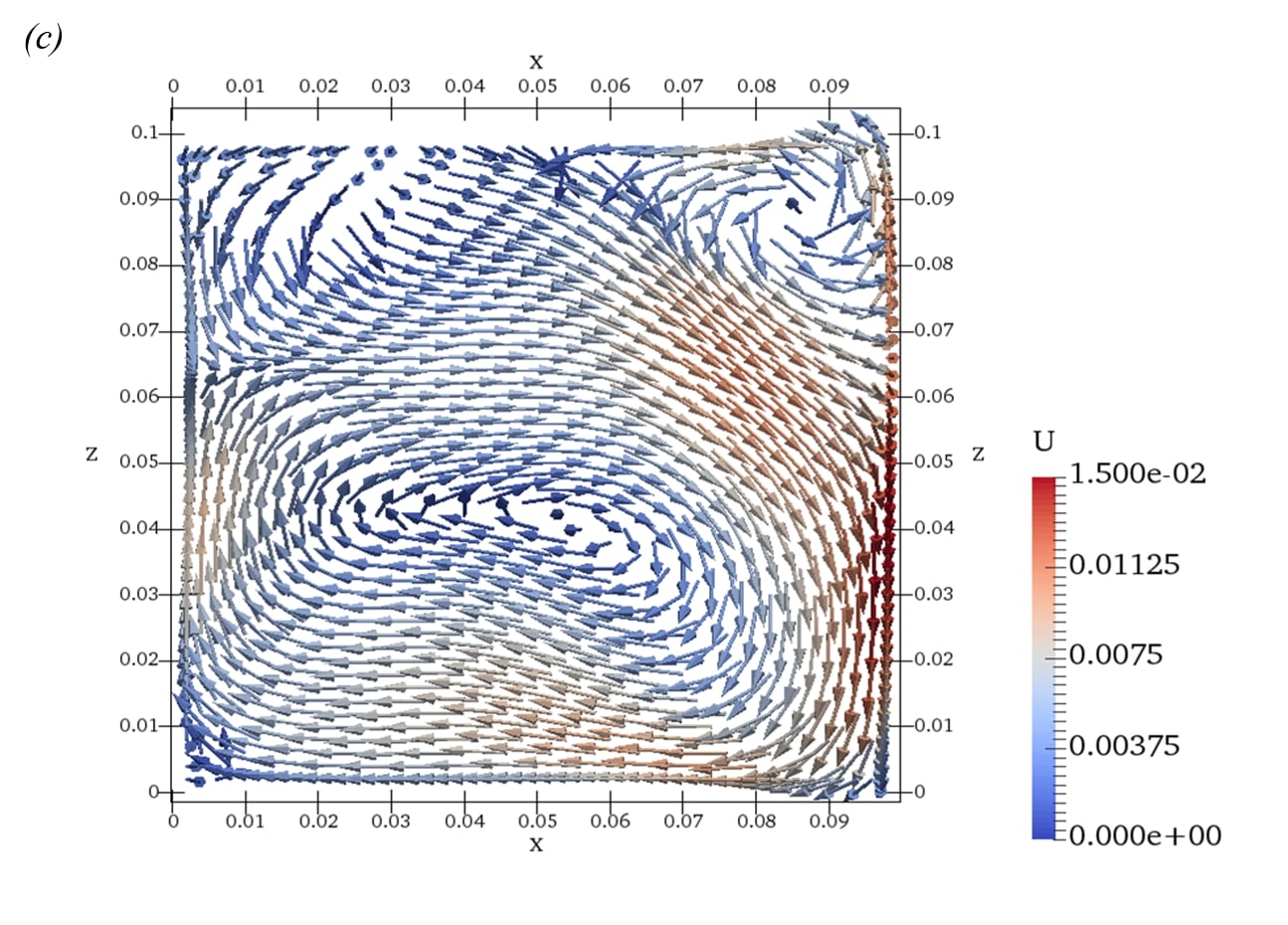}\includegraphics[width=0.5\textwidth]{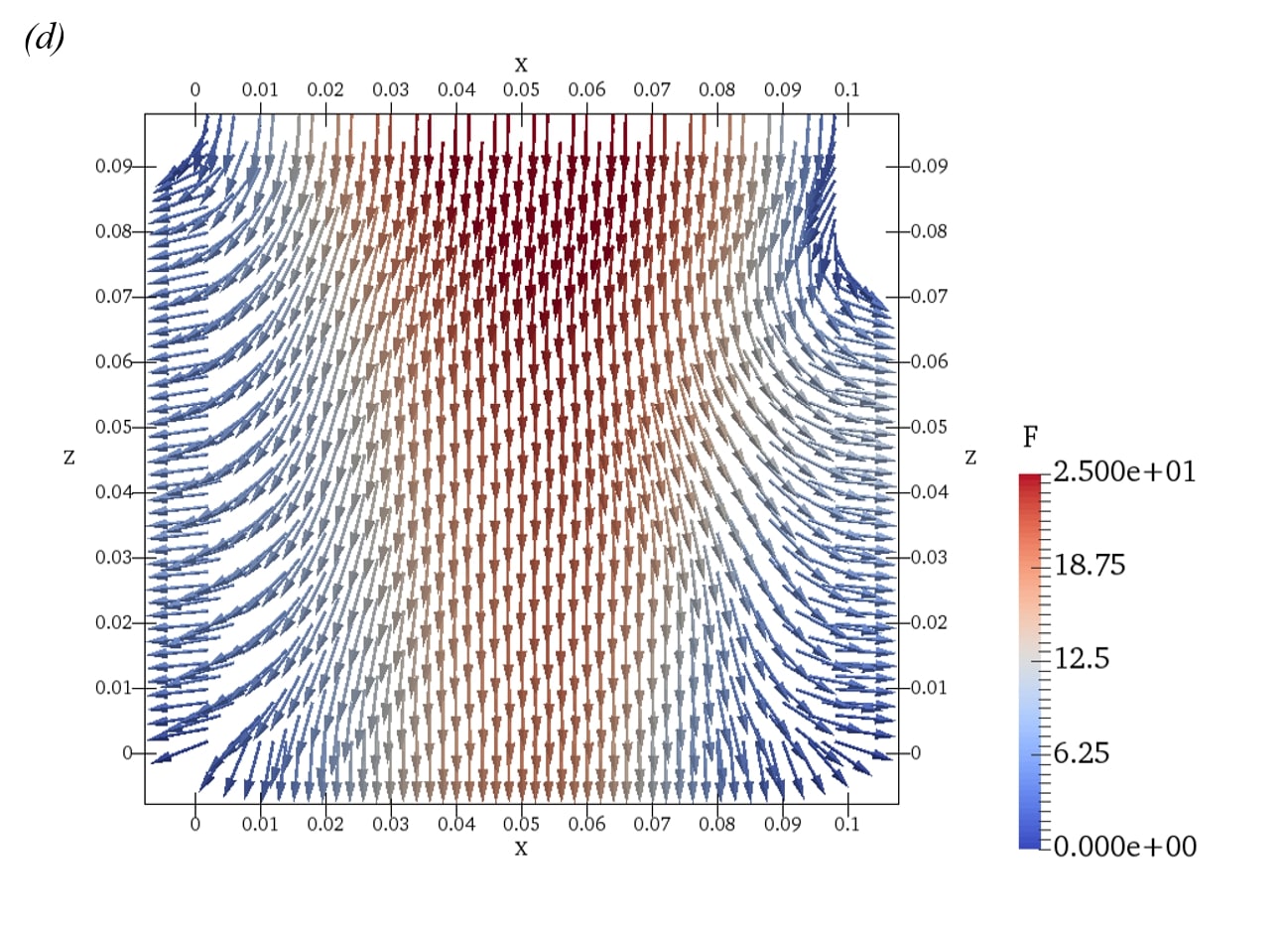}\\
\caption{Typical structure of the  flow at $H_E^0=5$ mm, $B_0=20$ mT, $\rho_E=1715$ kg/m$^3$ (the case 7 in table II). Instantaneous distributions of variables at $t=18$ s (about 0.5 s before rupture of electrolyte layer) are shown.  \emph{(a)}, profiles of upper ($\eta^A$) and lower ($\eta^B$) interfaces (scaled in the $y$-direction for better visibility);  \emph{(b)}, vertical component $J_y$ of electric current  in the electrolyte at $y=0.05$ m;  \emph{(e)}, velocity vector $(U_x,U_y,U_z)$, with color indicating the velocity magnitude, in the electrolyte at $y=0.05$ m;  \emph{(f)}, Lorentz force vector $(f_{L,x},f_{L,y},f_{L,z})$, with  color indicating the  force magnitude, in the electrolyte at $y=0.05$ m.}
\label{fig7}
\end{center}
\end{figure}

\begin{figure}
\begin{center}
\includegraphics[width=0.5\textwidth]{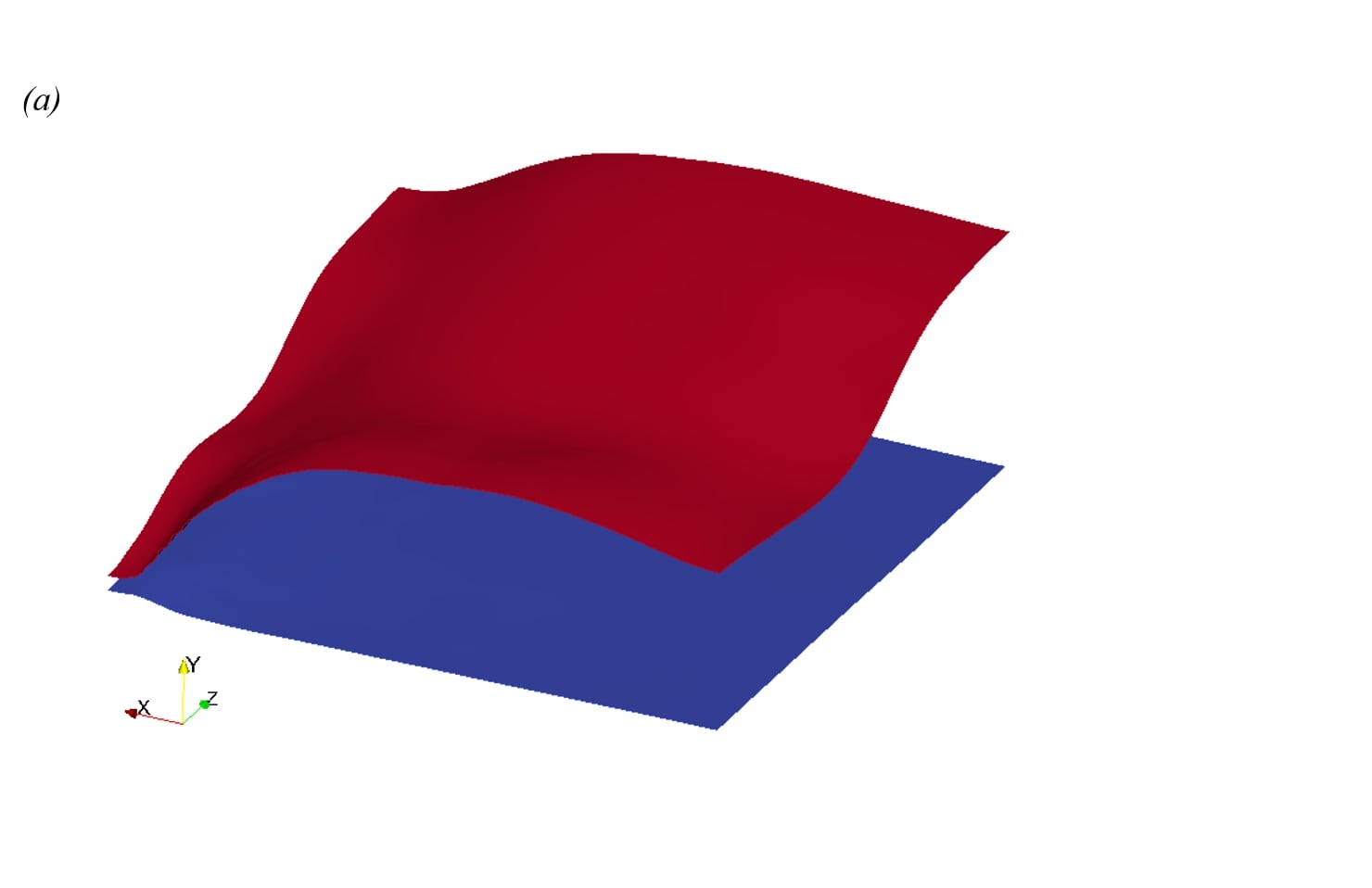}\includegraphics[width=0.5\textwidth]{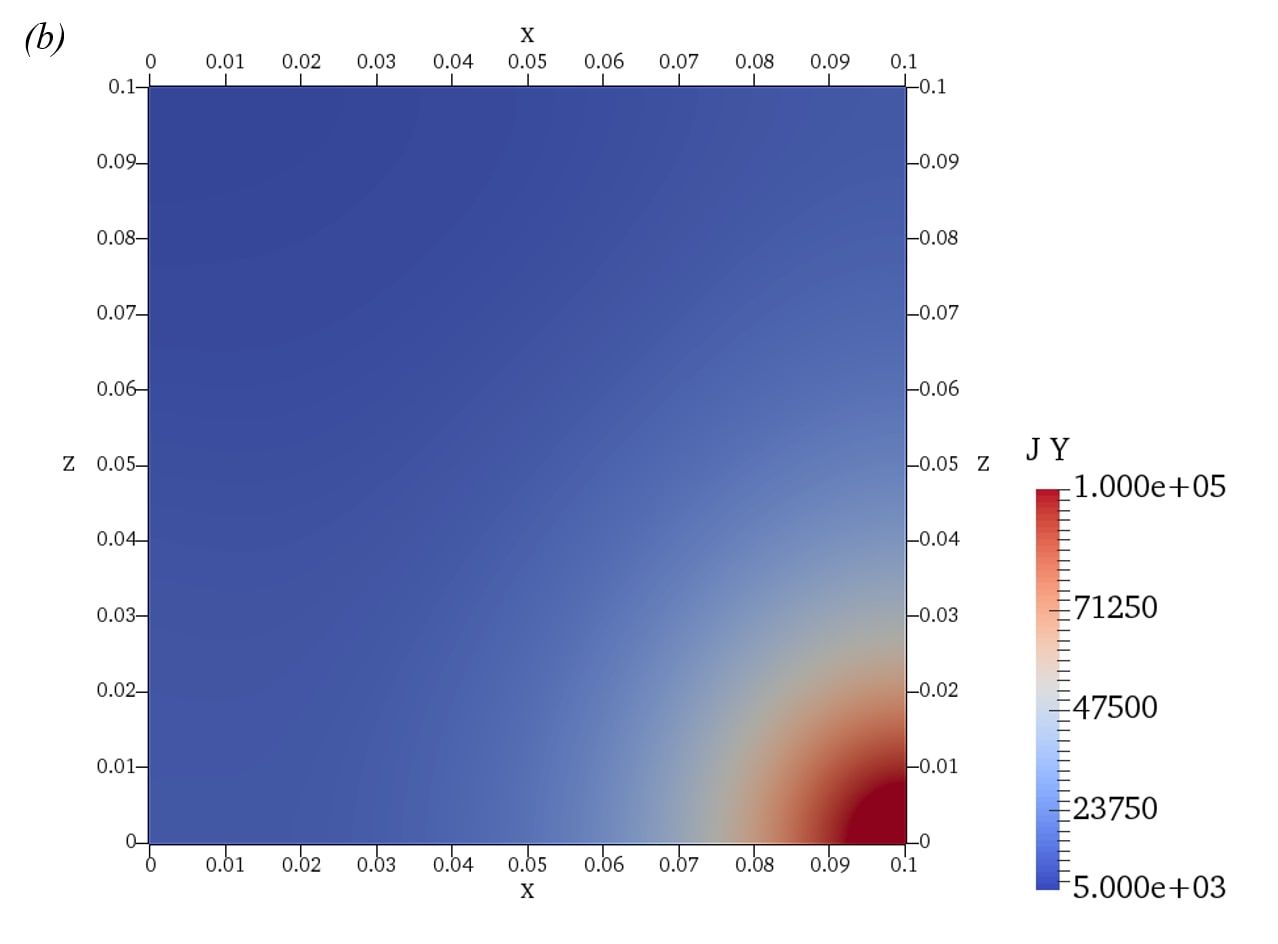}\\
\caption{Typical structure of the rupture flow at $H_E^0=5$ mm, $B_0=20$ mT, $\rho_E=1715$ kg/m$^3$ (case 7 in table II). Instantaneous  distributions of variables at $t=18.48$ s (immediately before rupture of electrolyte layer) are shown.  \emph{(a)}, profiles of upper ($\eta^A$) and lower ($\eta^B$) interfaces (the deformation is strongly exaggerated for better visibility);  \emph{(b)}, vertical component $J_y$ of electric current  in the electrolyte at $y=0.05$ m.}
\label{fig8}
\end{center}
\end{figure}

The structure of the flow about 0.5 s before the rupture is illustrated in Fig.~\ref{fig7}. We see current perturbations, velocity and velocity gradients of the amplitudes comparable to those in case 5. The situation changes shortly before the rupture. At illustrated in Fig.~\ref{fig8}, the interfaces nearly touch each other in a small area in the corner of the cell (the corner location was observed in all our simulations with rupture). The vertical electric current grows to about $10^6$ A/m$^2$ in this corner, which results in very strong localized horizontal current perturbations and Lorentz forces. 

At the moment of rupture, the vertical current increases further. The numerical model fails to converge, so the simulations must be stopped. It should be stressed that during the short time interval before the rupture, the small-scale, strong-gradient features appearing in the flow imply that the accuracy of the model's prediction is lower than estimated in our grid sensitivity studies. Additional simulations performed on a refined grid and with substantially reduced time step produced a qualitatively and quantitatively similar flow. The rupture event was delayed by a just fraction of a second. This indicates that the electrolyte rupture  observed in the simulations is a real physical effect and not an artifact of the numerical model.

\subsection{Effects of system parameters}
\label{sec:results_param}
In this section we analyze how the system's behavior is affected by the key parameters of the rolling pad instability: the strength of the vertical magnetic field $B_0$, the electrolyte thickness $H_E^0$, and the electrolyte density $\rho_E$. The discussion is based on the data summarized in Table II and the illustrations in Figs.~3-14. 

The following discussion will benefit from a reference to the recent analysis of interfacial waves in a stably-stratified three-layer system.\cite{Horstmann:2018} Flow in a cylindrical vessel without electromagnetic effects was considered. Two possible solution modes were identified. One was the fast mode with high oscillation frequency and symmetric (in-phase) coupling between the waves on the upper and lower interfaces. Another was the slow mode with lower frequency and antisymmetric (with 180$^\circ$ degree phase shift) coupling. The solutions of the two types also differ from each other by the typical ratios between the amplitudes of the upper and lower interface waves.

The system considered in this paper is evidently very different from that of Ref.~\onlinecite{Horstmann:2018} due to the  different geometry and the effects of viscosity and surface tension. The electromagnetic effects provide an additional coupling and forcing mechanisms and may influence the modes themselves and the selection between them in a particular solution. Nevertheless, as we will see below, the classification remains valid and useful in the analysis.

\begin{figure}
\begin{center}
\includegraphics[width=0.5\textwidth]{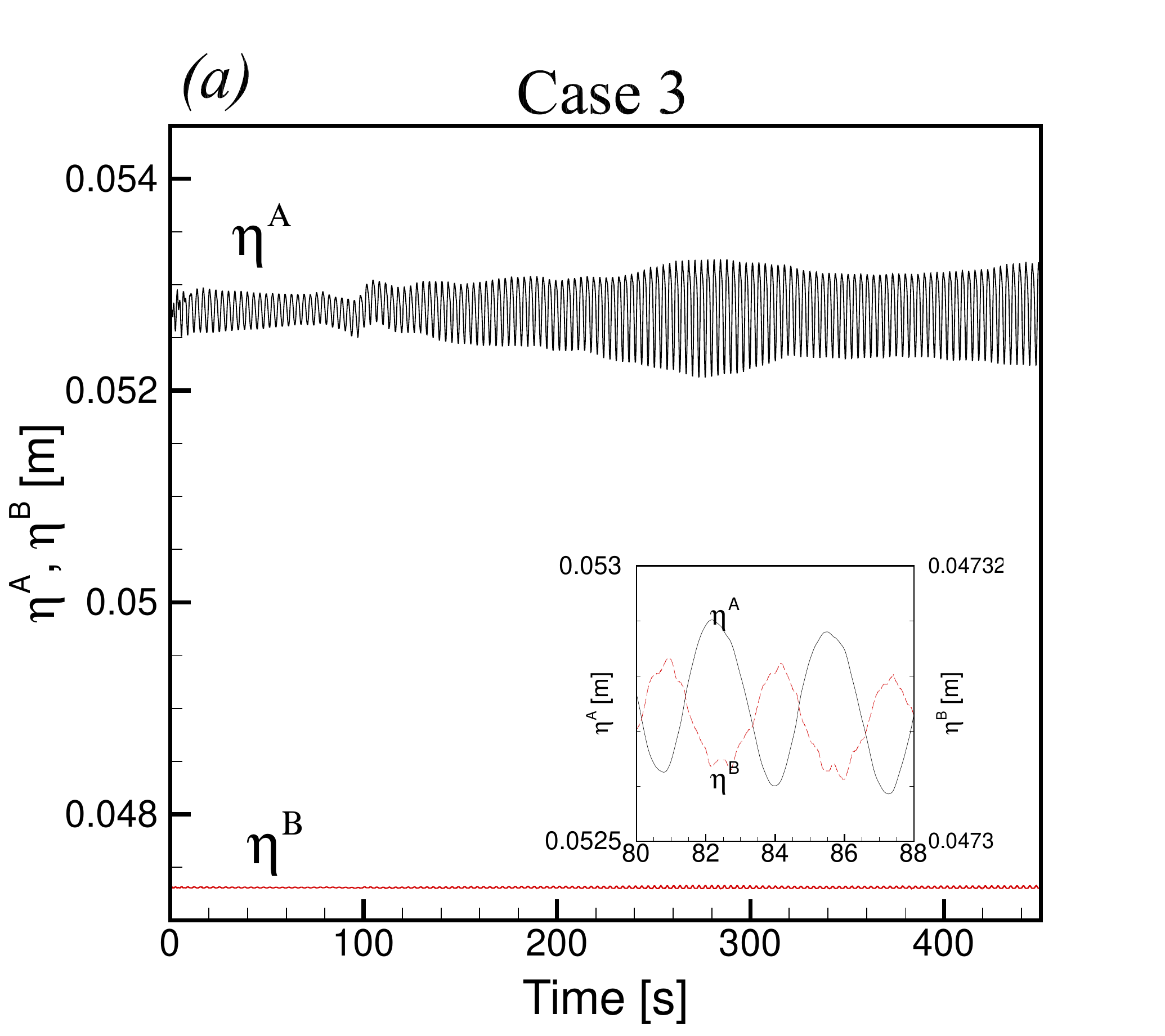}\includegraphics[width=0.5\textwidth]{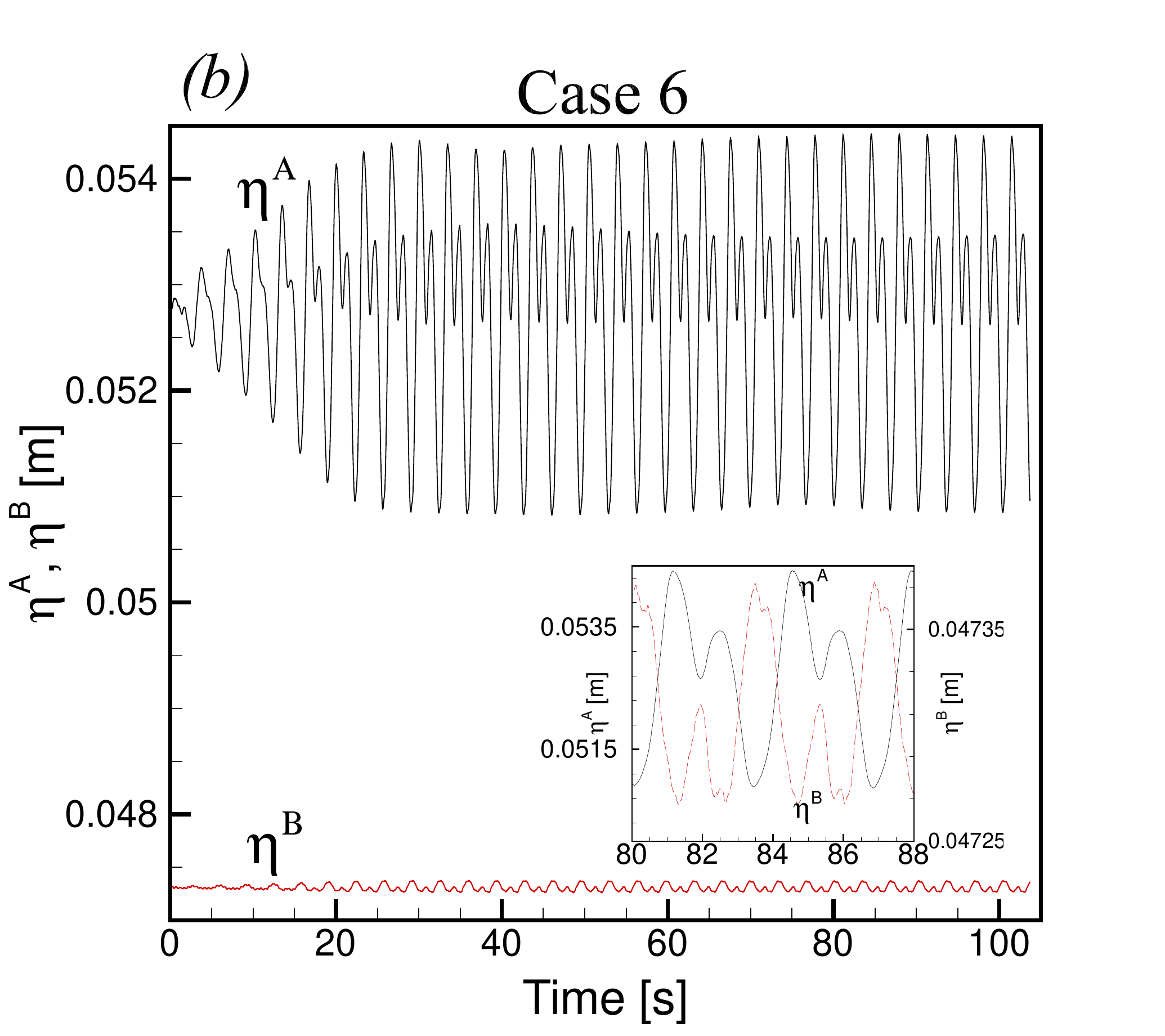}\\
\includegraphics[width=0.5\textwidth]{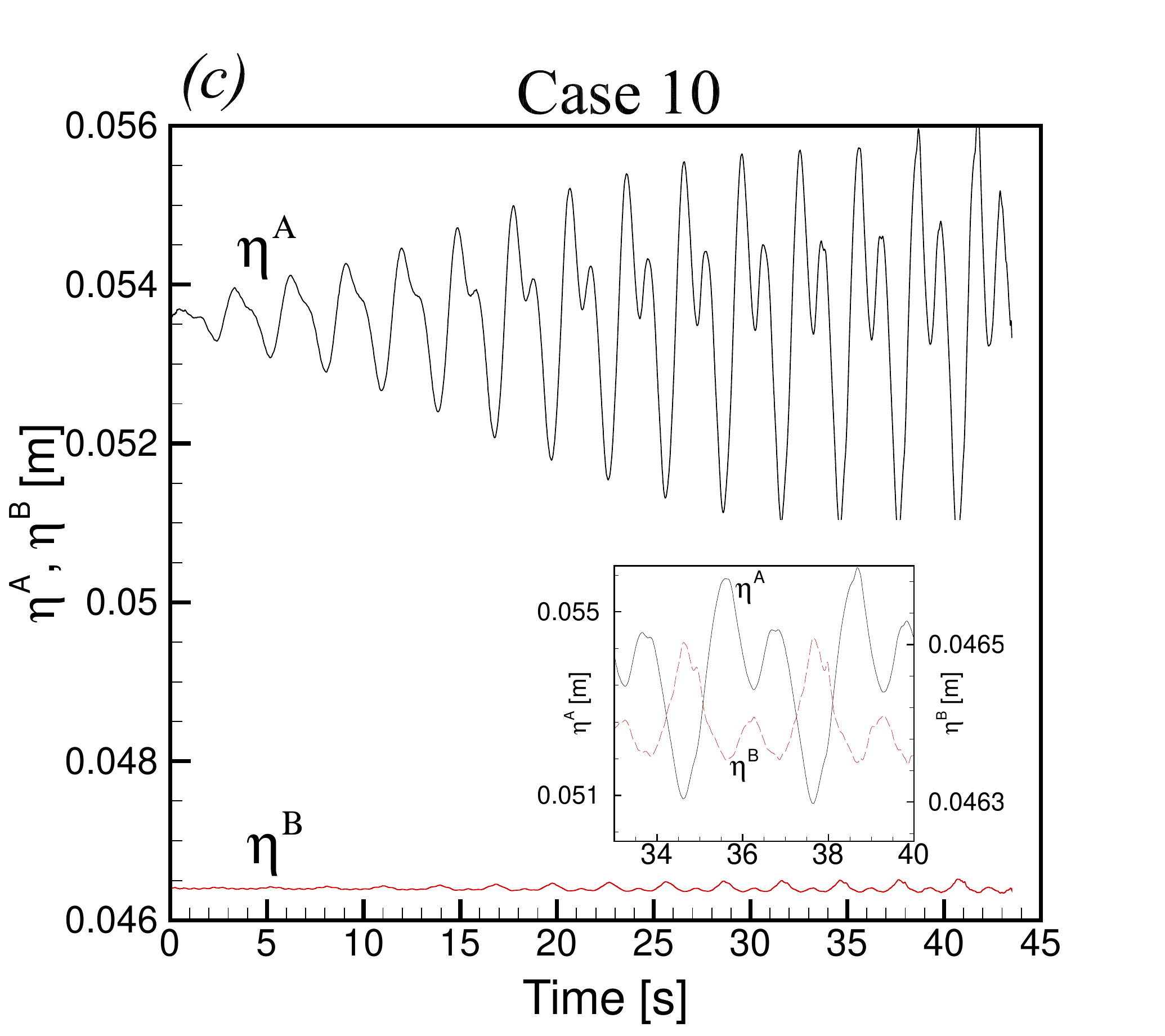}\includegraphics[width=0.5\textwidth]{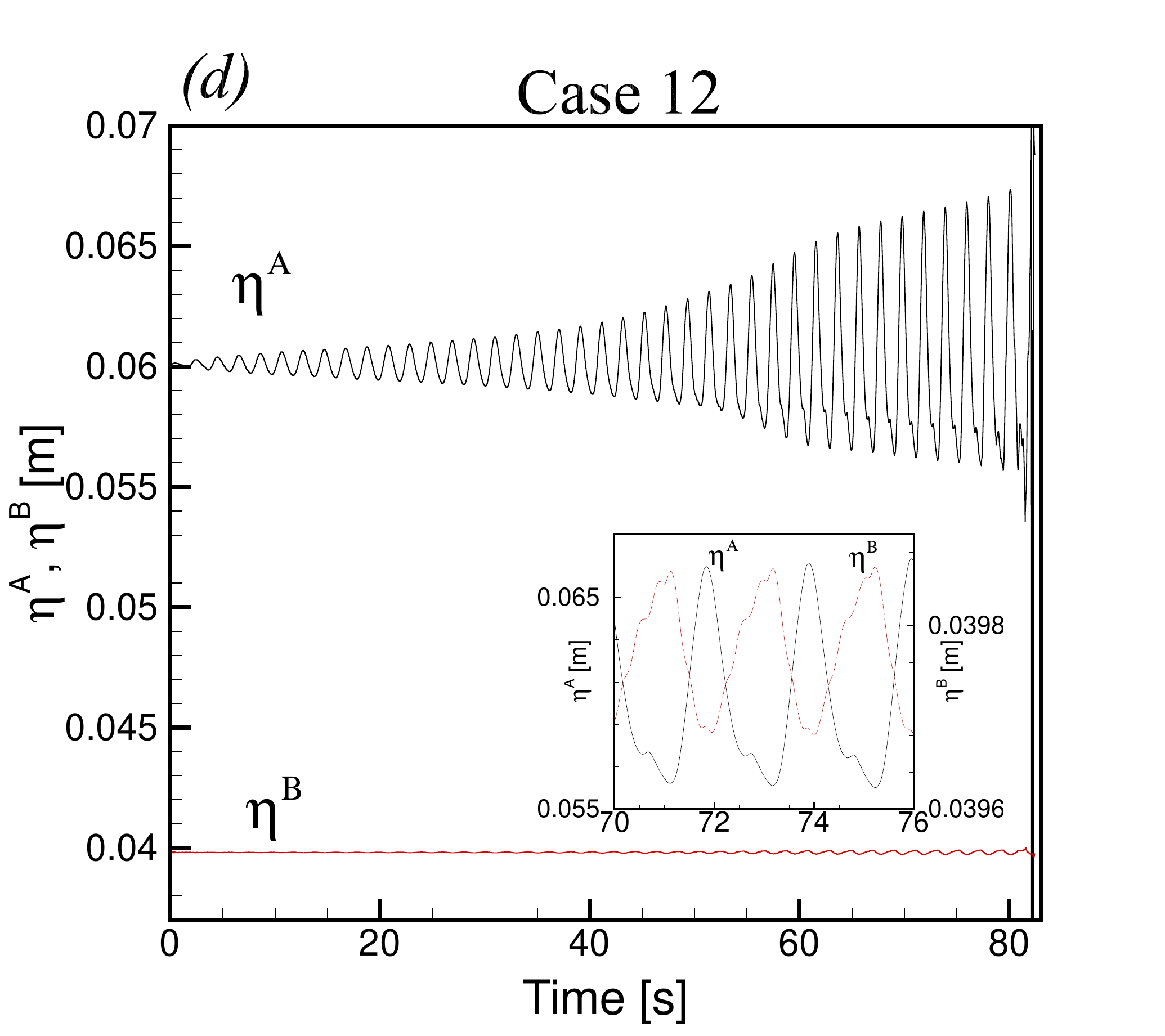}
\caption{Effect of $B_0$ and $H_E^0$ on flows with $\rho_E=1715$ kg/m$^3$ ($\Delta \rho_A \ll \Delta \rho_B$). Time signals of the locations of the upper ($\eta^A$) and lower ($\eta^B$) interfaces at $x=0.05$ m, $z=0.016$ m during the entire simulation are shown. Insets illustrate the behavior of fully developed waves during several wave periods. Note that the curves are shifted and different vertical scales are used in the insets for better comparison. Results obtained for the cases 3, 6, 10, and 12 are shown in \emph{(a)}, \emph{(b)}, \emph{(c)} and \emph{(d)} (see Table II for parameters and characteristics). }
\label{fig9}
\end{center}
\end{figure}

We start with the cases 1-13, in which $\Delta \rho_A\ll \Delta \rho_B$.  The three flows used as examples in section \ref{sec:results_examples} (see Figs.~\ref{fig3}-\ref{fig8}) belong to this group. Further information is provided in Table II and Fig.~\ref{fig9}. We clearly see that the solutions have the form of a slow mode in the classification of Ref.~\onlinecite{Horstmann:2018}. The typical time period $T$ is large in comparison to the other cases. The waves at the two interfaces are coupled antisymmetrically (see Figs.~\ref{fig3}, \ref{fig4}, \ref{fig6}, \ref{fig9}). Finally, the wave amplitude on the upper interface is much (20 to 60 times) larger than the amplitude of the lower interface wave. This is in agreement with the results of calculations for the material properties of a Mg-Sb cell presented in Fig.~4 of Ref.~\onlinecite{Horstmann:2018}. For the typical horizontal wavenumber $\sim 20 m^{-1}$ (observed in our simulations as illustrated in Figs.~\ref{fig5} and \ref{fig7}), the slow modes are expected to have $\Delta \eta^A$ one to two orders of magnitude larger than $\Delta \eta^B$. The growth of $\Delta \eta^A/\Delta \eta^B$ with increasing $H_E^0$ (see Table II) is also consistent with the behavior of the slow mode solutions illustrated in Fig.~4 of Ref.~\onlinecite{Horstmann:2018}. 

The combination of the material properties in the cases 1-12 corresponds to the Mg-Sb battery, which is relatively well studied in literature.\cite{Weber:2016,Horstmann:2018,Zikanov:2018shallow,Tucs:2018,Molokov:2018} It has been found (see, e.g. Refs.~\onlinecite{Weber:2016,Zikanov:2018shallow}) that the value of the parameter $\beta$ defined by (\ref{beta}) is an important, albeit not unique, factor of the instability in a cell of a given shape of the horizontal cross-section. The data in Table II are partially consistent with this conclusion. As we increase $B_0$, while keeping $H_E^0=5$ mm and $\rho_E=1715$ kg/m$^3$, the state of the system changes from stable at approximately $\beta$ less than one (cases 1 and 2) to saturated at $\beta$ up to about 4 (cases 3-6) and then to ruptured (cases 7 and 8). The threshold values of $\beta$ are reasonably close to those of Ref.~\onlinecite{Weber:2016} with the difference attributable to different cell geometries (a cylinder on Ref.~\onlinecite{Weber:2016} and a cube in our case).

The situation with the effect of the electrolyte thickness $H_E^0$ is less clear. Five simulations (cases 6, 9-12) are performed at $B_0=15$ mT, $\rho_E=1715$ kg/m$^3$, and $H_E^0$ increasing from 3 to 20 mm. This corresponds to decrease of $\beta$, and, if $\beta$ is the parameter controlling the instability, should lead to less unstable behavior.   This is observed in the transition from case 9 ($H_E^0=3$ mm, $\beta=5.984$, rupture) to case 6 ($H_E^0=5$ mm, $\beta=3.666$, saturation). At larger $H_E^0$, however, rupture of the electrolyte layer is observed (cases 10-12), even though increase of $H_E^0$ within this group  leads to slower growth of the unstable wave (compare the signals for cases 10 and 12 in Fig.~\ref{fig9}). We note that special care has been given to assure adequate grid resolution of the flow in these cases. For example, in the case 12, the fine mesh with $\Delta y=\Delta y_{min}=2\times 10^{-4}$ m was used at $0.02\le y\le 0.08$. This can be compared with the highest (just before the rupture) position of the upper interface at about $y=0.0668$. We conclude that the most plausible explanation of the behavior in the cases 6, 9-12 is that $\beta$ is an imperfect instability parameter for such systems.

The analysis\cite{Horstmann:2018} also shows that, for solutions with $\Delta \eta^A\gg \Delta \eta^B$,  the behavior of the wave on the upper interface is close to that of the wave developing in a two-layer system. The presence of the lower interface has only small impact. The fact that such solutions are found in our and other\cite{Weber:2016,Horstmann:2018,Zikanov:2018shallow,Tucs:2018,Molokov:2018,Herreman:2019} studies of the Mg-Sb cells is, therefore, consistent with the detected at least partial role of $\beta$ derived for a two-layer aluminum reduction cell as the instability parameter.

\begin{figure}
\begin{center}
\includegraphics[width=0.5\textwidth]{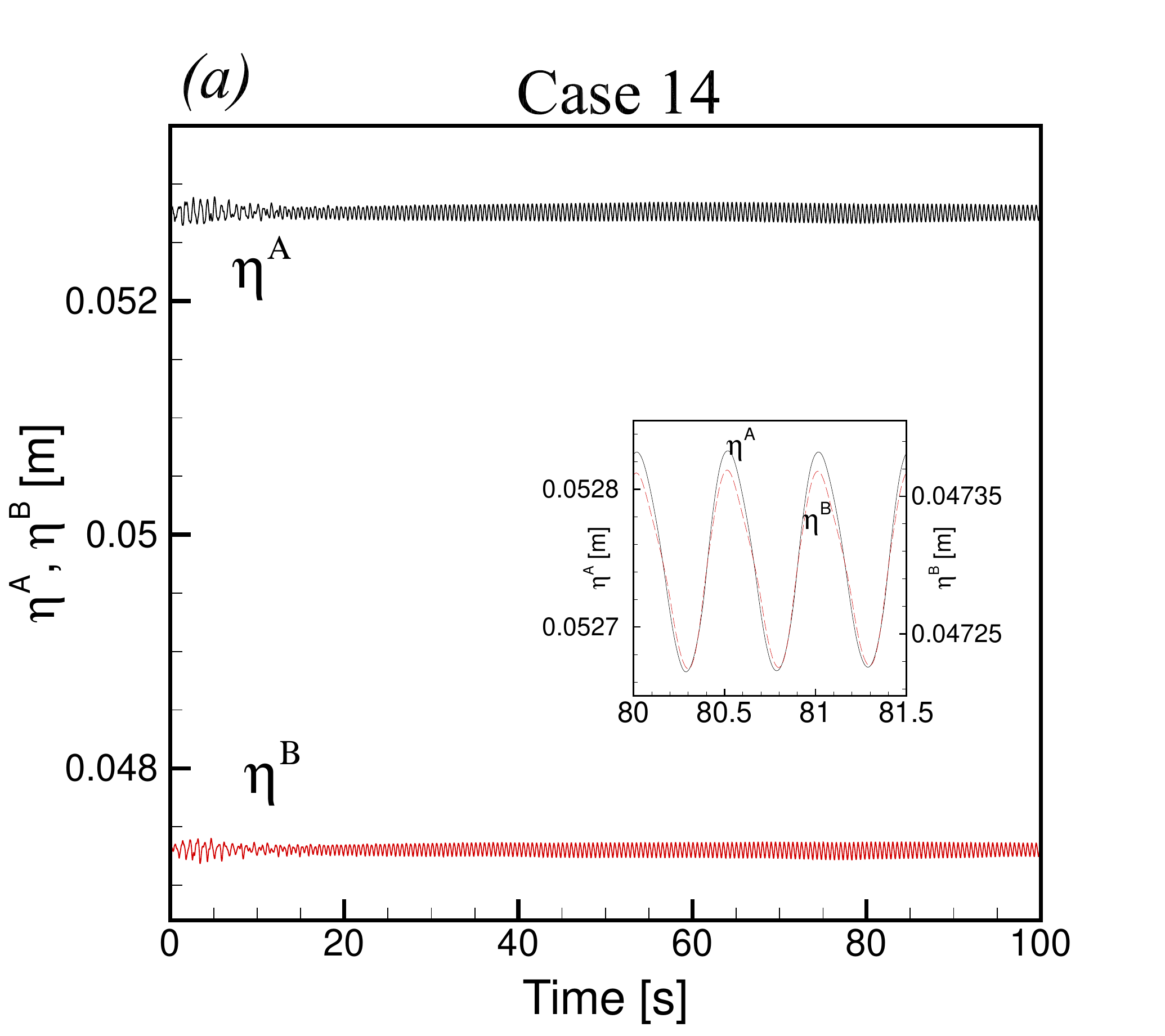}\includegraphics[width=0.5\textwidth]{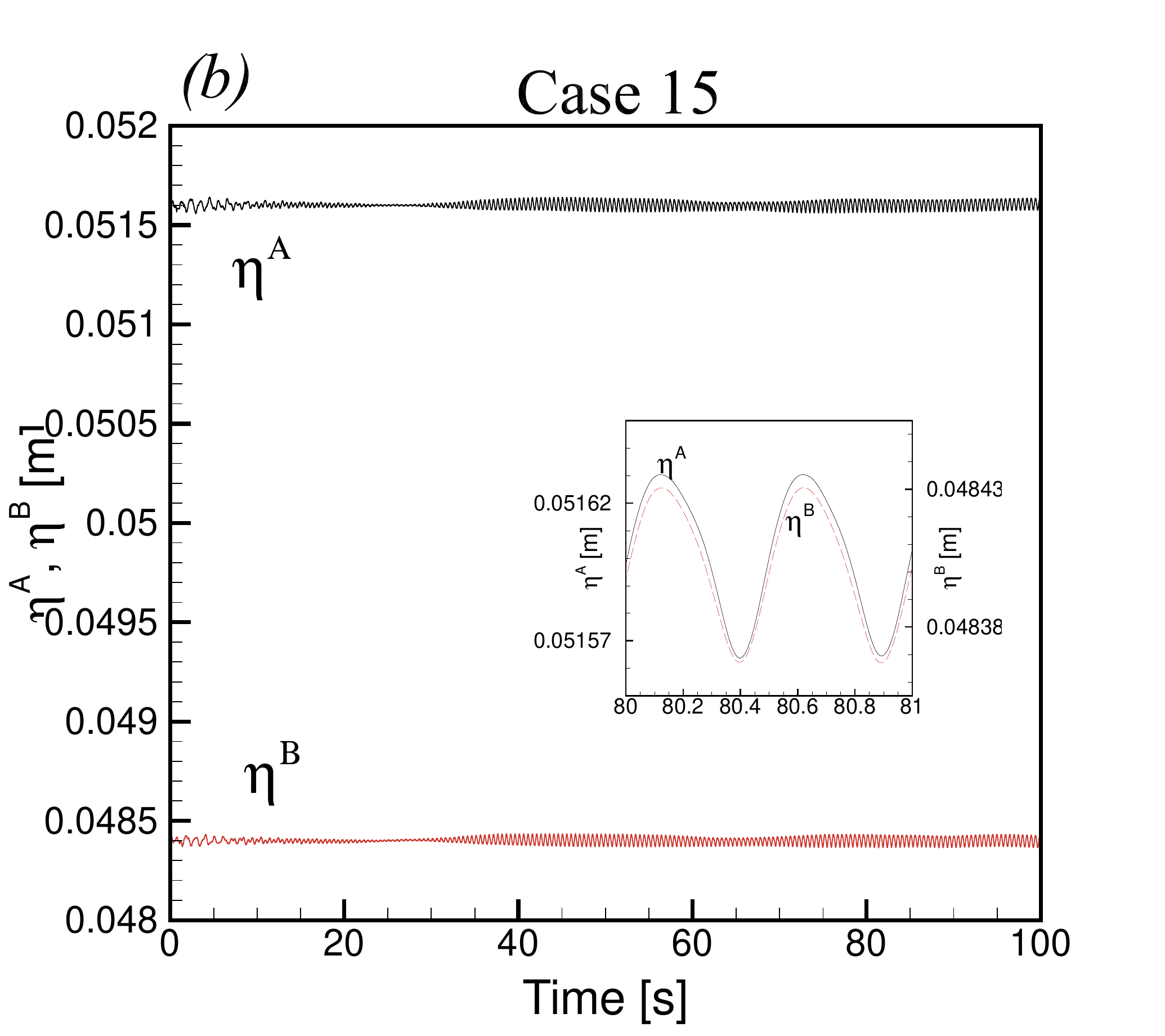}\\
\includegraphics[width=0.5\textwidth]{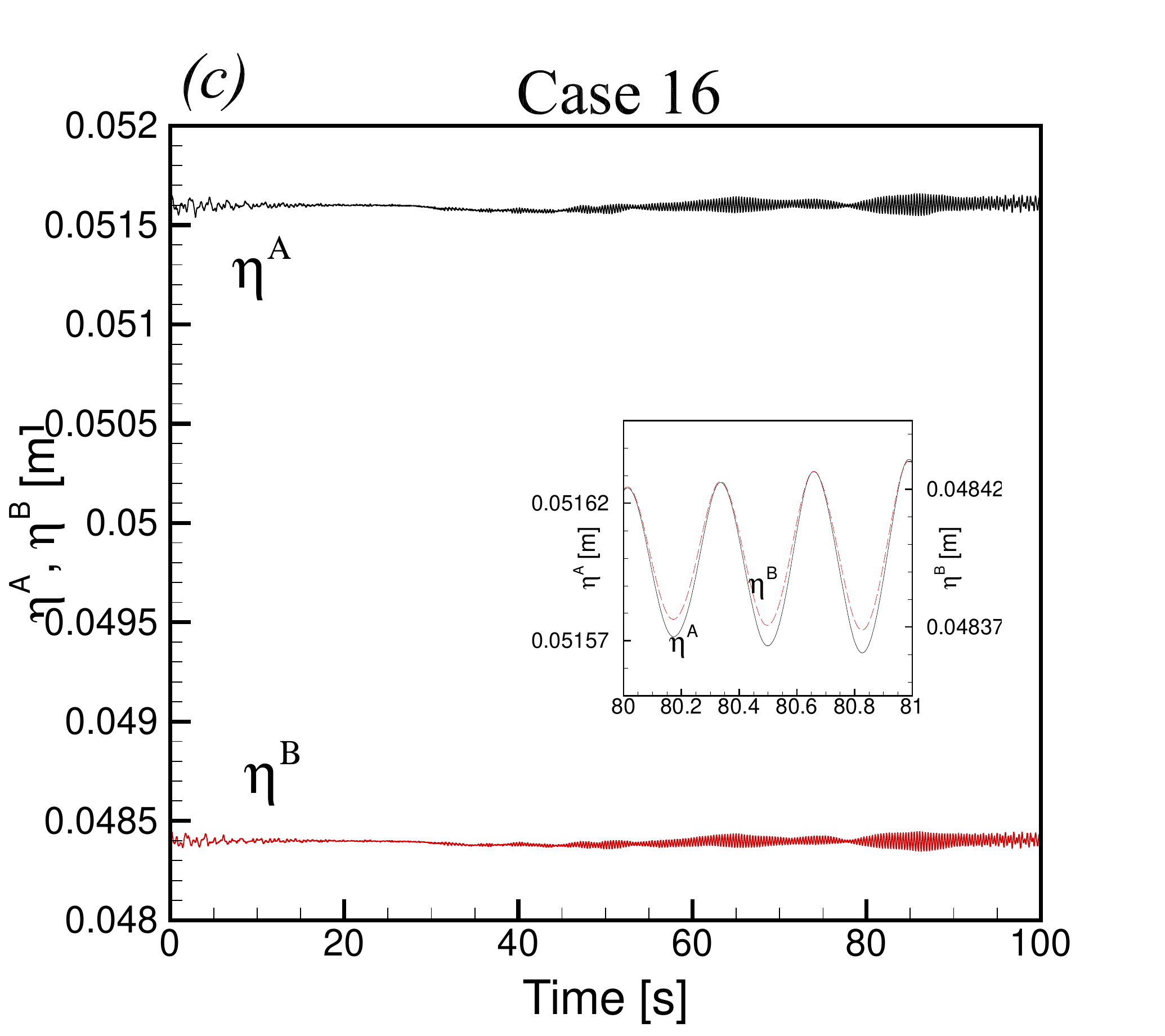}\includegraphics[width=0.5\textwidth]{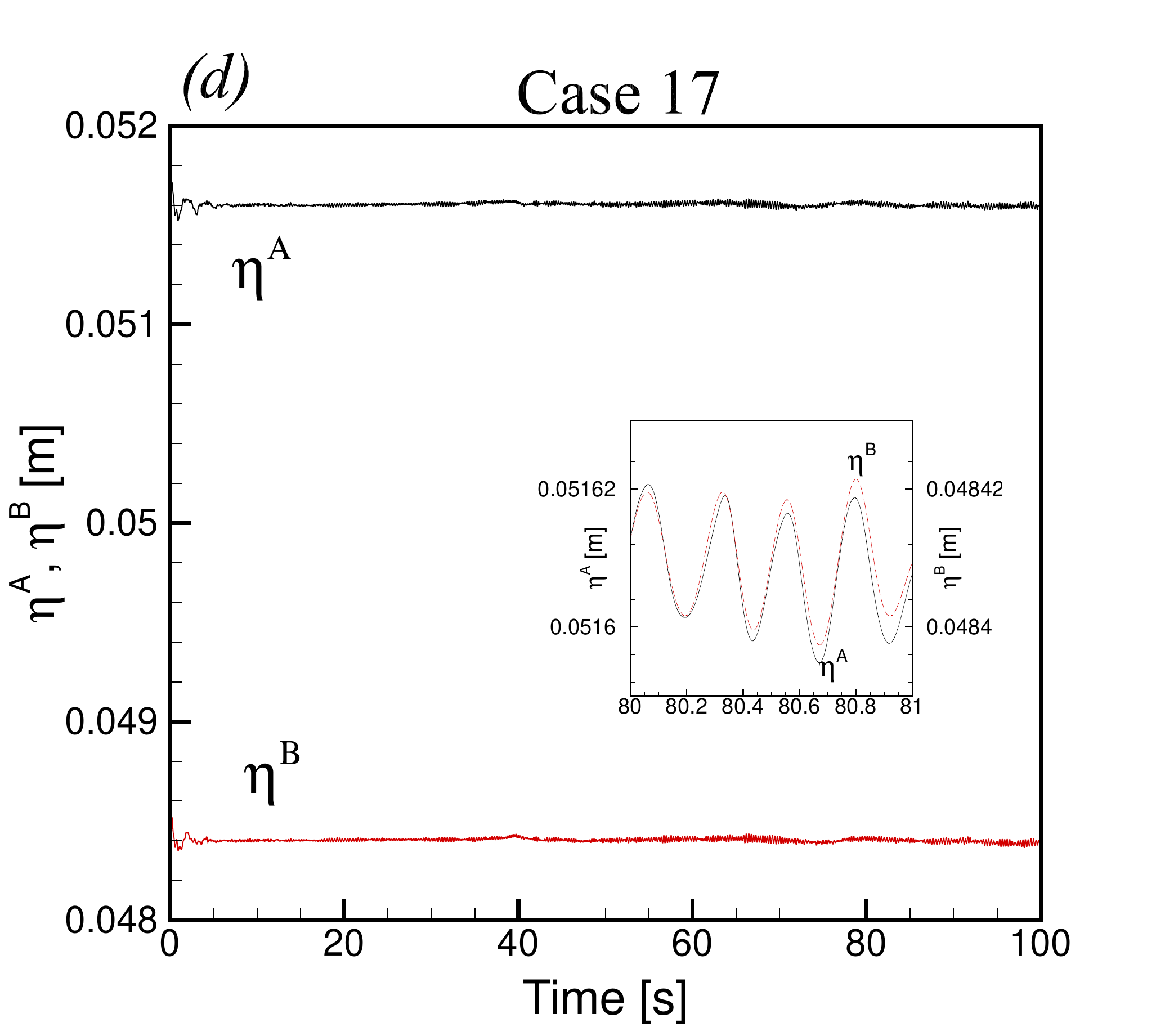}
\caption{Effect of $B_0$ on flows with $\rho_E=3452.2$ kg/m$^3$ ($\Delta \rho_A = \Delta \rho_B$). Time signals of the locations of the upper ($\eta^A$) and lower ($\eta^B$) interfaces at $x=0.05$ m, $z=0.016$ m during the entire simulation are shown. Insets illustrate the behavior of fully developed waves during several wave periods. Note that in the Insets the curves are shifted, but the vertical scales are the same for $\eta^A$ and $\eta^B$. Results obtained for the cases 14, 15, 16, and 17 are shown in \emph{(a)}, \emph{(b)}, \emph{(c)} and \emph{(d)} (see Table II for parameters and characteristics). }
\label{fig10}
\end{center}
\end{figure}

In the cases 14-17, the electrolyte density $\rho_E$ is such that the density differences across the lower and upper interfaces are the same. The system is stable in all these cases even though $\beta$ can be as large as 6 (in case 17). No consistent growth of perturbations is found. Instead, the perturbations decay initially and then the flow shows nearly periodic oscillations of approximately constant small amplitude. An illustration in the form of the signals of interface locations is provided in Fig.~\ref{fig10}. We see that the  waves are coupled symmetrically (in-phase with each other) and have much smaller time period (between 0.24 and 0.5 s) than the waves observed in the cases 1-13. This allows us to identify them as the fast modes according to the classification of Ref.~\onlinecite{Horstmann:2018}. We also see in Fig.~\ref{fig10} that the amplitudes of the upper and lower interfacial waves are nearly equal to each other. 
 
\begin{figure}
\begin{center}
\includegraphics[width=0.4\textwidth]{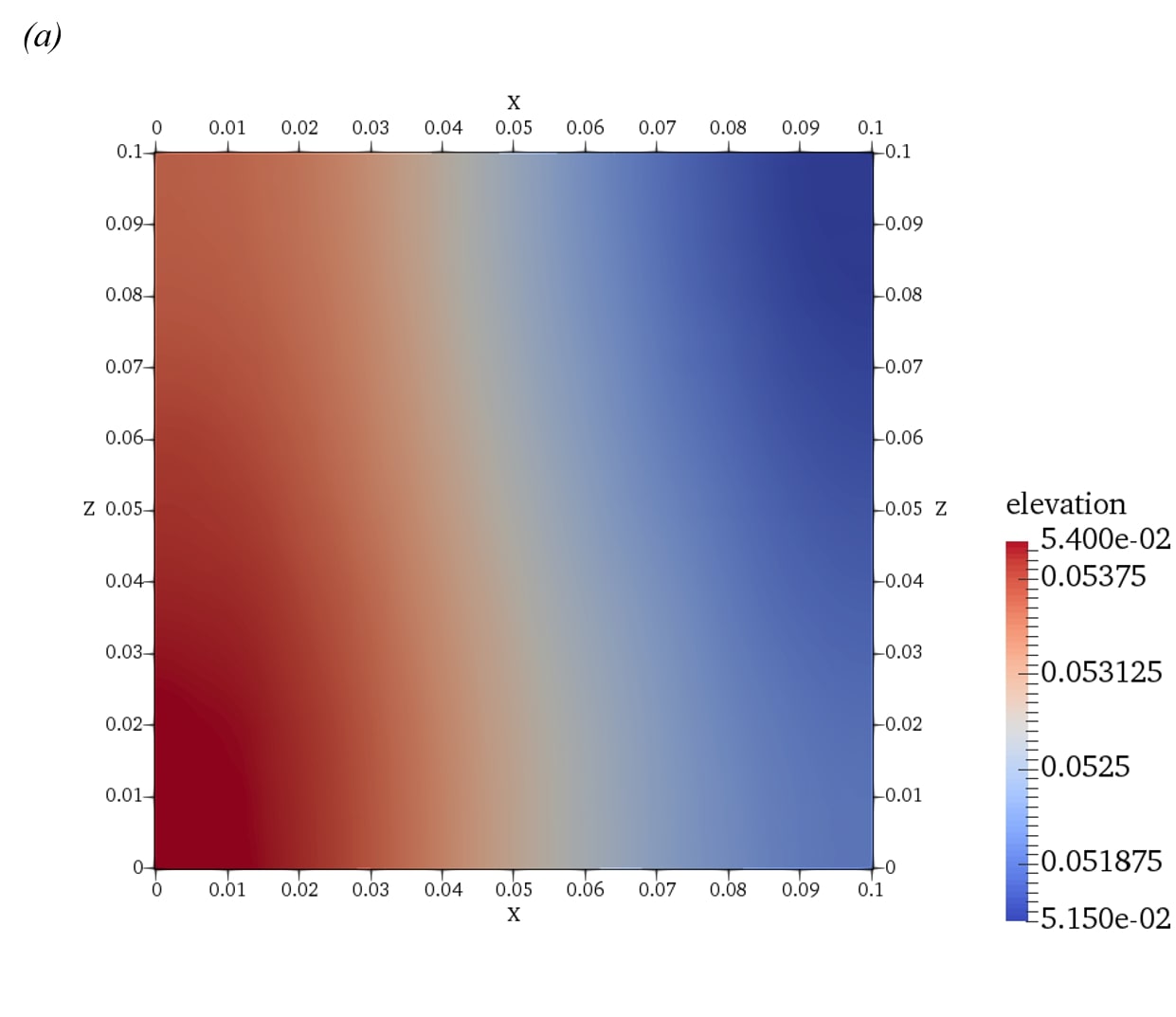}\includegraphics[width=0.4\textwidth]{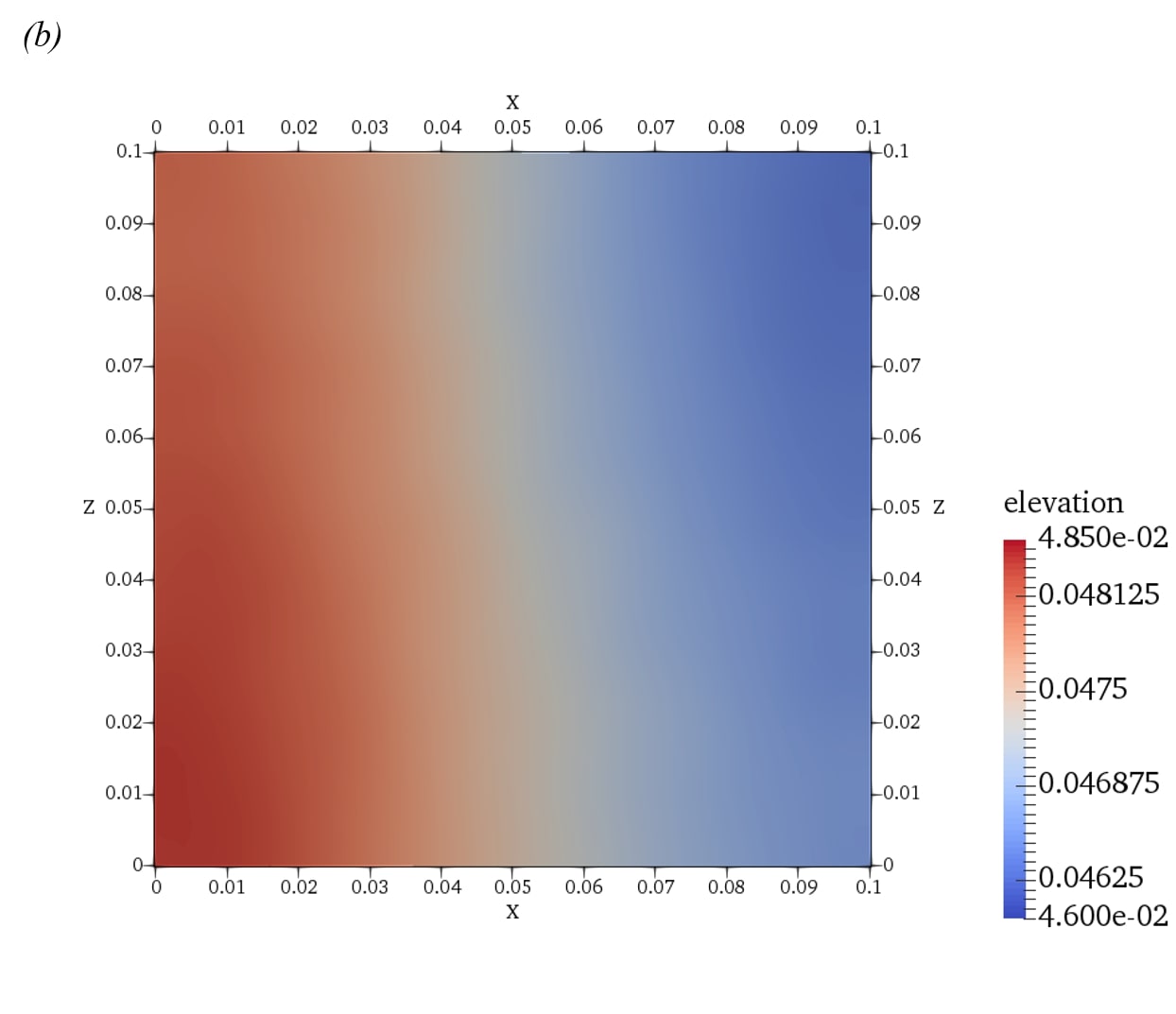}\\
\includegraphics[width=0.4\textwidth]{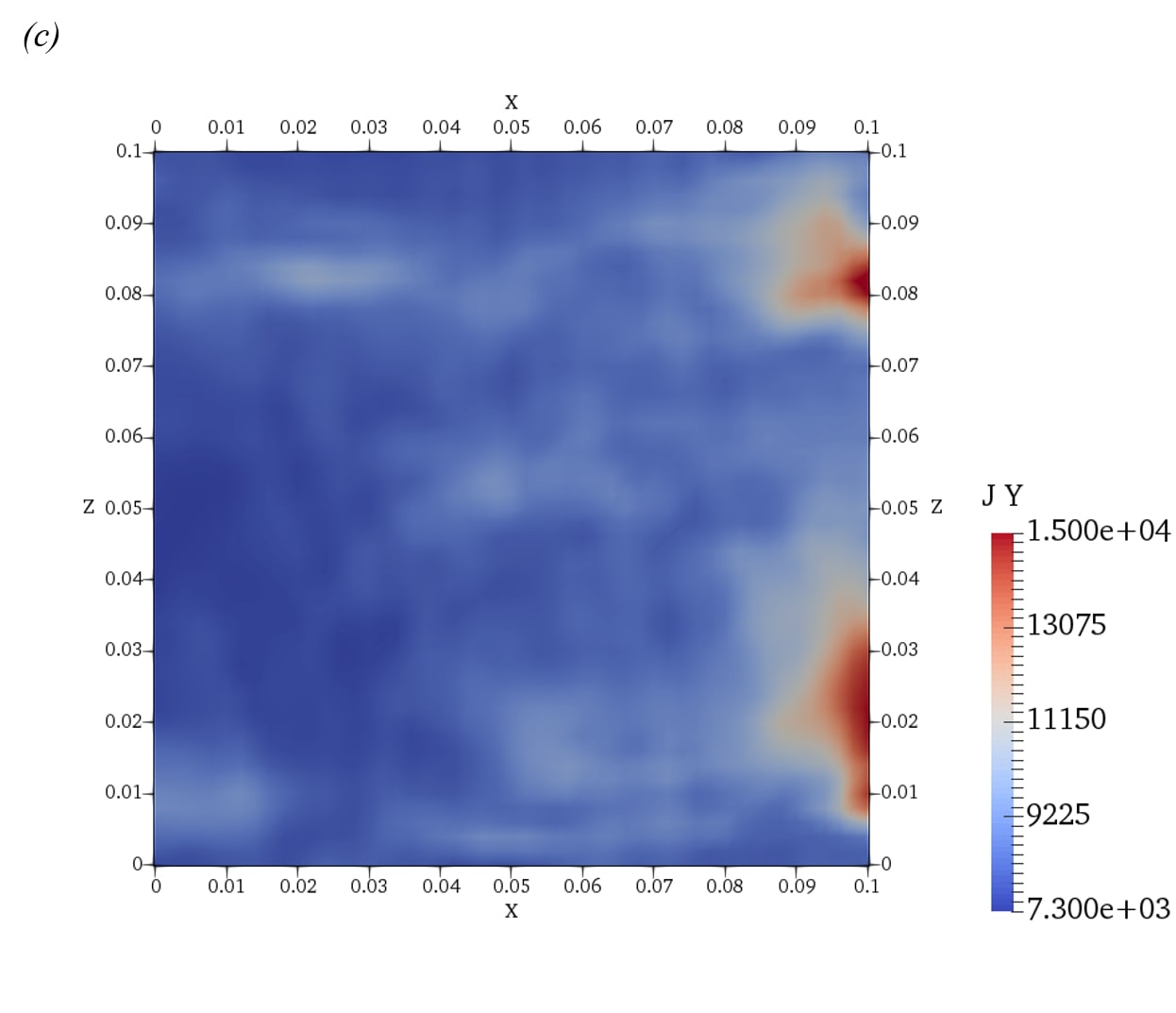}\includegraphics[width=0.4\textwidth]{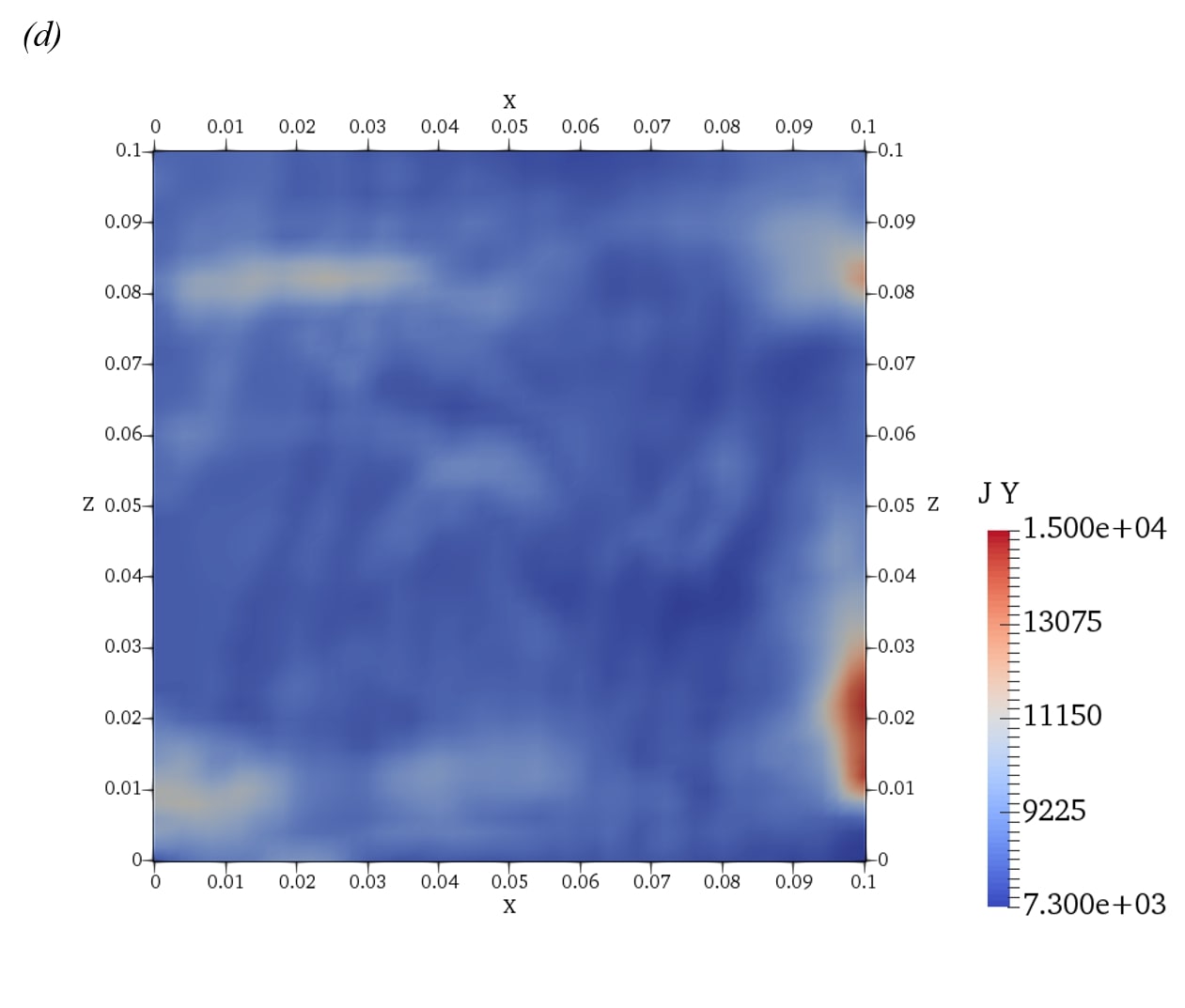}\\
\includegraphics[width=0.4\textwidth]{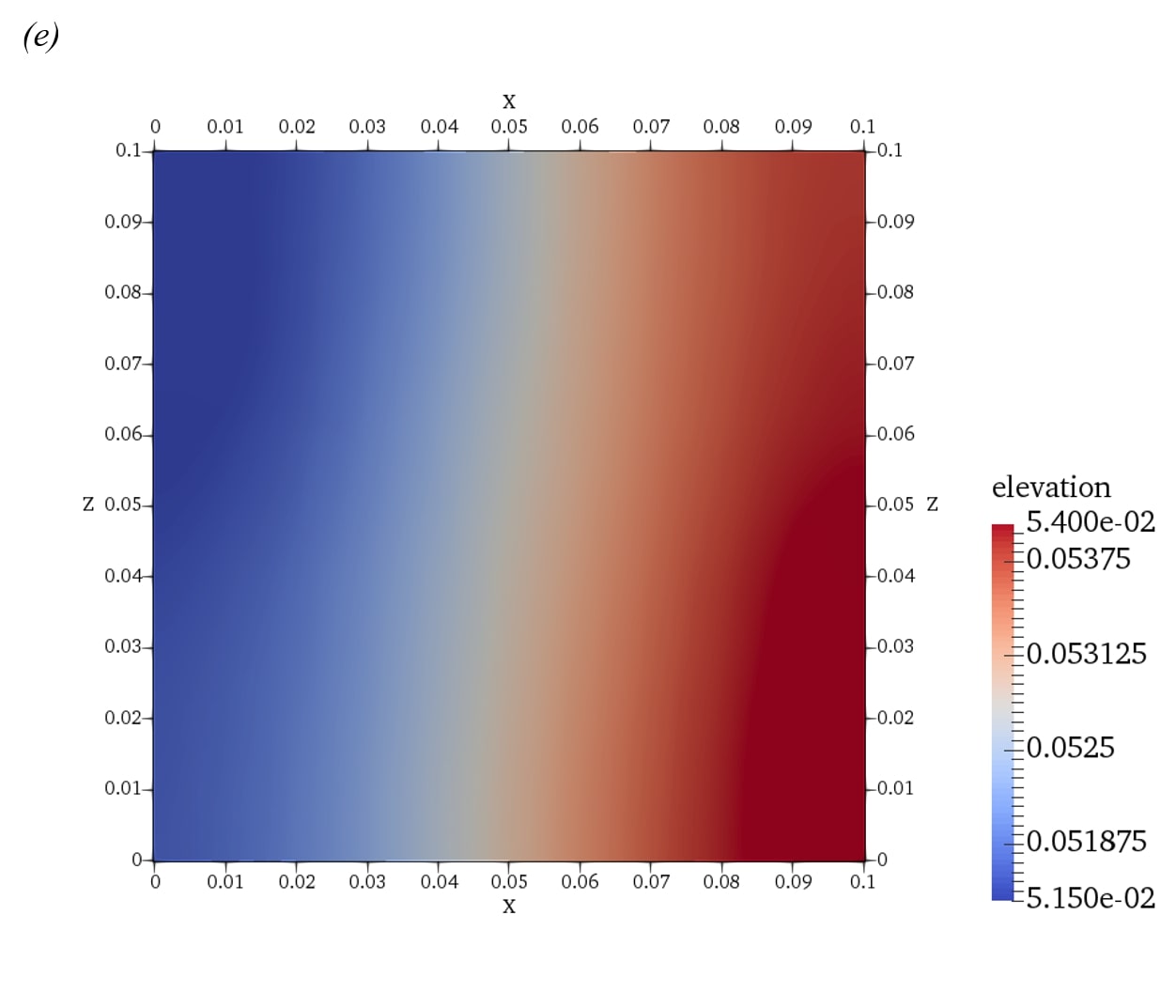}\includegraphics[width=0.4\textwidth]{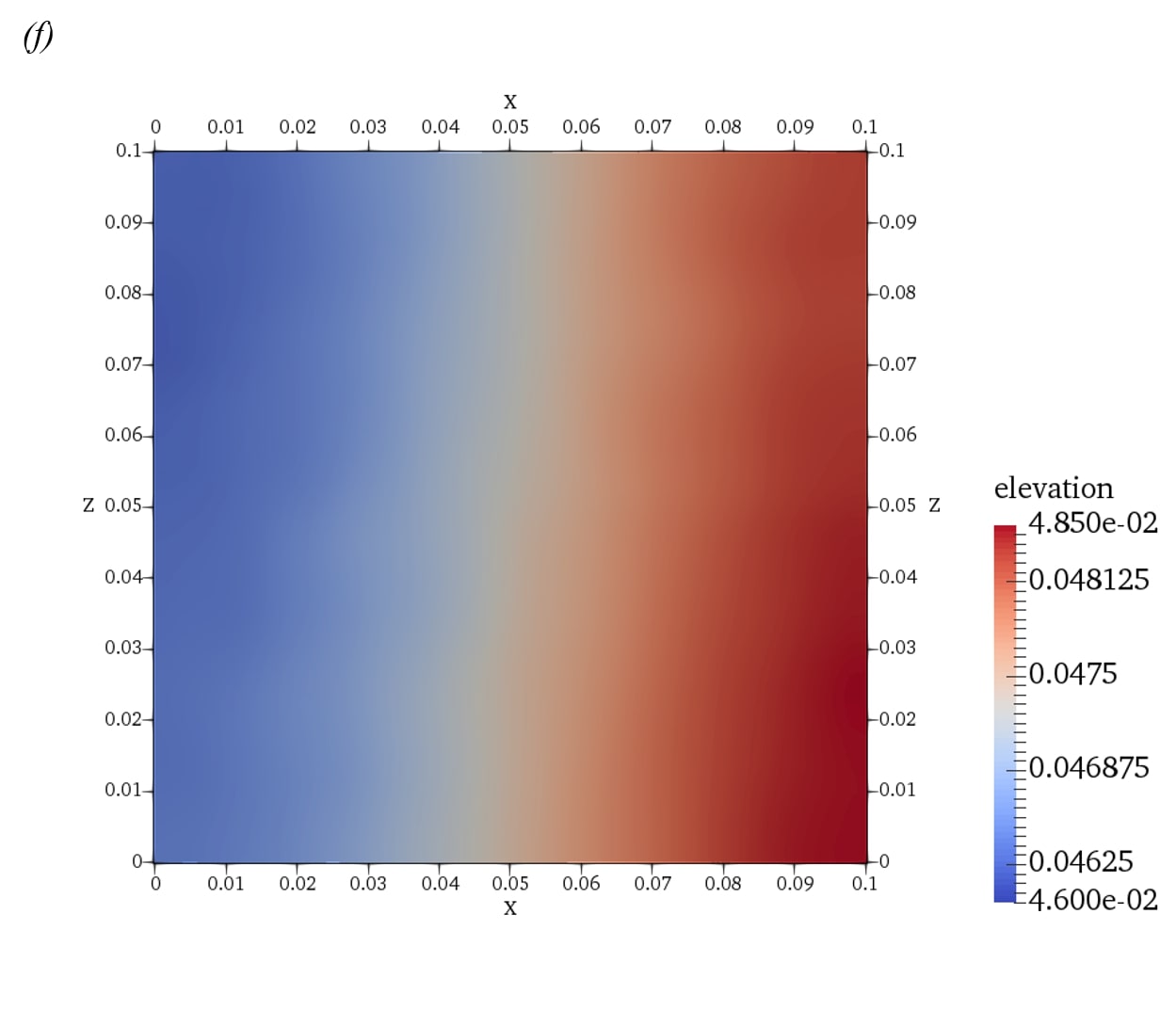}\\
\caption{Stable flow at $H_E^0=5$ mm, $B_0=15$ mT, $\rho_E=3452.2$ kg/m$^3$ (the case 14 in table II).  \emph{(a)} and \emph{(b)}, profiles of the upper ($\eta^A(x,z)$) and lower  ($\eta^B(x,z)$) interfaces at $t=50$ s;  \emph{(c)}, vertical component $J_y$ of electric current in the electrolyte (y = 0.05 m) at $t=50$ s;  \emph{(d)}, vertical component $J_y(x,z)$ of electric current in the electrolyte (y = 0.05 m) at $t=60$ s;  \emph{(e)} and \emph{(f)}, profiles of the upper ($\eta^A(x,z)$) and lower  ($\eta^B(x,z)$) interfaces at $t=60$ s. }
\label{fig11}
\end{center}
\end{figure}

The spatial structure of the flow is illustrated in Fig.~\ref{fig11} as spatial distributions of $H_E(x,z,t)$ on upper ($\eta^A$) and lower interface ($\eta^B$). The combination of symmetric coupling and nearly equal oscillation amplitudes implies that the local thickness of the electrolyte layer is nearly constant. Some variations of $H_E(x,z,t)$ and  $J_y(x,z,t)$ on the small length scale are found, but they are of small and not growing amplitude. We conclude that the fast mode solutions realized in our system at $\Delta \rho_A=\Delta \rho_B$ do not generate significant Lorentz forces and, thus, cannot lead to the rolling pad instability. 

Another feature of the interfacial waves in flows with $\Delta \rho_A=\Delta \rho_B$ is that, unlike in the unstable flows at $\Delta \rho_A \ll \Delta \rho_B$, the shape of the wave remains strongly correlated with the shape imposed as an initial perturbation. We see in Fig.~\ref{fig11} that, apart from effect of rotation and development of a wave at the lower interface, the shape resembles the initial constant slope wave 60 s after the start.

\begin{figure}
\begin{center}
\includegraphics[width=0.5\textwidth]{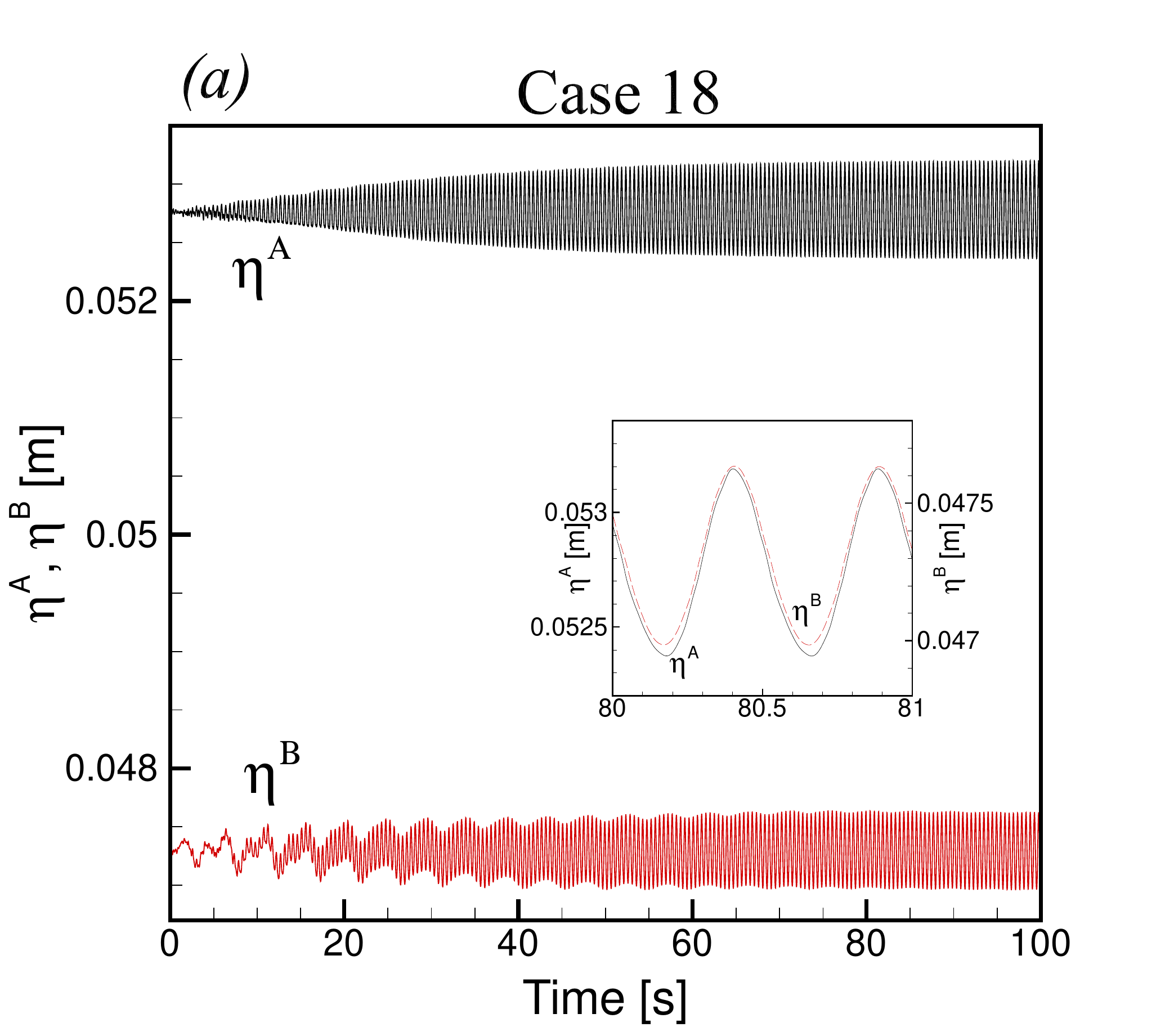}\includegraphics[width=0.5\textwidth]{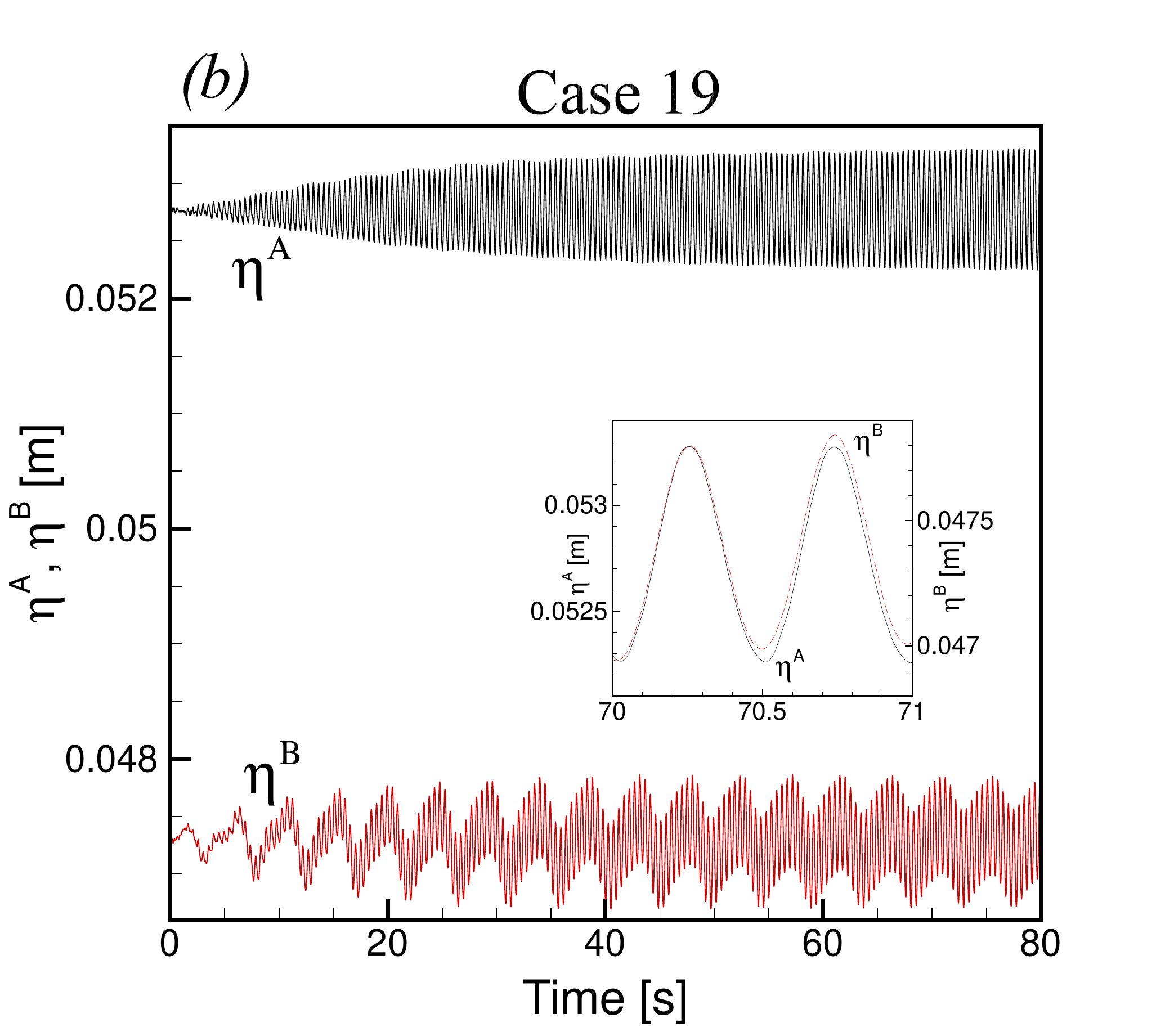}\\
\includegraphics[width=0.5\textwidth]{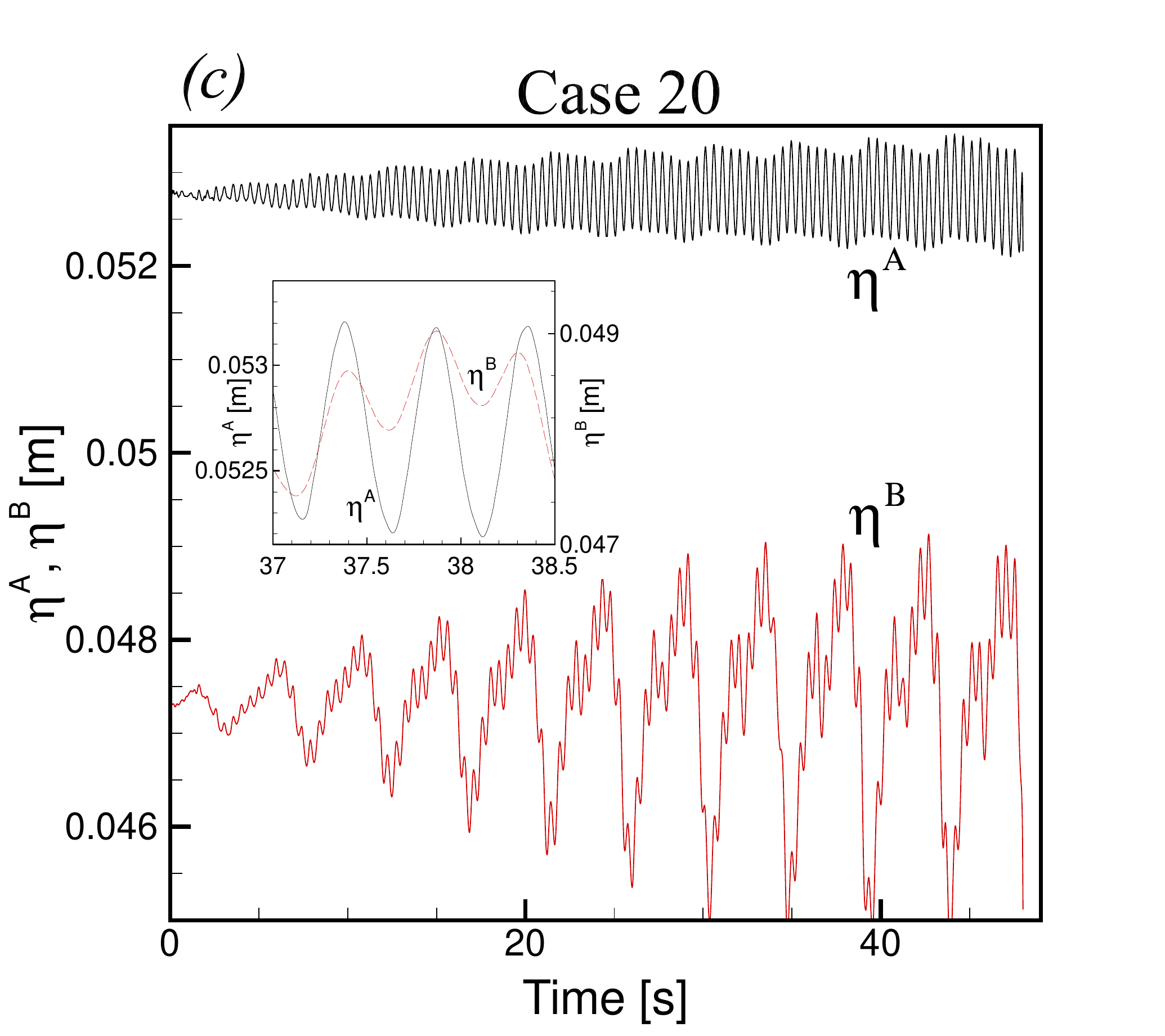}\includegraphics[width=0.5\textwidth]{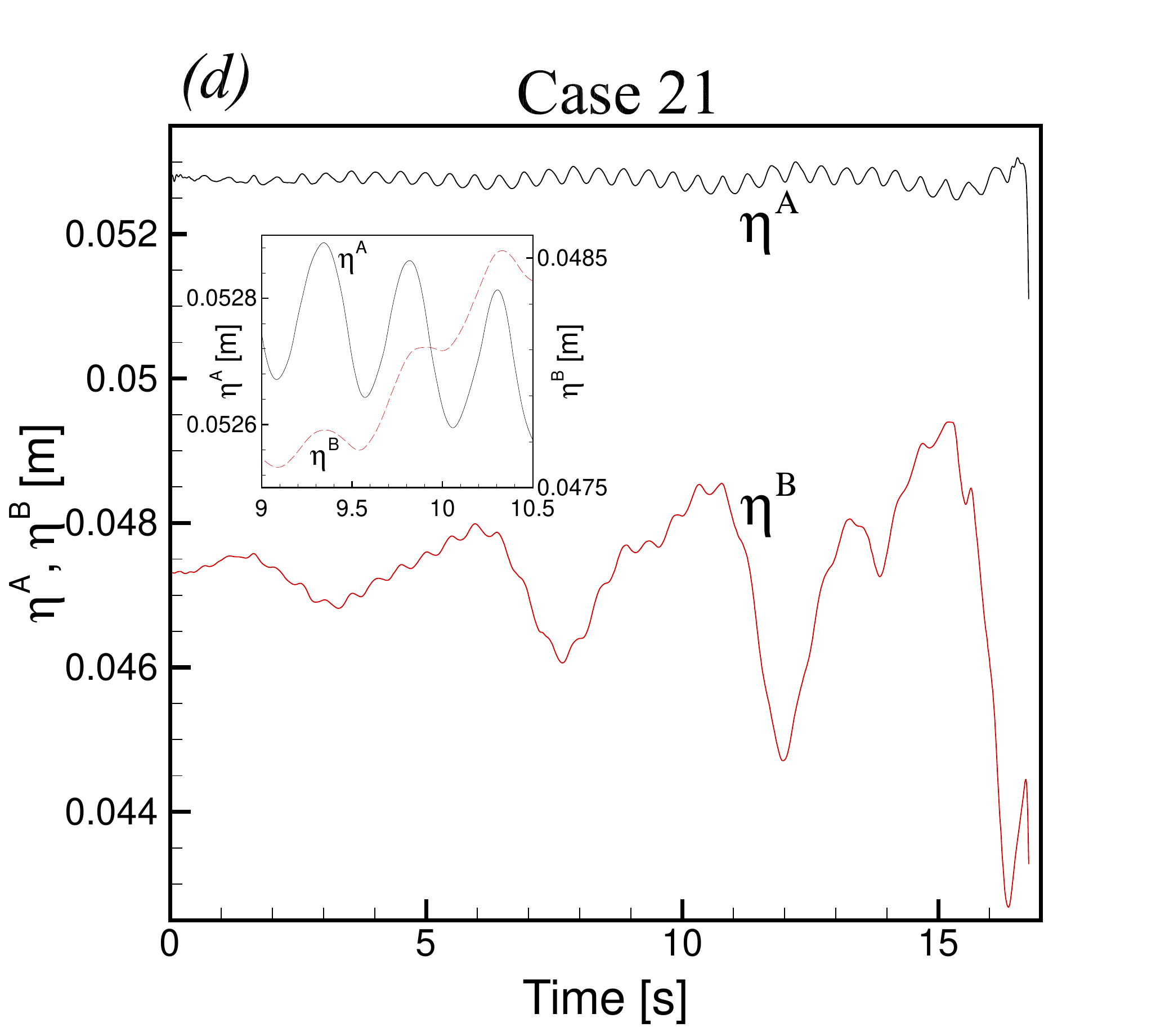}
\caption{Effect of $B_0$ and $H_E^0$ on flows with $\rho_E=5994$ kg/m$^3$ ($\Delta \rho_B \ll \Delta \rho_A$). Time signals of the locations of the upper ($\eta^A$) and lower ($\eta^B$) interfaces at $x=0.05$ m, $z=0.016$ m during the entire simulation are shown. Insets illustrate the behavior of fully developed waves during several wave periods. Note that in the insets the curves are shifted and different vertical scales are used for $\eta^A$ and $\eta^B$. Results for the cases 18, 19, 20, and 21 are shown in \emph{(a)}, \emph{(b)}, \emph{(c)} and \emph{(d)} (see Table II for parameters and characteristics). }
\label{fig12}
\end{center}
\end{figure}

The just reported findings can be compared with the results of the two-dimensional model of shallow  battery cells of rectangular cross-section,\cite{Zikanov:2018shallow} to our knowledge the only other study where the case $\Delta \rho_A=\Delta \rho_B$ is considered. Despite the evident differences in the geometry and modeling approach, the results of the two studies are in clear quantitative agreement with each other. Stable solutions in the form of symmetrically coupled waves of approximately equal amplitudes are found in both studies.

The situation when $\Delta \rho_A \gg \Delta \rho_B$ is represented by the cases 18-21. As illustrated in Fig.~\ref{fig12}, the flow can be unstable. Growth of perturbations occurs and ends in saturation in the flow with $\beta$ (based on $\Delta \rho_B$) equal to 0.917 (case 18) and 1.833 (case 19). Growth leading to rupture of electrolyte occurs at $\beta=3.666$ (case 20) and 6.11 (case 21). 
The analysis of the distributions of flow variables and time signals, such as those in  Fig.~\ref{fig12}, leads us to the conclusion that the solutions are superpositions of fast and slow modes. The fast modes cause the high-frequency (period of about 0.5 s) symmetric oscillations with approximately equal amplitudes on both interfaces. The slow modes are responsible for low-frequency (period between 4.4 and 4.7 s) antisymmetric oscillations. Their amplitudes cannot be accurately determined from the data in Fig.~\ref{fig12}. For this reason only the amplitude ratios of the fast mode oscillations are listed in Table II. Nevertheless we clearly see that the amplitude of the slow mode is significantly higher on the lower than on the upper interface. Interestingly, the slow modes are only noticeably strong when the perturbations are growing, i.e. during the  initial growth stage in the saturation cases 18 and 19 and during the entire flow evolution in the rupture cases 20 and 21. During the saturation phase in cases 18 and 19, the amplitude of the slow mode oscillations gradually decreases until they practically disappear. 

\begin{figure}
\begin{center}
\includegraphics[width=0.5\textwidth]{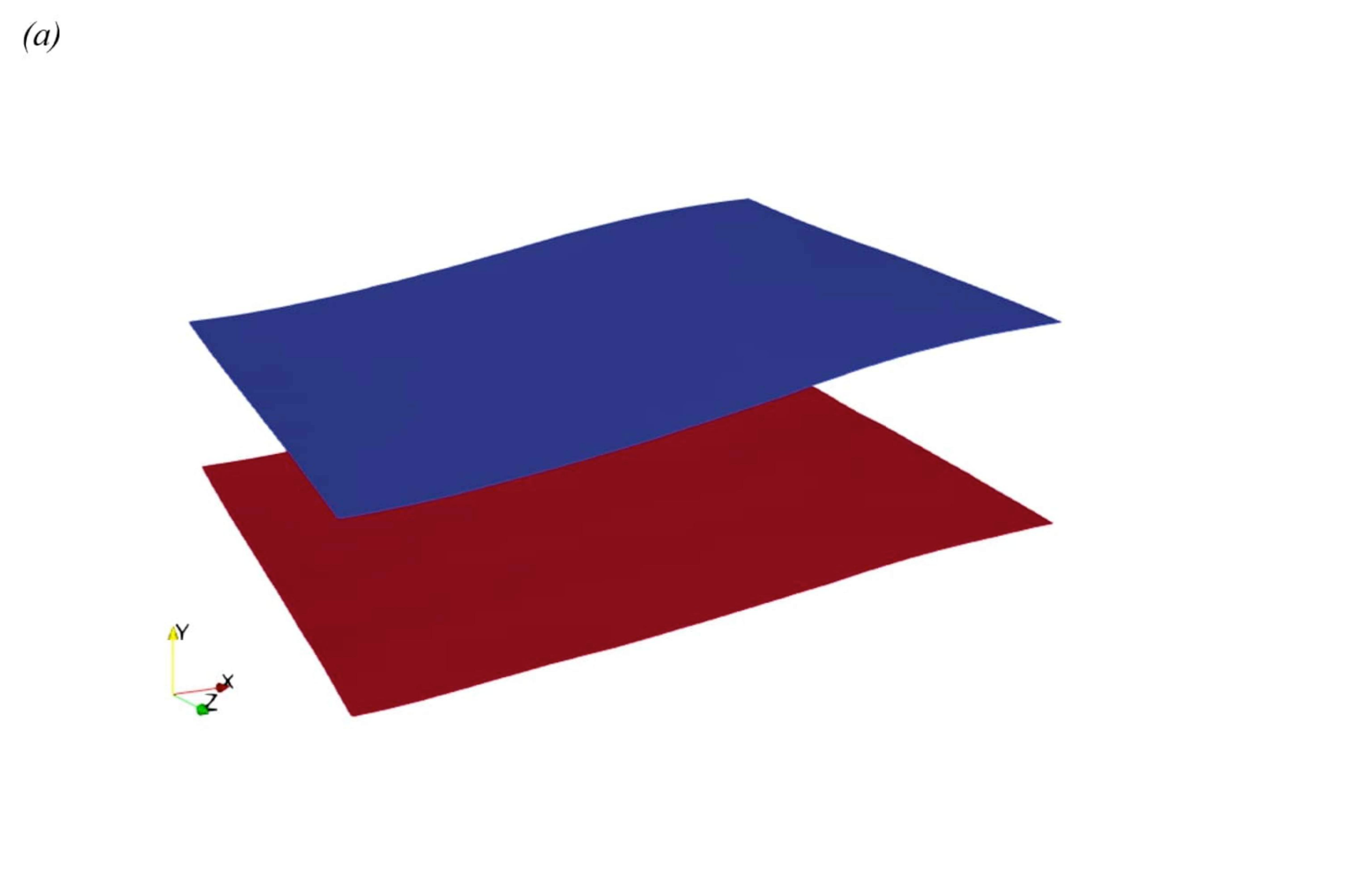}\includegraphics[width=0.5\textwidth]{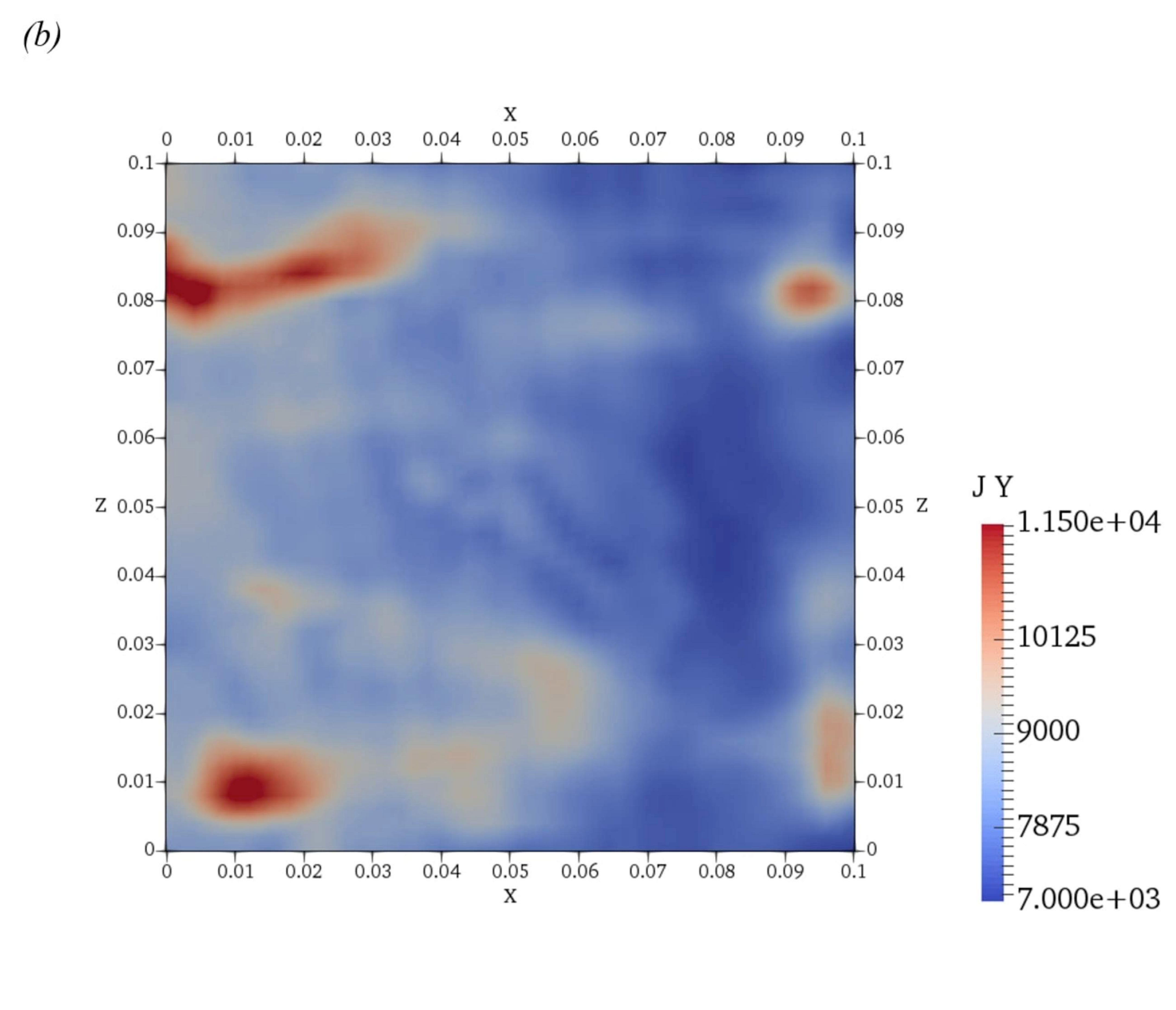}\\
\includegraphics[width=0.5\textwidth]{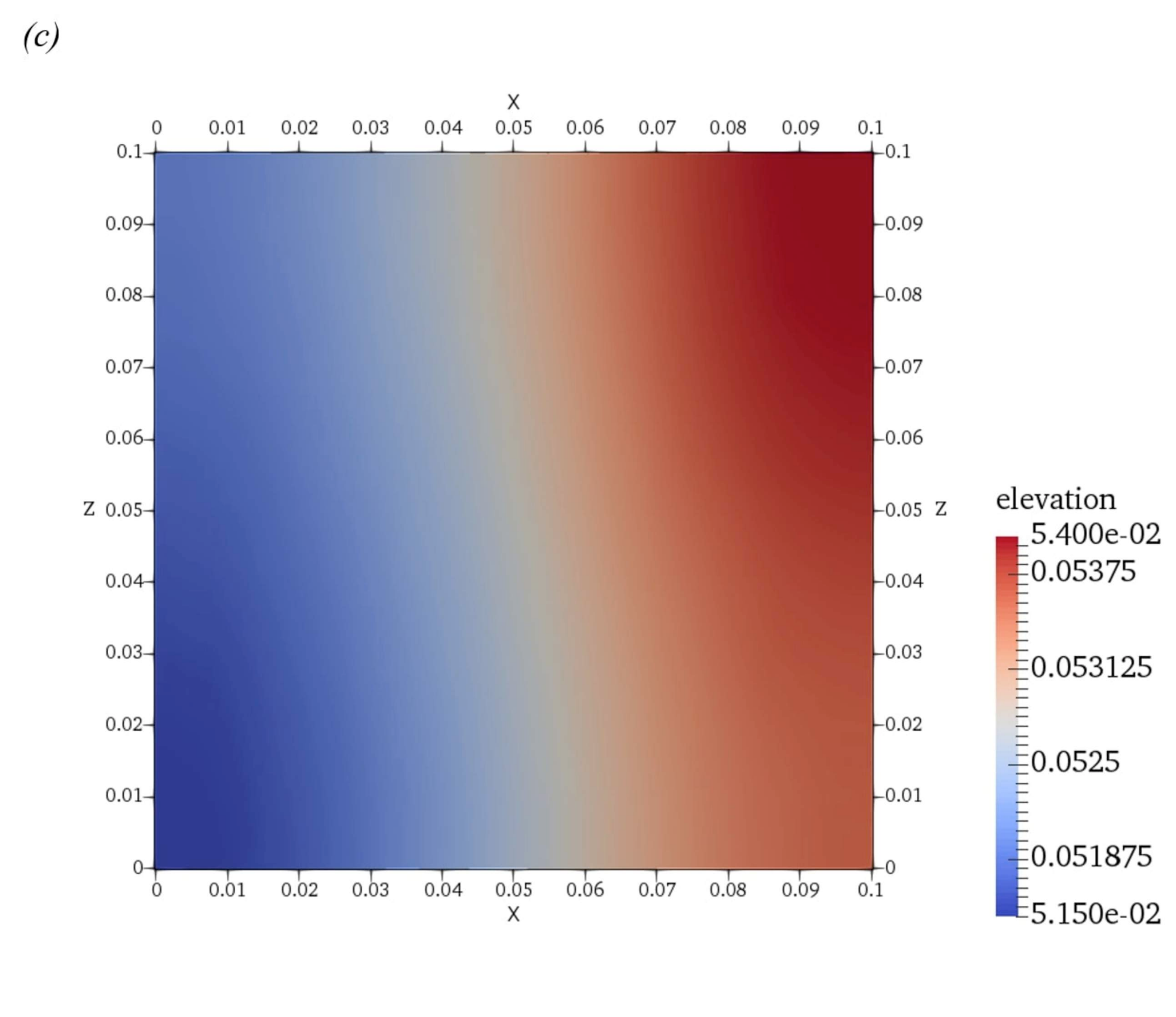}\includegraphics[width=0.5\textwidth]{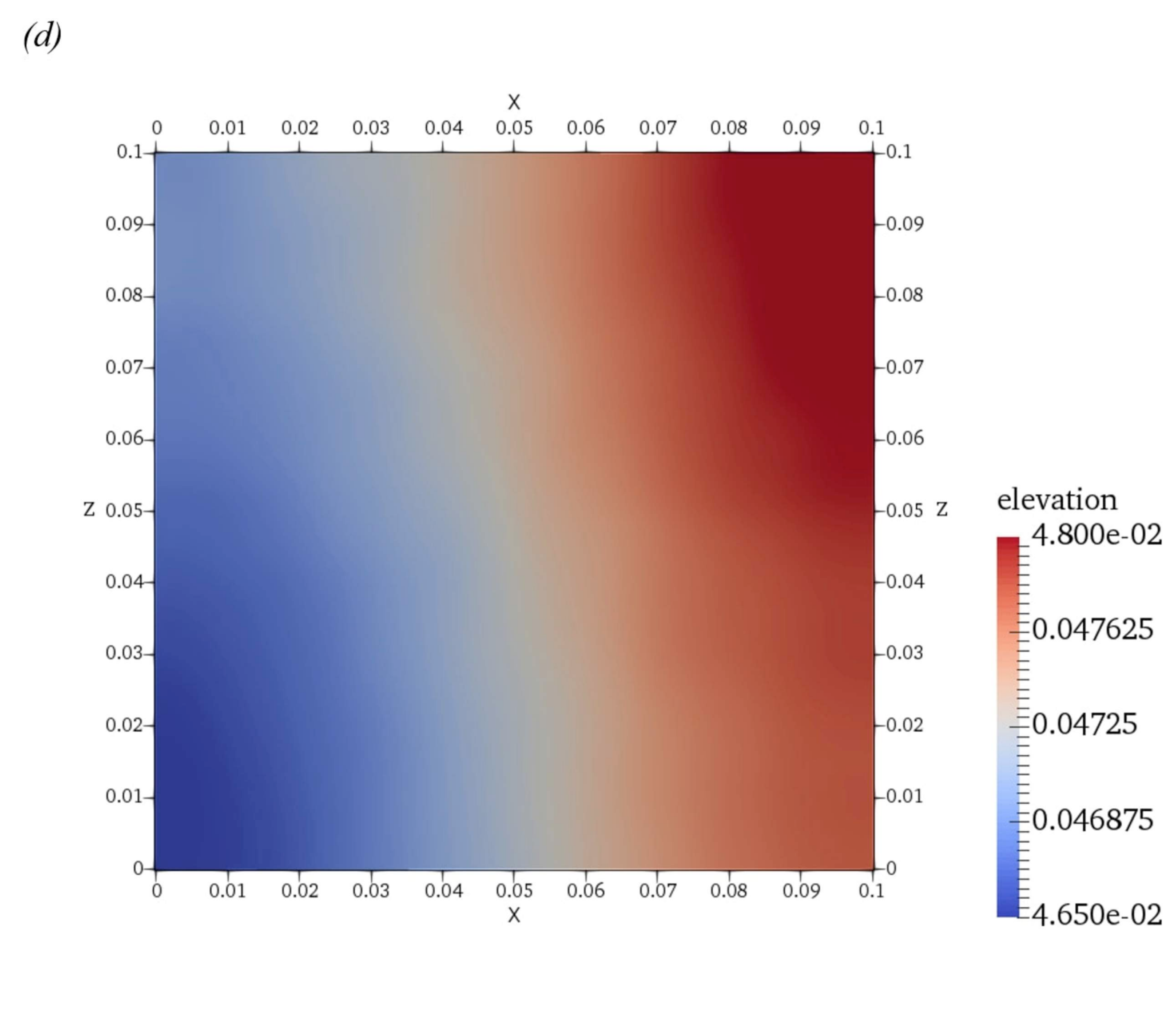}
\caption{Rolling pad instability caused by slow modes at $H_E^0=5$ mm, $B_0=15$ mT, $\rho_E=5994$ kg/m$^3$ (case 19 in table II). \emph{(a)}, flow structure of rolling pad instability caused by slow modes at $t=28$ s; \emph{(b)}, vertical component $J_y$ of electric current in the electrolyte (y = 0.05 m) at $t=28$ s; \emph{(c)} and \emph{(d)}, elevation plot on upper ($\eta^A$) and lower interface ($\eta^B$) at $t=28$ s. }
\label{fig13}
\end{center}
\end{figure}

The typical spatial structure of the flows in the cases 18-21 is illustrated in Fig.~\ref{fig13}. We see that the upper ($\eta^A$) and lower interface ($\eta^B$) are coupled and oscillated symmetrically while the amplitude of oscillation differs from upper and lower interface. This implies variable local electrolyte thickness and, thus,  Lorentz forces and the possibility of unstable behavior.
 
We conclude that in the simulations with $\Delta \rho_A \gg \Delta \rho_B$ (cases 18-21), the rolling pad instability is caused by the slow modes developing at the lower interface. Unlike the flows with $\Delta \rho_A \ll \Delta \rho_B$, however, the unstable waves evolve in the presence of strong fast mode oscillations.

\begin{figure}
\begin{center}
\includegraphics[width=0.5\textwidth]{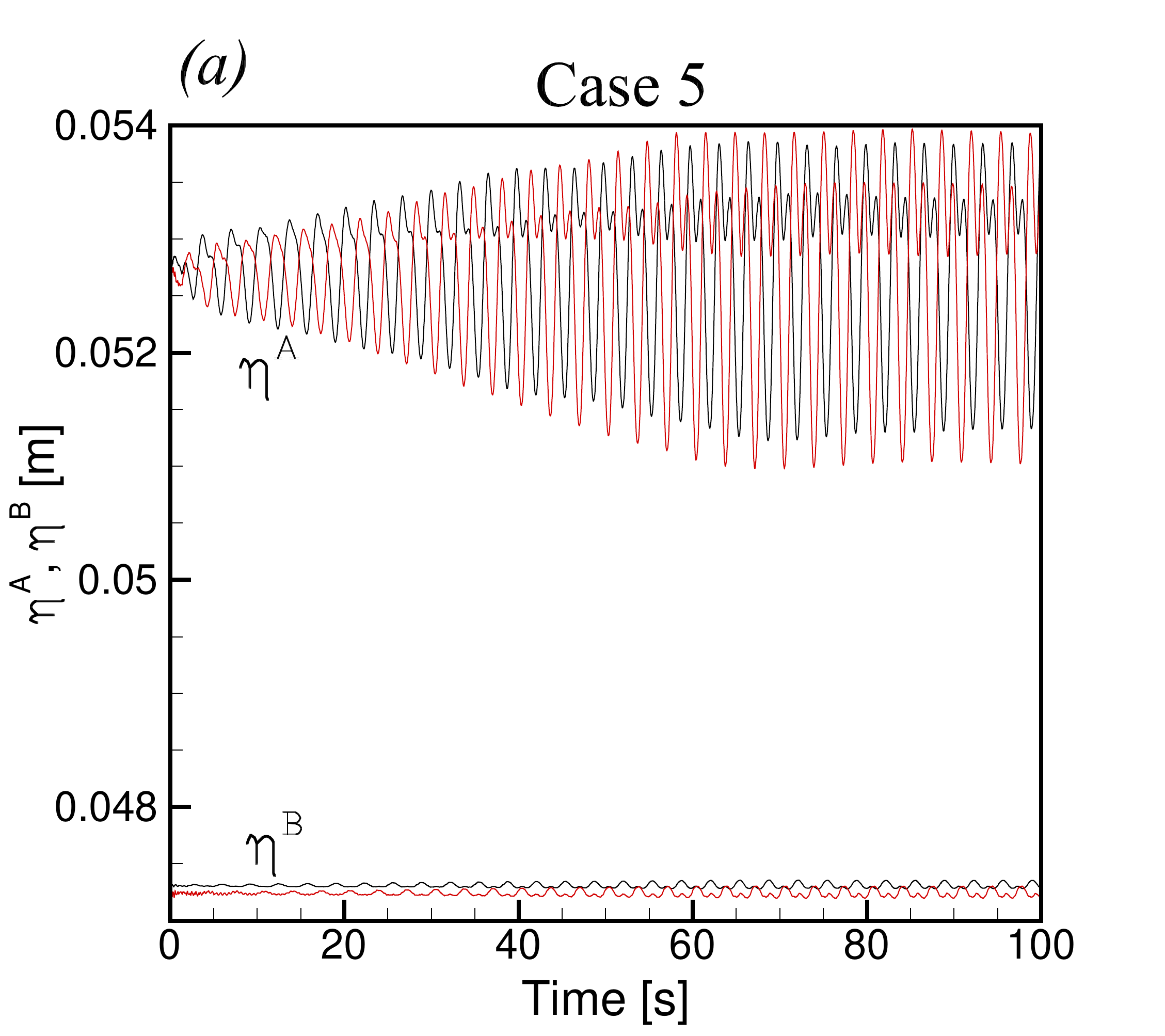}\includegraphics[width=0.5\textwidth]{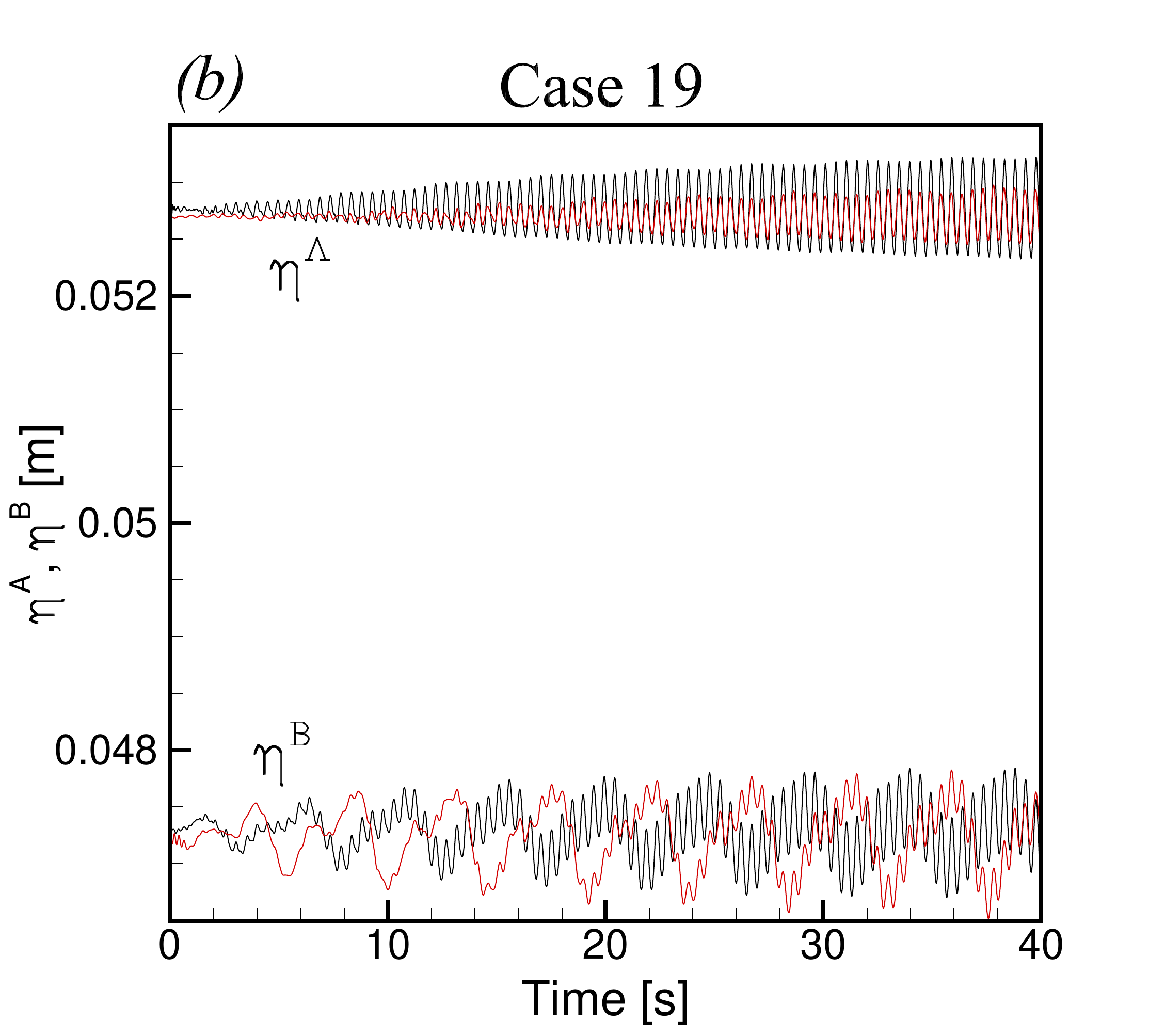}
\caption{Effect of the form of initial perturbations illustrated by the results for the case 5 ($H_E^0=5$ mm, $B_0=10$ mT, $\rho_E=1715$ kg/m$^3$) in \emph{(a)} and the case 19 ($H_E^0=5$ mm, $B_0=15$ mT, $\rho_E=5994$ kg/m$^3$) in \emph{(b)}.
Time signals of the locations of the upper ($\eta^A$) and lower ($\eta^B$) interfaces at $x=0.05$ m, $z=0.016$ m are shown for the simulations with identical initial incline given to the top (black curves) or bottom (red curves) interfaces. }
\label{fig14}
\end{center}
\end{figure}

We will now discuss the effect of the form of the initial perturbations. All the simulations presented so far were initialized by a slight incline of the upper interface (\ref{deta_1}). The simulations show that shape of the wave is retained by stable flows, in particular by the flows with $\Delta \rho_A=\Delta \rho_B$ (cases 14-17). The situation is quite different for unstable flows, in which the initial conditions only affect the early evolution. The final fully developed state is determined by the instability mechanism.

Our first illustration is the test simulations in which the  initial perturbation of the same shape are applied to the lower rather than upper interface. This does not change the initial distribution of the Lorentz forces, but affects the initial potential energy of the perturbations, specifically its distribution between the waves on the upper and lower interfaces and its total magnitude, which is proportional to the density difference across the perturbed interface. One may justifiably ask whether this would affect the selection between fast and slow modes or affect the solution in some other way.  

The results of the numerical experiments carried out for cases 5 and 19 are shown in Fig.~\ref{fig14}. We see that the change of the initial conditions does not influence the solution in any major way.

\begin{figure}
\begin{center}
\includegraphics[width=0.44\textwidth]{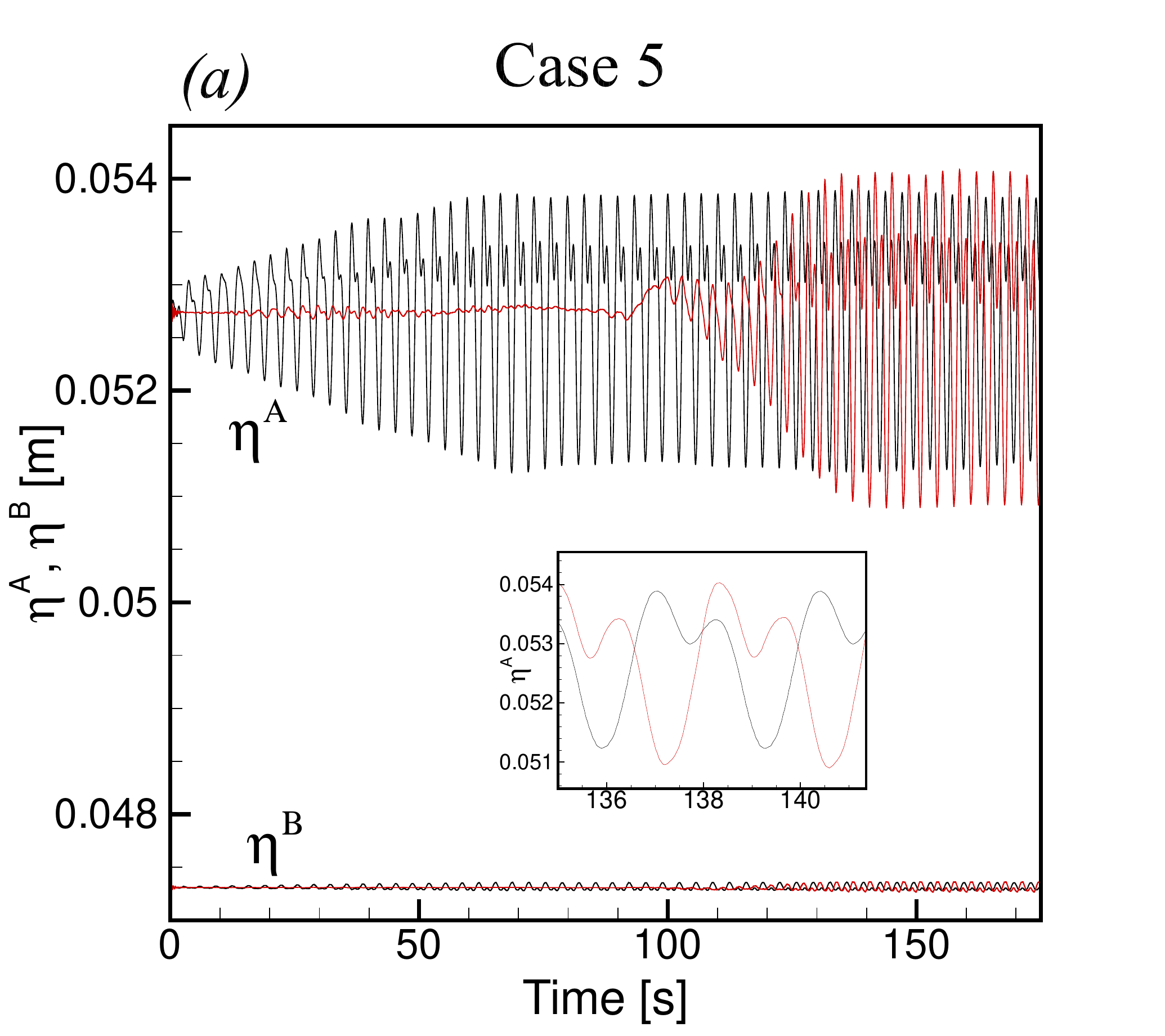}\includegraphics[width=0.5\textwidth]{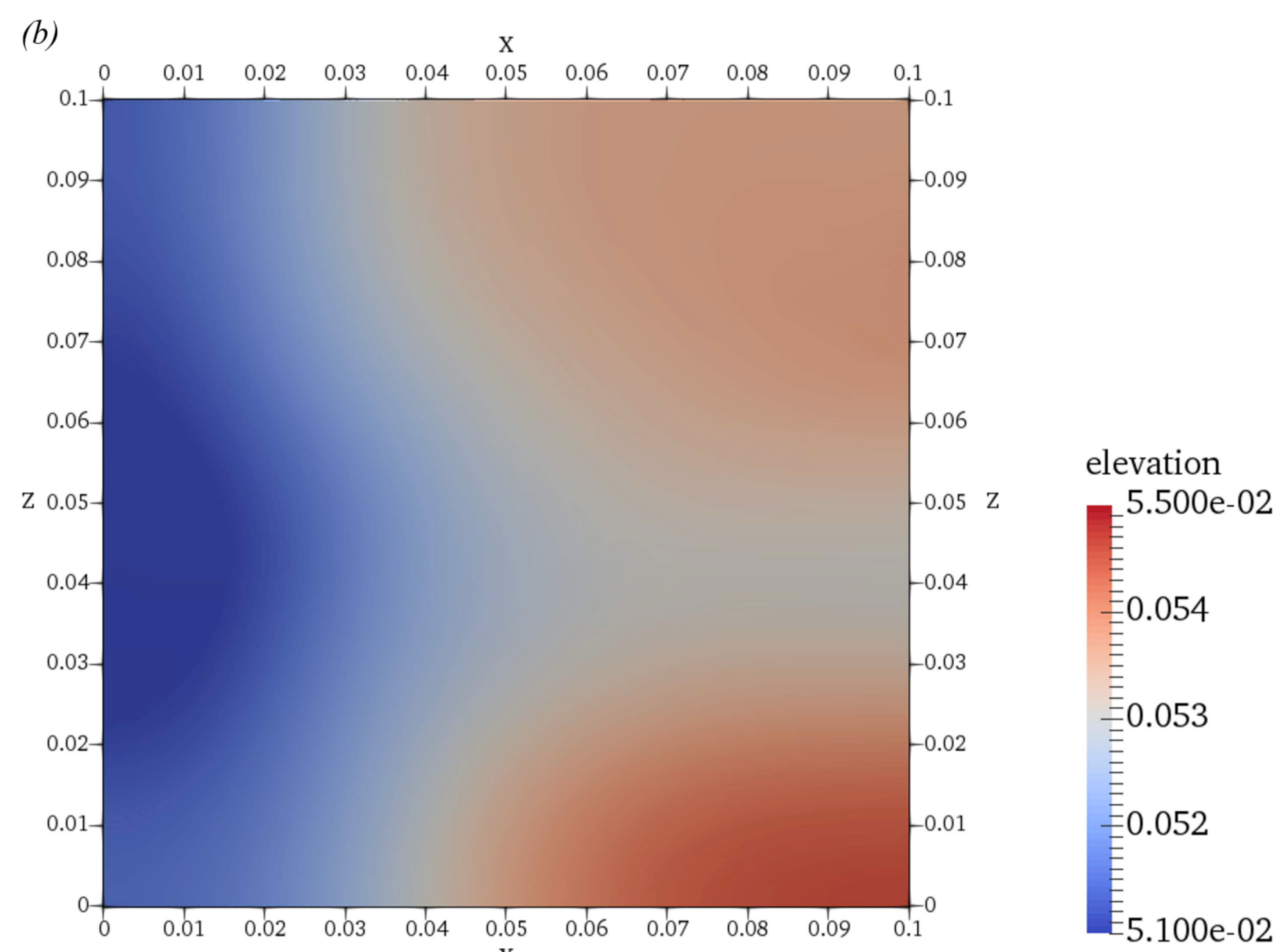}

\caption{Effect of the form of initial perturbations illustrated by the result for the case 5 ($H_E^0=5$ mm, $B_0=10$ mT, $\rho_E=1715$ kg/m$^3$).
\emph{(a)}, time signals of the locations of the upper ($\eta^A$) and lower ($\eta^B$) interfaces at $x=0.05$ m, $z=0.016$ m are shown for the simulations with initial perturbations (\ref{deta_1}) (black curves) and (\ref{deta_2}) (red curves) applied at the top interface. Inset illustrates the similar  behaviors of fully developed waves during several wave periods. \emph{(b)}, elevation plot of the upper interface ($\eta^A$) obtained in the simulations with the initial perturbations (\ref{deta_2}) at t=138s. (compare with Fig.~\ref{fig5}c)}
\label{fig15}
\end{center}
\end{figure}

In the second test, the  perturbations are imposed at the upper interface but have a different shape, smaller  horizontal length scale, and  smaller (with the maximum of 0.25 mm) amplitude: 
\begin{equation}\label{deta_2}
    \Delta \eta^A=0.00025 \cos(80\pi x) \cos(80\pi z).
\end{equation}
The results obtained for the case 5 are presented in Fig.~\ref{fig15}. We see that the instability takes much longer to develop at the new initial conditions. At the fully developed stage, however, the parameters of the wave, such as its period, amplitude, and the spatial shape (compare Figs.~\ref{fig15}b and \ref{fig5}c) remain practically the same.

\newpage

\section{Concluding remarks}
\label{sec:conclusion}
The simulation results reported in this paper confirm the conclusion reached earlier in the studies of other geometries\cite{Weber:2016,Horstmann:2018,Zikanov:2018shallow,Tucs:2018,Molokov:2018,Herreman:2019} that the rolling pad instability is a significant factor affecting performance of liquid metal batteries. In particular, batteries with much smaller density difference across the upper interface ($\Delta \rho_A\ll \Delta \rho_B$), for example the better studied Mg-Sb battery, become unstable and face the danger of a ruptured electrolyte layer if the vertical component of the magnetic field is too large. The instability is caused by growing antisymmetrically coupled interfacial waves, which can be identified as the slow modes in the classification of Ref.~\onlinecite{Horstmann:2018}. The wave amplitude is much larger on the upper than on the lower interface in such solutions, and the behavior is qualitatively similar to the behavior of unstable waves in two-layer aluminum reductions cells.

Much less expected and, to our knowledge, entirely new results are found for the much less explored configurations in which $\Delta \rho_A= \Delta \rho_B$ or $\Delta \rho_A\gg \Delta \rho_B$. We see that the analogy with the aluminum reduction cells would be misleading for such systems. In addition to the slow modes, the symmetrically coupled fast modes\cite{Horstmann:2018} appear in the solution. 
At $\Delta \rho_A= \Delta \rho_B$, only the fast modes are present. Having nearly equal amplitudes, the interfacial waves do not cause significant variations of electrolyte thickness and, therefore, cannot cause the rolling pad instability. 
The interfacial waves that appear in batteries with $\Delta \rho_A\gg \Delta \rho_B$ are superpositions of fast and slow modes, with the latter causing the instability.

The results are consistent with the findings of earlier works,\cite{Weber:2016,Horstmann:2018,Zikanov:2018shallow,Tucs:2018,Molokov:2018,Herreman:2019} in which batteries of other geometries (cylinders or shallow parallelepipeds) were considered. Specifically, qualitative agreement is found with Refs.~\onlinecite{Weber:2016,Horstmann:2018,Zikanov:2018shallow,Tucs:2018,Molokov:2018,Herreman:2019} for the case $\Delta \rho_A\ll \Delta \rho_B$ and with Ref.~\onlinecite{Zikanov:2018shallow} for the case $\Delta \rho_A = \Delta \rho_B$.

Our parametric study is incomplete, since only a small sample of possible combinations of material properties could be covered. Furthermore, comparison of our results with the results of earlier studies confirms the rather evident conclusion that the interfacial wave solutions are critically affected by the shape of the horizontal cross-section of the battery cell. An exhaustive parametric study of the rolling pad instability would be unjustifiably difficult and costly to complete. Analyzing specific shapes and material combinations of the battery versions selected for commercialization appears to be a more promising path of future research.

A practically relevant numerical model of a working battery cell would require accurate description of a large number of diverse physical effects. In addition to the  effects addressed in this paper, the list includes the thermal and concentration convection, diffusive and convective mass transport, surface tension, Tayler instability, electrovortex effect, etc. Full three-dimensional magnetic field has to be included.  No such comprehensive simulations have been done so far. The question whether such a simulation is possible at the current level of computer modeling techniques remains open. The OpenFOAM seems a promising tool due to the abundance of already developed and verified solver modules and the relative easiness of implementation of physical models. Our work and the earlier studies\cite{Weber:2016,Horstmann:2018} suggest that one  computationally particularly challenging task, namely the description of the interface motion and  the associated electric currents, can be accomplished with the help of the OpenFOAM. Acceptable, albeit not perfect accuracy is achieved on a properly refined grid if appropriate modifications of the discretization scheme (see section \ref{sec:method}) are made.

\begin{acknowledgments}
The authors are thankful to Norbert Weber for help with the numerical method and to Norbert Weber, Tom Weier, Valdis Bojarevics, and Gerrit Horstmann  for interesting and stimulating discussions. Financial support was provided by the US NSF (Grants CBET 1435269 and CBET 1803730) and the University of Michigan - Dearborn.
\end{acknowledgments}

%

\end{document}